\documentclass[journal=jctcce,manuscript=article,layout=twocolumn]{achemso}

\usepackage[version=3]{mhchem} % Formula subscripts using \ce{}
 \SectionNumbersOn
\usepackage{graphicx}

\usepackage{bm}
\usepackage{braket}
\usepackage{amsfonts}
\usepackage{amsmath}
\usepackage{amssymb}
\usepackage{cleveref}
\usepackage[varg]{txfonts}
\usepackage[font=footnotesize,labelfont=bf]{caption}
\usepackage[small]{titlesec}

\usepackage{xcolor}
\usepackage[section]{placeins}
\definecolor{colorbg}{RGB}{139, 233, 253}
\definecolor{colorkeyone}{RGB}{183, 28, 28}
\definecolor{colorkeytwo}{RGB}{25, 118, 210}
\definecolor{colorkeythree}{RGB}{46, 125, 50}
\definecolor{colorkeyfour}{RGB}{197, 96, 0}
\definecolor{colorkeyfive}{RGB}{64, 36, 26}
\usepackage{tikz}
\usetikzlibrary{arrows, shapes, backgrounds}

\definecolor{frgreen}{RGB}{0, 165, 0}

\usepackage{algorithm2e}
\SetInd{0.1em}{0.8em}

\SetKwFor{ForEach}{\textcolor{colorkeytwo}{foreach}}{\textcolor{colorkeytwo}{do}}{\textcolor{colorkeytwo}{end}}%
\SetKwFor{For}{\textcolor{colorkeyfour}{for}}{\textcolor{colorkeyfour}{do}}{\textcolor{colorkeyfour}{end}}%
\SetKwFor{While}{\textcolor{colorkeythree}{while}}{\textcolor{colorkeythree}{do}}{\textcolor{colorkeythree}{end}}%
\SetKwIF{If}{ElseIf}{Else}{\textcolor{colorkeyone}{if}}
{\textcolor{colorkeyone}{then}}{\textcolor{colorkeyone}{else if}}
{\textcolor{colorkeyone}{else}}{\textcolor{colorkeyone}{end if}}

\SetCommentSty{mycommfont}

\tikzstyle{startstop} = [rectangle, rounded corners, 
minimum width=4cm, 
minimum height=1cm,
text centered, 
draw=black, 
fill=red!30]

\tikzstyle{io} = [trapezium, 
trapezium stretches=true, % A later addition
trapezium left angle=70, 
trapezium right angle=110, 
minimum width=4cm, 
minimum height=1cm, text centered, 
text width=4cm, 
draw=black, fill=blue!30]

\tikzstyle{process} = [rectangle, 
minimum width=4cm, 
minimum height=1cm, 
text centered, 
text width=3cm, 
draw=black, 
fill=orange!30]

\tikzstyle{decision} = [diamond, 
minimum width=3cm, 
text centered, 
draw=black,
text width=2cm, 
fill=green!20]
\tikzstyle{arrow} = [very thick,->,>=stealth]
\usepackage{booktabs}

\usepackage{float}
\author{Jannis Kockl{\"a}uner}
\affiliation{Faculty of Chemistry and Food Chemistry, Technische Universit\"at Dresden, 01062 Dresden, Germany}
\email{jannis.kocklaeuner@tu-dresden.de}
\author{Dorothea Golze}
\affiliation{Faculty of Chemistry and Food Chemistry, Technische Universit\"at Dresden, 01062 Dresden, Germany}
\email{dorothea.golze@tu-dresden.de}

\title[gwc]{\textit{GW} plus cumulant approach for predicting core-level shake-up satellites in large molecules.}

\abbreviations{GW,HPC,FHI-aims}

\let\oldmaketitle\maketitle
\let\maketitle\relax
\begin{document}
\linespread{1.1}
\fontsize{10}{12}\selectfont
\twocolumn[
  \begin{@twocolumnfalse}
    \oldmaketitle
    \begin{abstract}
\fontsize{10}{12}\selectfont
Recently, the $GW$ approach has emerged as a valuable tool for computing deep core-level binding energies as measured in X-ray photoemission spectroscopy. However, $GW$ fails to accurately predict shake-up satellite features, which arise from charge-neutral excitations accompanying the ionization. In this work, we extend the $GW$ plus cumulant ($GW+C$) approach to molecular 1s excitations, deriving conditions under which $GW+C$ can be reliably applied to shake-up processes. We present an efficient implementation with $O(N^4)$ scaling with respect to the system size $N$, within an all-electron framework based on numeric atom-centered orbitals. We demonstrate that decoupling the core and valence spaces is crucial when using localized basis functions. Additionally, we meticulously validate the basis set convergence of the satellite spectrum for 65 spectral functions and identify the importance of diffuse augmenting functions. To assess the accuracy, we apply our $GW+C$ scheme to $\pi$-conjugated molecules containing up to 40 atoms, predicting dominant satellite features within 0.5~eV of experimental values.
For the acene series, from benzene to pentacene, we demonstrate how $GW+C$ provides critical insights into the interpretation of experimentally observed satellite features. %Our $GW+C$ implementation scales as $O(N^4)$ with respect to the system size $N$ and produces spectral functions with negligible computational overhead after the quasiparticle evaluation.% This enables satellite calculations for molecules with more than 100 atoms.
    \end{abstract}
  \end{@twocolumnfalse}
  ]
\begin{tocentry}
\centering
\includegraphics[width=1.0\columnwidth]{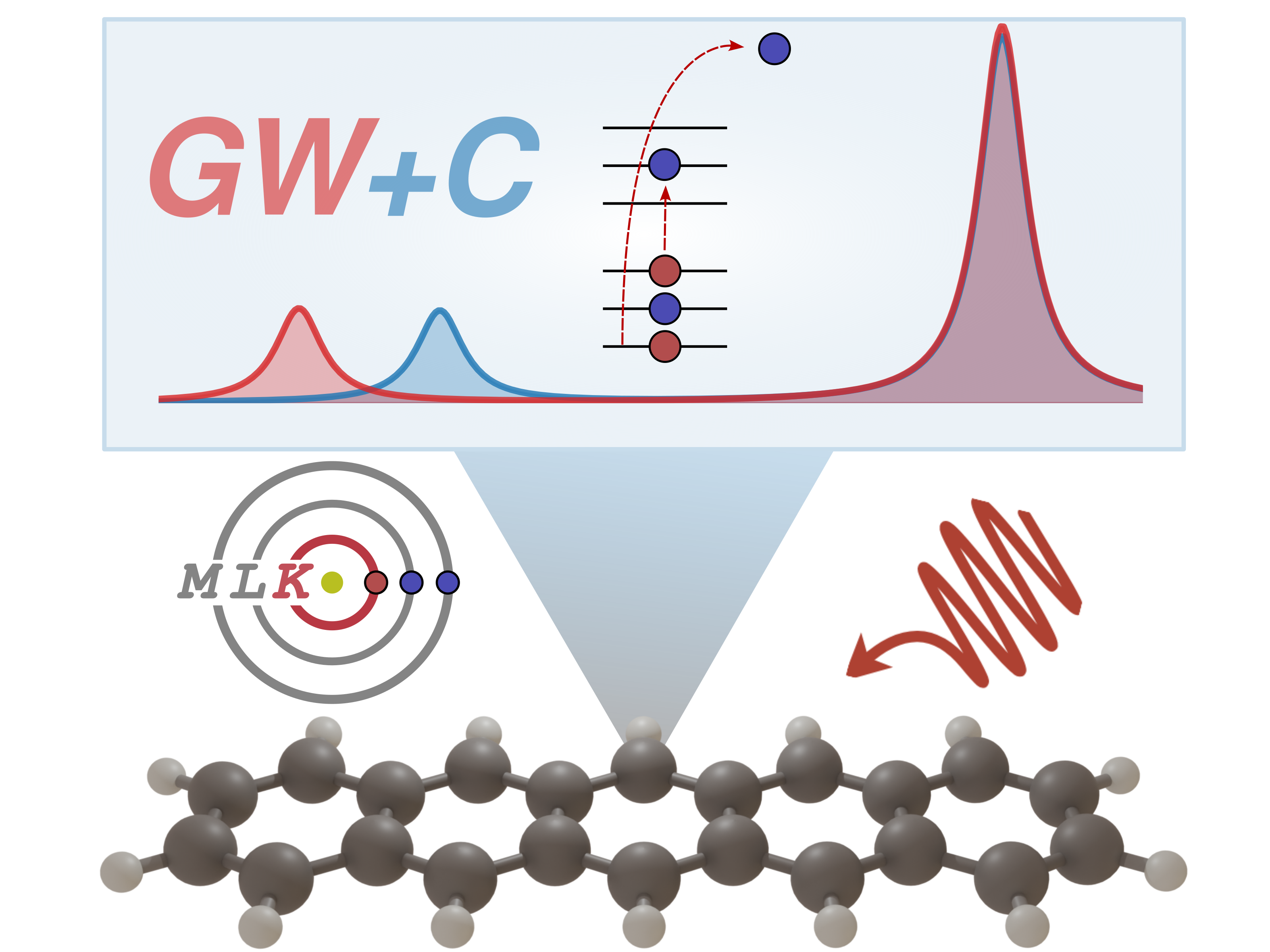}
\end{tocentry}
\section{Introduction}
X-ray photoelectron spectroscopy (XPS) probes strongly localized deep core-levels and is routinely used for chemical analysis.\cite{fadley2010x, bagus2013interpretation, bagus2018extracting}
In XPS spectra, satellite features appear alongside the main lines, which correspond to core-level binding energies. We distinguish between  shake-off, plasmon, or shake-up satellites. Shake-offs are additional charged excitations accompanying the photoionization process, giving rise to broad spectral features.
Plasmons are collective electronic oscillations predominantly found in solid-state systems, while shake-up excitations involve coupled electron-hole pairs and are typically observed in molecular spectra. However, for large molecules the distinction between the two phenomena can be difficult.\cite{bernadotte2013plasmons}
In this work we focus on shake-up satellites. Particularly strong shake-up features have been reported in transition metal complexes and in molecules containing large $\pi$-electron systems.\cite{van1993nonlocal, de2008core, martin1976theory}
\par
Shake-up excitations involve coupled many-electron interactions, requiring an explicit treatment of correlation effects. These effects are not captured by mean-field methods like Hartree-Fock (HF) or Kohn-Sham density functional theory (DFT).\cite{cederbaum1974application, martin1976theory} Instead, various correlated methods have been used to simulate shake-up satellites. Cederbaum and co-workers pioneered the use of Green's function methods for molecular systems, applying techniques like the algebraic diagrammatic construction (ADC) scheme\cite{cederbaum1977theoretical, cederbaum1986correlation, norman2018simulating} and the two-particle-hole Tamm-Dancoff approximation (2ph-TDA) to study valence- and core-level ionization spectra of small molecules, though with mixed success.\cite{schirmer1978two, cederbaum1980many, von1984computational}
\par
Wave function-based approaches have proven to be highly accurate for modeling shake-up processes, including configuration interaction\cite{brisk1975shake, arneberg1982configuration, lunell1989semiempirical,chatterjee2019second, marie2024reference}, symmetry-adapted cluster configuration interaction (SAC-CI)\cite{nakatsuji1991description, kuramoto2005theoretical, ehara2005sac} and quasi-degenerate perturbation theory configuration interaction (QDPTCI).\cite{lisini1992calculation, lisini1993quasidegenerate, fronzoni1999theoretical}
In addition, coupled-cluster techniques like (real-time) equation-of-motion coupled cluster Green's function approaches\cite{rehr2020equation,vila_real-time_2020, vila2021equation, vila2022real} have been developed, using a mix of wave function and propagator techniques.
Despite the variety of methodologies available, accurate \textit{ab-initio} simulations for shake-up satellites have only been reported for small molecular systems, such as water\cite{vila_real-time_2020, sankari2006high, cederbaum1980many}, carbon monoxide\cite{schirmer1987high, angonoa1987theoretical, fronzoni1999theoretical, ehara2006c1s}, and formaldehyde\cite{kuramoto2005c, angonoa1989theoretical, pathak2023real}.
\par
As an alternative, the $GW$ approximation, introduced by Hedin in 1965,\cite{hedin1965new} offers a framework for the computation of the one-particle propagator, including correlation effects in a perturbative way.\cite{onida_electronic_2002, martin_interacting_2016, golze_gw_2019}
The $GW$ method is a powerful and efficient tool for predicting ionization energies and is widely regarded as the gold standard for simulating photoemission spectra in both solid-state\cite{golze_gw_2019} and, more recently, molecular systems.\cite{van2015gw,bruneval2021gw}
While the $GW$ method excels at predicting ionization energies, it falls short for satellite features, predicting too few satellites and with incorrect energies.\cite{langreth1970singularities, martin_interacting_2016}
This limitation of $GW$ is attributed to the neglect of vertex corrections, necessitating approaches beyond $GW$ for accurately simulating shake-up satellites.\cite{martin_interacting_2016}
\par
A systematic way to add vertex corrections for the satellites is to combine the $GW$ self-energy with a cumulant $ansatz$ for the propagator, resulting in the $GW+C$ approximation.
The $GW+C$ method has been successfully applied to plasmonic satellites in the valence spectra of various materials, such as sodium\cite{aryasetiawan1996multiple, zhou_dynamical_2015, vigil2016dispersion, mcclain2016spectral, zhou2018cumulant}, aluminium\cite{aryasetiawan1996multiple,vos1999determination, vos2002quantitative, kas2016particle}, silicon\cite{guzzo2011valence, kheifets2003spectral, lischner2015satellite, caruso2015band, gumhalter2016combined}, graphene\cite{lischner2013physical, guzzo2014multiple} and more\cite{kas2016particle, nakamura2016ab}, showing good agreement with experiment.
Despite its success, we are aware of only one attempt to apply the $GW+C$ approach to molecules, targeting the valence states of benchmark systems like methane, water and neon.\cite{loos2024cumulant}
In this work, we advance the $GW+C$ approach for the computation of satellites in  deep core-level spectra of molecular systems.
\par
The application of $GW$ to deep core-levels is a rather recent development.\cite{golze2018core, golze2020accurate, li2022benchmark,golze2022accurate,zhu2021all, mejia2021scalable, mejia2022basis, bruneval2024fully, aoki2018accurate, van2018assessing, keller2020relativistic, duchemin2020robust,galleni2022modeling, galleni2024c1s, tolle2024ab} We showed that 
the following points are important for core-level $GW$ calculations:
The complex structure of the self-energy in the core region demands the use of a highly accurate technique for the frequency integration, e.g. the contour deformation (CD) approach or an fully analytical approach.\cite{golze2018core, golze2020accurate, li2022benchmark}
Furthermore, self-consistency in $G$ is required to ensure a distinct quasiparticle solution, as the one-shot scheme shifts the quasiparticle into the satellite region, causing the satellites to gain artificial spectral weight.\cite{golze_gw_2019,golze2020accurate,li2022benchmark}
\par
Core-level $GW$ calculations with CD scale $O(N^5)$ with respect to system size $N$.\cite{golze2018core} We recently proposed the CD-WAC method (CD with $W$ analytic continuation),\cite{panades2023accelerating} achieving a  scaling reduction to $O(N^4)$ for core-levels, while retaining the CD accuracy.
In addition, we introduced the $G_{\Delta H}W_0$ variant, which mimics eigenvalue self-consistency $G$ with a constant Hedin shift $\Delta H$. $G_{\Delta H}W_0$  effectively restores the quasiparticle peak,\cite{li2022benchmark} at a fraction of the computational cost of a partial self-consistent scheme.
\par
%Building up on the computational framework we established for molecular core-levels and the recent algorithmic advances in the calculation of the self-energy, we
Building on our core-level $GW$ developments,\cite{golze2018core,golze2020accurate,li2022benchmark,panades2023accelerating} we introduce a $GW+C$ scheme within an all-electron framework using localized basis functions. This work reports the first application of $GW+C$ to molecular shake-up satellites accompanying 1s excitations. We derive conditions for applying $GW+C$ to shake-up processes, along with numerical settings that include recommendations on basis sets, starting points, and other factors affecting numerical stability. To the best of our knowledge, we also present the largest full \textit{ab initio} calculations of core-level shake-up satellites to date, applying our $GW+C$ approach to organic molecules with up to 40 atoms.
\par
The paper is structured as follows: In Section~\ref{sec:theory} and \ref{sec:gwc}, we summarize the key equations of $GW$ and $GW+C$ theory for deep core-levels.
In Section~\ref{sec:technical_details}, we present the final working equations and implementation together with a brief overview of the low-scaling CD-WAC method. 
The relevant technical settings are summarized in Section\ref{sec:ComputationalDetails}.
In Section~\ref{sec:results}, we provide a thorough discussion of technical aspects of $GW+C$. We then assess the accuracy of the method by comparing the $GW+C$ predictions to experiment and $GW$. 
Finally, we summarize our key findings and give an outlook on future developments in Section~\ref{sec:Conclusion}.

\section{\textit{GW} theory}
\label{sec:theory}

\subsection{$\boldsymbol{G_0W_0}$ approximation}
\label{subsec:gw}
The $GW$ approximation is a popular many-body perturbation theory approach derived from Hedin's equations by neglecting all vertex corrections.\cite{hedin1965new, onida_electronic_2002, martin_interacting_2016, golze_gw_2019}
In $G_0W_0$, the $GW$ solution is approximated by using only the first iteration of the $GW$ equations starting from a mean-field reference, e.g. a DFT or HF reference.
In the time-ordered formalism, the frequency-dependent mean-field propagator $G_{0, n}(\omega)$ is defined as 
\begin{align}    
  \label{propagator_basis}
    G_{0, n}(\omega)&=\bra{\psi_{n}}G_{0}(\omega)\ket{\psi_{n}} \\
    \label{dft_propagator}
    G_{0, n}(\omega)&=\frac{1}{\omega - \epsilon_n - \mathrm{i}\eta \,\mathrm{sgn}(\epsilon_\mathrm{F} - \epsilon_n)}
\end{align}
Here, $\epsilon_\mathrm{F}$ is the Fermi level and $\eta$ is an infinitesimal broadening parameter.
$G_{0, n}(\omega)$ is diagonal in the basis of molecular orbitals $\psi_{n}(\mathbf{r})$, where $n$ is the index of a molecular orbital with an energy $\epsilon_n$.\cite{golze_gw_2019}
The molecular orbitals are expanded in a finite set of atomic orbitals $\varphi_{m}$
\begin{equation}
    \label{basis_set_expansion}
    \psi_{n}(\mathbf{r})=\sum_m c_{mn}\varphi_{m}(\mathbf{r})
\end{equation}
where $c_{mn}$ are expansion coefficients determined by a mean-field calculation.
In the following, we use the common notation that arbitrary molecular orbitals are indexed as $m, n$, while occupied (unoccupied) orbitals are specified by indices $i, j$ ($a, b$).
\par
The central quantity in $G_0W_0$ is the dynamic self-energy $\Sigma^{G_0W_0}(\omega)$, which is computed from a mean-field propagator $G_0$ and the screened Coulomb interaction $W_0$. 
In the basis of molecular orbitals, the diagonal elements $\Sigma_{n}^{G_0W_0}(\omega)$ are given by% obtained as
\begin{equation}
  \label{gw_sigma}
    \Sigma_{n}^{G_0W_0}(\omega)=\frac{\mathrm{i}}{2\pi}\sum_{m}\int \mathrm{d}\omega^\prime G_{0, m}(\omega+\omega^\prime)W_{0, mn}(\omega^\prime)
\end{equation}
The screened Coulomb interaction $W_0$ is computed in the random-phase approximation (RPA).
The correlation part $W^c = W-v$ is given by
\begin{align}
\label{W_spectral}
W_{mn}^c(\omega) &=  \sum_\nu  \frac{\rho^\nu_{mn}\rho^\nu_{nm}}{\omega - \Omega^\nu + \mathrm{i}\eta} - \frac{\rho^\nu_{mn}\rho^\nu_{nm}}{\omega + \Omega^\nu - \mathrm{i}\eta} 
\end{align}
where $v$ denotes the bare Coulomb interaction. $\Omega^{\nu}$ and $\rho_{mn}^{\nu}$ are the RPA excitations energies and transition moments.\par

The RPA excitation energies are obtained as solution of an eigenvalue equation in the subspace of single particle excitations from occupied orbitals $i$ to virtual orbitals $a$. 
\begin{equation}
  \label{rpa_casida}
  \begin{bmatrix}
    A & B \\
    -B & -A
  \end{bmatrix}
  \begin{bmatrix}
    X^\nu \\
    Y^\nu 
  \end{bmatrix} = \Omega^\nu  \begin{bmatrix}
    X^\nu \\
    Y^\nu 
  \end{bmatrix}
\end{equation}
The matrices $A$ and $B$ couple singly excited states and are defined as Eq.~\eqref{rpa_matrices}.
\begin{equation}
  \label{rpa_matrices}
  \begin{split}
    A_{ia,jb} &= \delta_{ij}\delta_{ab} (\epsilon_a - \epsilon_i) + (ia|jb) \\
    B_{ia,jb} &= (ia|bj)
  \end{split}
\end{equation}
The integrals $(ia|jb)$ are the standard two-electron Coulomb repulsion integrals in chemists notation.
The transition moments $\rho_{mn}^\nu$ are computed from the excitation vectors $X^\nu,Y^\nu$.
\begin{equation}
  \label{density_shifts}
  \rho_{mn}^\nu = \sum_{ia}(mn|ia)(X^\nu_{ia} + Y^\nu_{ia})
\end{equation}\par
%Within the RPA, the correlation part $W^c = W-v$ of the screened Coulomb interaction is obtained as
%\begin{align}
%\label{W_spectral}
%W_{mn}^c(\omega) &=  \sum_\nu  \frac{\rho^\nu_{mn}\rho^\nu_{nm}}{\omega - \Omega^\nu + \mathrm{i}\eta} - \frac{\rho^\nu_{mn}\rho^\nu_{nm}}{\omega + \Omega^\nu - \mathrm{i}\eta} 
%\end{align}
%where $v$ denotes the bare Coulomb interaction.
Using Eqs.~\eqref{dft_propagator} and ~\eqref{W_spectral}  , the frequency integral in Eq.~\eqref{gw_sigma} is obtained in the form that we will refer to as fully analytical approach.
The self energy is split in a static exchange part $\Sigma^{x, G_0W_0}_n$ and a frequency-dependent correlation part $\Sigma^{c, G_0W_0}_n(\omega)$.
\begin{align}
    \label{sigma_parts}
  \Sigma_{n}^{G_0W_0}(\omega)&=\Sigma^{x, G_0W_0}_n + \Sigma^{c, G_0W_0}_n(\omega) \\
  \label{sigma_x}
  \Sigma^{x, G_0W_0}_n &= -\sum_{m}\left(nm|mn\right)\\
   \label{sigma_spectral}
  \Sigma^{c, G_0W_0}_n(\omega)&=\sum_{m, \nu} \, 
  \frac{\rho^\nu_{mn}\rho^\nu_{nm}}{\omega - \epsilon_m + (\Omega^\nu - \mathrm{i}\eta) \mathrm{sgn}(\epsilon_\mathrm{F} - \epsilon_{m})}
\end{align}\par
The interacting propagator $G_n$ is obtained from a Dyson equation:
\begin{align}
  \label{g_dyson}
  G_n(\omega) &= G_{0, n}(\omega) +  G_{0, n}(\omega) \Sigma_{n}^{G_0W_0}(\omega)G_n(\omega) \\
  \label{g_dyson_expanded}
  &= G_{0, n}(\omega) +  G_{0, n}(\omega) \Sigma_{n}^{G_0W_0}(\omega)G_{0, n}(\omega) + \dots 
\end{align} 
By inverting Eq.\eqref{g_dyson} and inserting Eq.~\eqref{dft_propagator}, the following compact expression for the propagator is derived
\begin{equation}
  \label{gw_full_propagator}
  G_n(\omega) = \left[\omega - \epsilon_n + v_{n}^{xc} - \Sigma^{G_0W_0}_n(\omega)\right]^{-1}
\end{equation}
where $v_n^{xc}=\bra{\psi_n}v^{xc}\ket{\psi_n}$ is the approximate mean-field exchange-correlation potential which has to be removed to avoid double-counting of exchange- and correlation effects.
\par
The poles of $G_n(\omega)$ are the charged excitations of the system within the $G_0W_0$ approximation and can be directly linked to the experiment using the spectral function $A_n^{G_0W_0}(\omega)$.
\begin{equation}
  \label{A_definition}
  A_n^{G_0W_0}(\omega) = \frac{1}{\pi}\left|\mathrm{Im} \, G_n(\omega)\right|
\end{equation}
By inserting Eq.~\eqref{gw_full_propagator} in Eq.~\eqref{A_definition} and dividing the self-energy in its real- and imaginary part, the spectral function can be rewritten as 
\begin{equation}
\label{A_explicit_real_imaginary}
% Doublecheck
A_n^{G_0W_0}(\omega)=\frac{1}{\pi}\frac{\left|\mathrm{Im}\Sigma_n^{G_0W_0}(\omega)\right|}{\left[\omega - \epsilon_n + v_{xc} - \mathrm{Re}\Sigma_n^{G_0W_0}(\omega)\right]^2 + \left[\mathrm{Im}\Sigma_n^{G_0W_0}(\omega)\right]^2}
\end{equation}
The main peaks in Eq.~\eqref{A_definition} appear at the quasiparticle (QP) energies $\epsilon^\mathrm{QP}_n$. The spectral weight of the QP peaks in Eq.~\eqref{A_explicit_real_imaginary} is large, as $\mathrm{Im}\,\Sigma_n(\omega)$ is typically small in this region.
\par
Instead of evaluating $A_n^{G_0W_0}(\omega)$ over a broad frequency range, $\epsilon^\mathrm{QP}_n$ can be obtained as solution of a non-linear equation.
\begin{equation}
  \label{quasiparticle_equation}
  \epsilon_n^{\mathrm{QP}} = \epsilon_n + \mathrm{Re}\Sigma_{n}^{G_0W_0}\left( \epsilon_n^{\mathrm{QP}} \right) - v_{n}^\mathrm{xc}
\end{equation}
Eq.~\eqref{quasiparticle_equation} is solved iteratively, which typically takes around 10 steps and is computationally much cheaper compared to the computation of $A_n^{G_0W_0}(\omega)$ over a large frequency range.

\subsection{Satellites in the $\boldsymbol{G_0W_0}$ approximation}
\label{subsec:gw_satellites}

\begin{figure}[!h]
  \centering
  \includegraphics[width=\linewidth]{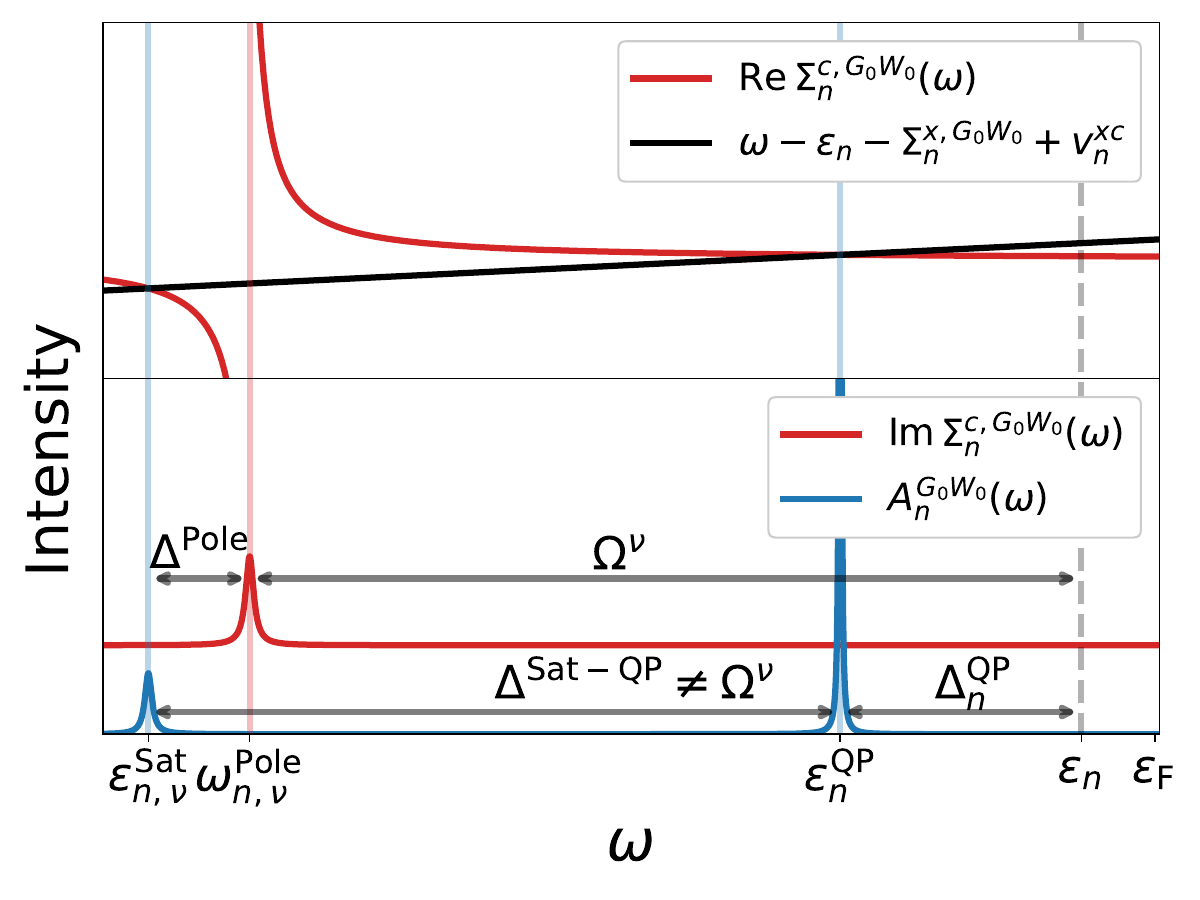}
  \caption{Sketch of the $G_0W_0$ solution for a system with a single satellite. The upper part shows the approximate graphical solution of Eq.~\eqref{quasiparticle_equation} (see Ref.~\citenum{golze_gw_2019}), with the two solutions highlighted in blue. The bottom part displays the resulting spectral function along with $\mathrm{Im}\,\Sigma_n^{c, G_0W_0}(\omega)$.}
  \label{fig:gw_sketch}
\end{figure}
The $G_0W_0$ spectral function exhibits QP peaks corresponding to the ionization process, along with additional satellite features of plasmonic or shake-up character. Satellites are expected to appear roughly at $\epsilon_n^\mathrm{QP} - \Omega^\nu$ because these satellites result from charge-neutral excitations $\Omega^{\nu}$ coupling to the ionization process. $G_0W_0$ provides a poor qualitative and quantitative description of the satellites. To facilitate the discussion, we plot in Figure~\ref{fig:gw_sketch} the real and imaginary parts of $\Sigma_n^{c, G_0W_0}(\omega)$, along with the spectral function $A^{G_0W_0}_n(\omega)$, for a system featuring one QP peak and one satellite.\par 
First, the $G_0W_0$ satellites are not defined relative to the QP peak positions $\epsilon_n^{\mathrm{QP}}$, but relative to the DFT positions $\epsilon_n$. This is because, in $G_0W_0$, satellites occur near frequencies where the real part of $\Sigma^c$ has poles, and its imaginary part exhibits complementary peaks, see Figure~\ref{fig:gw_sketch}. If $n$ corresponds to an occupied level, these poles in $\mathrm{Re}\,\Sigma^c_n$ are located at $\omega^\mathrm{Pole}_{n,\nu} = \epsilon_n - \Omega^\nu$ rather than $\epsilon_n^{\mathrm{QP}} - \Omega^\nu$. Using a functional based on the generalized gradient approximation (GGA) or a hybrid functional with low amount of exact exchange as starting point, the QP correction $\Delta^\mathrm{QP}_n=\epsilon_n^\mathrm{QP}-\epsilon_n$ is negative and the satellites are generally too close to the QP peak.\par
Second, the satellites in the spectral function do not occur at $\omega^\mathrm{Pole}_{n,\nu} = \epsilon_n - \Omega^\nu$, but at more negative frequencies, as shown in the lower part of Figure~\ref{fig:gw_sketch}. In other words, peaks in $\text{Im}\,\Sigma^c_n$ do not directly translate into peaks in $A_n^{G_0W_0}$. This can be seen in Eq.~\eqref{A_explicit_real_imaginary}, where the denominator introduces an offset, labeled as $\Delta^\mathrm{Pole}$ in Figure~\ref{fig:gw_sketch}. Consequently, the interpretation of a charge-neutral excitation coupling to an ionization —while still present in the self energy, albeit with an incorrect reference — is lost in the spectral function.\par
Third, $G_0W_0$ includes only single RPA excitations, while higher-order excitations require additional vertex corrections. As a result, it has been shown for solid-state systems that $G_0W_0$ does not produce the expected series of plasmonic satellites but instead yields a spurious single 'plasmaron'.\cite{langreth1970singularities,aryasetiawan1996multiple, guzzo2011valence,guzzo2012dynamical} In Section~\ref{subsec:validity_gwc}, we will demonstrate that $G_0W_0$ also generates too few peaks in the shake-up case.\par
In summary, $G_0W_0$ predicts too few satellites at too negative frequencies (or equivalently too high binding energies) with a satellite QP-splitting of $\Delta^\mathrm{Sat-QP}=\epsilon^\mathrm{Sat}_{n,\nu}-\epsilon^\mathrm{QP}_{n} = -\Omega + |\Delta_n^{\mathrm{QP}}| - |\Delta^{\mathrm{Pole}}|$ with generally $ |\Delta_n^{\mathrm{QP}}| > |\Delta^{\mathrm{Pole}}|$ for core levels.

%%%%
\subsection{\textit{GW} for core-levels: Beyond $\boldsymbol{G_0W_0}$}
\label{subsec:corelevels}
We demonstrated that $G_0W_0$, when starting from a GGA reference such as the Perdew-Burke-Ernerzhof (PBE)~\cite{perdew1996generalized} functional, fails to produce a unique QP solution for molecular 1s excitations.~\cite{golze2020accurate,li2022benchmark} Instead, multiple solutions with equal spectral weight are obtained.~\cite{golze2020accurate} This is in striking contrast to experimental results~\cite{gelius1970molecular}, which show a strong 1s main excitation. For small molecules, the intensities of the satellite features are typically orders of magnitude smaller compared to the the main line.\cite{lunell1978theoretical, schirmer1987high, sankari2006high}
\par
The failure of $G_0W_0@$PBE for deep core levels arises from the incorrect positioning of the satellites. While $\left|\Delta^\mathrm{QP}_n\right|$ typically ranges from 1–2~eV for valence levels\cite{golze_gw_2019}, it increases to 20–35~eV for 1s excitations of second-row elements\cite{golze2018core}. For core levels, $\left|\Delta^\mathrm{QP}_n\right|$ can approach or even exceed $\Omega^\nu$. In the former case, the satellite-QP separation becomes very small, whereas in the latter, satellites may appear at lower binding energies (more positive frequencies) than the QP peak — or, in other words, the QP peak is located between the satellites. Both scenarios result in an artificial transfer of spectral weight from the QP peak to the satellites. We observed the second scenario for every single 1s excitation in the CORE65 benchmark set~\cite{golze2020accurate} because the optical gap ($\approx \Omega^1$) is in the range of $5-15$~eV and thus always smaller than $\left|\Delta^\mathrm{QP}_n\right|$. 
\par
We showed that including eigenvalue self-consistency in the propagator $G$ solves this problem.\cite{golze2020accurate, li2022benchmark}
In the eigenvalue self-consistent $\mathrm{ev}GW_0$ method, the DFT orbital energies in Eq.~\eqref{sigma_spectral} are replaced with the current approximation for the QP energies $\epsilon_m^\mathrm{QP}$ in each iteration, while the excitations $\Omega^\nu$ are kept at the RPA level.\cite{golze_gw_2019, golze2020accurate}
The correlation part of the ev$GW_0$ self-energy is given by
\begin{equation}
  \label{ev_sigma_spectral}
  \Sigma_{n}^{\mathrm{c, \mathrm{ev}}GW_0}(\omega) = \sum_{m, \nu} \, \frac{\rho^\nu_{mn}\rho^\nu_{nm}}{\omega - \epsilon_m^\mathrm{QP} + \left(\Omega^\nu - \mathrm{i}\eta \right) \mathrm{sgn}\left(\epsilon_\mathrm{F} - \epsilon_m^\mathrm{QP}\right)}
\end{equation}
The poles of $\Sigma_{n}^{\mathrm{c, \mathrm{ev}}GW_0}(\omega)$ are at $\epsilon_m^\mathrm{QP} \pm \Omega^\nu$, thus the satellites are placed relative to the QP energies $\epsilon_m^\mathrm{QP}$.
This effectively restores the QP peak in core-levels, yielding absolute core-level binding energies within $0.3\,\mathrm{eV}$ of the experiment for $\mathrm{ev}GW_0$@PBE.\cite{golze2020accurate, li2022benchmark}
The $\mathrm{ev}GW_0$ scheme comes at the price of increased computational cost compared to $G_0W_0$, since \textit{all} $\epsilon_m^\mathrm{QP}$ need to be recalculated for every iteration of Eq.~\eqref{quasiparticle_equation}.
\par
The $\mathrm{ev}GW_0$ scheme can be approximated by applying a global so-called ``Hedin shift" $\Delta H$ to the self-energy, which is calculated once and is kept constant during the iterations of Eq.~\eqref{quasiparticle_equation}.\cite{hedin1999correlation}
For deep core-levels, we previously demonstrated that a level-specific shift $\Delta H_n$ instead the global shift is required to approximate the $\mathrm{ev}GW_0$ solution adequately.\cite{li2022benchmark}
$\Delta H_n$ is obtained from the first iteration of $\mathrm{ev}GW_0$ as
\begin{equation}
  \label{hedin_shift}
  \Delta H_n = \mathrm{Re}\Sigma^{c, G_0W_0}_{n}(\epsilon_n) + \Sigma^{x, G_0W_0}_{n} - v_{n}^\mathrm{xc}
\end{equation}
The Hedin shift enters the correlation part of the self-energy $\Sigma_{n}^{c, G_{\Delta H}W_0}(\omega)$ as
\begin{equation}
  \label{hedin_sigma_spectral}
  \begin{split}
  &\Sigma_{n}^{c, G_{\Delta H}W_0}(\omega) = \Sigma_{n}^{c, G_0W_0}(\omega - \Delta H_n)\\
  &=\sum_{m,\nu} \, \frac{\rho^\nu_{mn}\rho^\nu_{nm}}{\omega - \epsilon_m - \Delta H_n + \left(\Omega^\nu - \mathrm{i}\eta\right) \mathrm{sgn}\left(\epsilon_\mathrm{F} - \epsilon_m^\mathrm{QP}\right)}
  \end{split}
\end{equation}
The poles in $\Sigma_{n}^{c, G_{\Delta H}W_0}(\omega)$, and consequently the satellites, are shifted by $\Delta H_n\approx \Delta^{\mathrm{QP}}_n$.
We showed that $\Sigma_{\mathrm{1s}}^{c, G_{\Delta H}W_0}$ closely resembles $\Sigma_{\mathrm{1s}}^{\mathrm{c, \mathrm{ev}}GW_0}$ and that the 1s QP energies from $G_{\Delta H}W_0@$PBE are comparable to those from $\mathrm{ev}GW_0@$PBE, with absolute core-level binding energies deviating by $0.25~\mathrm{eV}$ from experiment.\cite{li2022benchmark}
\par
Using an approximate self-consistent scheme for $G$ resolves the first problem discussed in Section~\ref{subsec:gw_satellites}: the poles in the self-energy are positioned relative to the QP energies rather than the DFT energies. However, the second and third problems remain. There are too few satellites, and they appear at more negative frequencies than the poles in the self-energy, with $\Delta^\mathrm{Sat-QP} = -\Omega - |\Delta^\mathrm{Pole}|$. A consistently better description of the satellite region requires additional vertex corrections in $G$, such as those provided by $GW+C$.
\section{\textit{GW+C} for molecules}
\label{sec:gwc}    

\subsection{Cumulant expansion}
\label{subsec:gwc}
In the following, we summarize the key equations of the $GW+C$ approach, typically presented in the literature~\cite{aryasetiawan1996multiple, kas_cumulant_2014, zhou_dynamical_2015, zhou2018cumulant} using a numerical scheme based on the spectral representation of the self-energy. We adopt this numerical scheme in our implementation (see Section~\ref{sec:Implementation}). However, for this section, we use the analytical expressions derived in Ref.~\citenum{ loos2024cumulant}, employing Eq.~\eqref{sigma_spectral}, to facilitate the discussion on the validity of the method for shake-up satellites. The retarded Green's function formalism is employed for the $GW+C$ equations~\cite{kas_cumulant_2014, loos2024cumulant}.\par
The cumulant Green's function $G_n^C(t)$ for a hole takes the form\cite{langreth1970singularities, aryasetiawan1996multiple}
\begin{equation}
\label{cumulant_propagator}
G_{n}^C(t) = G_{0,n}(t)e^{C_{n}(t)}
\end{equation}
where
\begin{equation}
G_{0, n}(t)=-\mathrm{i} \Theta(t) e^{-\mathrm{i}\epsilon_n t}
\label{eq:G0time}
\end{equation}
is the retarded, time-dependent free propagator. $\Theta$ is the Heaviside function and $C_n$ is the cumulant. The combination of Eq.~\eqref{cumulant_propagator} with the $G_0W_0$ self-energy yields the $G_0W_0+C$ approach, which we will refer to as $GW+C$ for brevity.
\par
We start by a Taylor expansion of the exponential ansatz given in Eq.~\eqref{cumulant_propagator}
\begin{equation}
        G_{n}^C(t) = G_{0, n}(t) + C_{n}(t)G_{0,n}(t) + \frac{1}{2} C_{n}(t)^2 G_{0,n}(t) + \dots 
            \label{cumulant} 
\end{equation}
Next, we assume that the cumulant is linear in $W$.\cite{ aryasetiawan1996multiple, zhou_dynamical_2015, vila_real-time_2020}
An approximate expression for the cumulant $C_n(t)$ is derived by comparing Eq.~\eqref{cumulant} with the Dyson equation Eq.~\eqref{g_dyson_expanded} for the terms that are linear in the self-energy. Specifically, we compare the second term of Eqs.~\eqref{cumulant} and \eqref{g_dyson_expanded}:
\begin{align}
        \label{cumulant_gw}
        C_{n} G_{0, n} &\approx G_{0, n} \Sigma^{G_0W_0}_{n} G_{0, n}  
\end{align}
Substituting Eq.~\eqref{eq:G0time} into Eq.~\eqref{cumulant_gw}, the time-dependent cumulant $C_n(t)$ can be linked to $\Sigma^{G_0W_0}_{n}(\omega)$ via a Fourier transform.
\begin{equation}
    \label{cumulant_fourier}
    C_{n}(t) = \mathrm{i}\int \frac{\mathrm{d}\omega}{2\pi} e^{-\mathrm{i}(\omega - \epsilon_n) t} G_{0, n}(\omega) \Sigma^{G_0W_0}_{n}(\omega)G_{0, n}(\omega)
\end{equation}
We insert the retarded $G_{0,n}(\omega)$, which corresponds to Eq.~\eqref{dft_propagator} replacing $ -\mathrm{i}\eta \,\mathrm{sgn}(\epsilon_\mathrm{F} - \epsilon_n)$ by $\mathrm{i}\eta $, and shift the integration variable by $\epsilon_n$ to obtain
\begin{equation}
\label{cumulant_shifted_fourier}
    C_n(t) = \mathrm{i}\int \frac{d\omega}{2\pi} e^{-\mathrm{i} \omega t} \frac{\Sigma^{G_0W_0}_{n}(\omega + \epsilon_n)}{\left(\omega + \mathrm{i}\eta\right)^2}.
\end{equation}
\par
We separate the cumulant into a correlation $C_n^c$ and exchange $C_n^x$ contribution
\begin{equation}
C_n=C_n^c + C_n^x
\label{cumulant_split}
\end{equation}
$C_n^x$ is derived from the exchange part of the self-energy $\Sigma_n^{x, G_0W_0}$. 
\begin{align}
    C_n^x(t) &= \mathrm{i}\int \frac{\mathrm{d} \omega}{2\pi} e^{-\mathrm{i} \omega t} \frac{\Sigma_n^{x, G_0W_0}}{\left(\omega + \mathrm{i}\eta\right)^2}\\
    &= -\mathrm{i}t\Sigma_n^{x, G_0W_0}
    \label{cumulant_sigma_x}
\end{align}
The correlation part $C_n^c(t)$ is derived by inserting the retarded formulation of $\Sigma_n^{c, G_0W_0}$ (same as Eq.~\eqref{sigma_spectral} replacing $ -\mathrm{i}\eta \,\mathrm{sgn}(\epsilon_\mathrm{F} - \epsilon_n)$ by $\mathrm{i}\eta$)
\begin{equation}
  \label{analytic_cumulant}
  % TODO: Find bbreviation for the sign convention to fit the equation here
  C_{n}^{c}(t) = \mathrm{i}\sum_{m, \nu}\int \frac{\mathrm{d}\omega}{2\pi} \frac{e^{-\mathrm{i}\omega t}}{(\omega + \mathrm{i}\eta)^2}  \frac{\rho^\nu_{mn}\rho^\nu_{nm}}{\omega - \Delta_{mn}^\nu}
\end{equation}
where
\begin{equation}
    \label{cumulant_deltas}
    \Delta_{mn}^\nu = \epsilon_m - \epsilon_n - \mathrm{sgn}\left(\epsilon_\mathrm{F}-\epsilon_m\right)\Omega^{\nu}-\mathrm{i}\eta
\end{equation}
We solve the integral in Eq.~\eqref{analytic_cumulant} analytically,\cite{vila_real-time_2020, loos2024cumulant} resulting in the formulation known as the Landau form of the cumulant,\cite{landau1944energy}
\begin{align}
    \label{analytical_fourier}
    C^c_{n}(t) = \sum_{m,\nu}\gamma_{mn}^\nu\left(e^{-\mathrm{i}t \Delta_{mn}^\nu} + \mathrm{i}\Delta_{mn}^\nu t -1\right)%\\
  %  \label{cumulant_gammas}
  %  \gamma_{mn}^\nu &= \frac{\rho_{mn}^\nu\rho_{nm}^\nu}{\left(\Delta_{mn}^\nu\right)^2}
\end{align}
where the excitation weights $\gamma_{mn}^\nu$ are given by
\begin{equation}
  \label{cumulant_gammas}
    \gamma_{mn}^\nu = \frac{\rho_{mn}^\nu\rho_{nm}^\nu}{\left(\Delta_{mn}^\nu\right)^2}  
\end{equation}\par
The cumulant propagator is obtained by inserting Eq.~\eqref{eq:G0time} and Eq.~\eqref{cumulant_split} into Eq.~\eqref{cumulant_propagator} 
\begin{align}
        G_{n}^C(t) & =  -\mathrm{i} \Theta(t) e^{-i\epsilon_n t +  C_n^c(t) + C_n^x(t)}\\
         &= Z_n G_{0, n}^\mathrm{QP}(t) e^{C_n^\mathrm{S}(t)}\label{cumulant_gwc_propagator}
\end{align}
where we used Eqs.~\eqref{analytical_fourier} and Eq.~\eqref{cumulant_sigma_x} to obtain Eq.~\eqref{cumulant_gwc_propagator}. $Z_n$ is a renormalization factor which originates from the time-independent term of $C_n^c(t)$ (Eq.~\eqref{analytical_fourier}).
\begin{equation}
    Z_n = e^{-\sum_{m,\nu}\gamma_{mn}^\nu}
\end{equation}
$G_{0, n}^\mathrm{QP}(t)$ denotes the QP propagator and is given by
\begin{align}
\label{G_QPGWC}
   G_{0, n}^\mathrm{QP}(t) &= -\mathrm{i}\Theta(t)e^{-\mathrm{i}t\epsilon_n^\mathrm{QP}} \\
    \epsilon_n^\mathrm{QP} &= \epsilon_n + \Sigma_n^{x, G_0W_0} + \Sigma_n^{c, G_0W_0}(\epsilon_n) - v_n^{xc} \label{QPGWC}\\
    &= \epsilon_n + \Delta H_n  
\end{align}
where $\Delta H_n$ corresponds to the level-specific Hedin shift defined in Eq.~\eqref{hedin_shift}.
$\epsilon_n^{\mathrm{QP}}$ denotes the QP energy at the $GW+C$ level and corresponds to the $G_{\Delta H}W_0$ QP energy (see Supporting Information (SI) Section~S1). The $G_{0, n}^\mathrm{QP}(t)$ propagator arises from the sum of three terms: i) the contribution $-it\epsilon_n$ from the free propagator $G_{n,0}(t)$ (Eq.~\eqref{eq:G0time}), ii) $C_n^x$ ( Eq.~\eqref{cumulant_sigma_x})  and iii) the term linear in $t$ in Eq.~\eqref{analytical_fourier}, for which we used the identity
\begin{equation}  \label{cumulant_qp_correlation}
    \mathrm{i}t \sum_{m,\nu}\gamma_{mn}^\nu \Delta_{mn}^\nu = -\mathrm{i}t\Sigma^{c, G_0W_0}_n(\epsilon_n)
\end{equation}
Note that we must also subtract $v_n^{xc}$ in Eq.~\eqref{QPGWC} when $\epsilon_n$ is a DFT eigenvalue. Furthermore, $C_n^\mathrm{S}(t)$ in Eq.~\eqref{cumulant_gwc_propagator} is given by 
\begin{align}
   \label{cumulant_satellites}
        C_n^\mathrm{S}(t) &= \sum_{m,\nu}\gamma_{mn}^\nu e^{-i\Delta_{mn}^\nu t} 
\end{align}
which is derived from the exponential term in Eq.~\eqref{analytical_fourier} and is the term which creates the satellites.
\par
Next, we expand $\exp{\{C_n^\mathrm{S}(t)\}}$ in a Taylor series
\begin{equation}
    \label{cumulant_taylor}
    e^{C_{n}^\mathrm{S}(t)} \approx 1 + C_{n}^\mathrm{S}(t) + \frac{1}{2}C_{n}^\mathrm{S}(t)^2 + \dots
\end{equation}
Each individual order of $C_{n}^\mathrm{S}(t)$ gives rise to a series of features in the frequency-dependent propagator. Using Eq.~\eqref{cumulant_taylor}, we obtain the Fourier transform of Eq.~\eqref{cumulant_gwc_propagator} as follows:%, each individual order of $C_{n}^S(t)$ gives rise to a series of features in the frequency-dependent propagator.
\begin{align}
    G_{n}^C(\omega) &\approx  Z_n\int \mathrm{d}te^{i\omega t} G_{0,n}^{\mathrm{QP}}(t)\left(1+C_n^\mathrm{S}(t) + \frac{1}{2}C_n^\mathrm{S}(t)^2+\cdots\right)\\
    &=
    Z_n\left(G_{0, n}^\mathrm{QP}(\omega) + G_n^1(\omega) + \frac{1}{2}G_n^2(\omega) + \dots\right) \label{expanded_propagator}
\end{align}
with
\begin{align}
    G_{0, n}^\mathrm{QP}(\omega) &= \frac{1}{\omega - \epsilon_n^\mathrm{QP} + \mathrm{i}\eta} \label{eq:GQP}\\
    G_{n}^1(\omega) &= \sum_{m\nu} \frac{\gamma_{mn}^\nu}{\omega - \epsilon_n^\mathrm{QP} - \Delta_{mn}^\nu + \mathrm{i}\eta}\label{gwc_first_order}  \\
    \label{gwc_second_order}
    G_{n}^2(\omega) &= \sum_{mo, \nu \mu} \frac{\gamma_{mn}^\nu \gamma_{on}^\mu}{\omega - \epsilon_n^\mathrm{QP} - \Delta_{mn}^\nu - \Delta_{on}^\mu + \mathrm{i}\eta} 
\end{align}
The spectral function is given by 
\begin{equation}
A_n^{GW+C}(\omega)= -\frac{1}{\pi} \text{Im}\,G_n^C(\omega)
\label{eq:AGWC}
\end{equation}
which is slightly different compared to Eq.~\eqref{A_definition} because we formulated the $GW+C$ approach in the retarded and not time-ordered formalism.
\par
The poles in the real part of Eq.~\eqref{expanded_propagator} correspond to peaks in $A_n^{GW+C}$. The analytic expanded form of $G_n^C(\omega)$ directly reveals where and what type of features are expected in the spectral function: $G_{0,n}^{\mathrm{QP}}$, as defined in Eq.\eqref{eq:GQP}, generates the QP peak at $\omega = \epsilon_n^{\mathrm{QP}}$. The first-order term, $G_{n}^1(\omega)$, Eq.\eqref{gwc_first_order}, gives rise to satellite features at $\omega = \epsilon_m - \Omega^\nu + \Delta H_n$, corresponding to a single excitation $\Omega^{\nu}$ coupling to the charged excitation. The second-order term introduces higher-order satellite features, where two uncorrelated charge-neutral excitations, $\Omega^{\nu}$ and $\Omega^{\mu}$, couple to the charged excitation. Triple and higher-order excitations are generated by the propagators $G_n^{3,\cdots}(\omega)$. The intensities of the first, second, third, and higher-order satellites are distributed in a Poisson-like manner, as products of single-excitation intensities. Consequently, second and higher-order satellites only appear if strong single-excitations satellites are present.
\par
The $GW+C$ approach solves the shortcomings $G_0W_0$ discussed in Section~\ref{subsec:gw_satellites}.
First, for $n=m$, satellites now appear at positions relative to the QP peak, rather than relative to the DFT energies. Specifically, the first-order satellites associated with the level $n$ are located at $\omega = \epsilon_n - \Omega^\nu + \Delta H_n = \epsilon_n^{\mathrm{QP}} - \Omega^\nu$. Second, the poles at $\omega=\epsilon_n^{\mathrm{QP}} -\Omega^{\nu}$ appear directly in the propagator $G^C_n$, which means that they directly show up in the spectral function $A_n^{GW+C}$, and we have $\Delta^\mathrm{Sat-QP} = -\Omega^{\nu}$. The third point, concerning the number of satellites, is discussed in Section~\ref{subsec:validity_gwc}.

%\subsection{Identifying bosonic effects in $GW+C$ for a minimal model}
\subsection{Assessing the validity of \textit{GW+C} for a shake-up process in a minimal model}
\label{subsec:validity_gwc}
\begin{figure*}[!ht]
  \centering
  \includegraphics[width=\linewidth]{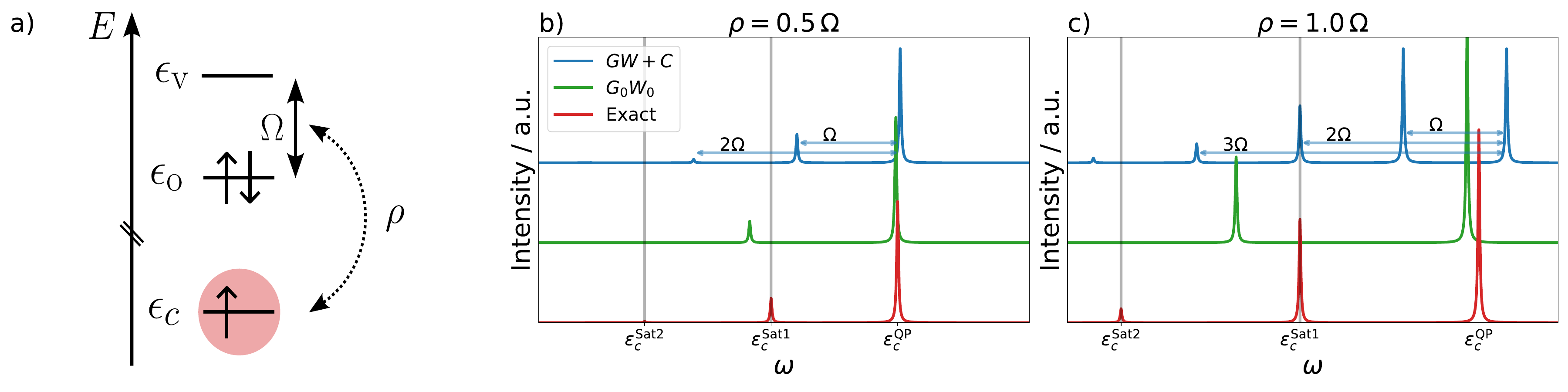}
  \caption{Three-level core-hole model system used by Cederbaum and co-workers in Ref. \citenum{cederbaum1980many}. Figure~\ref{fig:model_system}a sketches the molecular orbital diagram of the model, containing one occupied ($\epsilon_\mathrm{O})$ and one virtual ($\epsilon_\mathrm{V}$) valence level and an ionized core-level $\epsilon_c$. In the CVS approximation, the system has only a single excitation $\Omega$ coupled to the core level with a transition moment $\rho$. Figure~\ref{fig:model_system}a,b shows the $G_0W_0$ and $GW+C$ spectral functions with various relative coupling strengths compared to the exact solution of the system.}
 \label{fig:model_system}
\end{figure*}
The cumulant expansion is based on the exact solution of an infinite bosonic system~\cite{langreth1970singularities}. Collective bosonic excitations (plasmons) are predominantly found in solids, whereas $\Omega^{\nu}$ rather represent fermionic electron-hole pairs in the molecular case.\cite{bernadotte2013plasmons} For fermions, the higher order terms $G_n^{2, 3,\cdots}(\omega)$ contain non-physical excitations which violate the Pauli principle.
Here, we qualitatively discuss this fact for the three-level model system, sketched in Figure~\ref{fig:model_system}a.
This model was already introduced by Cederbaum and co-workers \cite{cederbaum1980many} in 1980. We refer the reader to Ref.~\citenum{cederbaum1980many} for an in-depth discussion of the exact and cumulant solutions.%, similar to $GW+C$.
\par
In the model system, a single core hole is created in an orbital with an unperturbed energy $\epsilon_c$.
The core hole couples to a valence single-particle excitation $\Omega$ from an occupied orbital with energy $\epsilon_o$ to a virtual orbital with energy $\epsilon_v$.
Interactions of the valence space are not included and thus $\Omega = \epsilon_v - \epsilon_o$.
The coupling is mediated by a transition moment $\rho_{cc} =\rho$.
The transition moments mixing core- and valence levels, i.e. $\rho_{co}$ and $\rho_{cv}$, are neglected following the core-valence separation (CVS) approximation.\cite{cederbaum1980many, norman2018simulating}
In the same spirit, charge-neutral core-valence excitations from $c$ to $v$ are omitted.
The exact, $GW+C$ and $G_0W_0$ spectral functions of this model system are shown in Figure~\ref{fig:model_system}b and c for two different ratios of $\rho/\Omega$ (0.5 and 1.0). The ratio $\rho/\Omega$ determines the effective coupling strength between the core hole and the excited state in all three approaches. For $GW+C$, this relationship is directly evident from Eq.~\eqref{cumulant_gammas}, where the excitation weight for the model system is given by $\gamma = \rho^2/\Omega^2$.
\par
Starting with the exact solution, we expect two satellites in addition to the QP peak, $\epsilon_c^{\mathrm{QP}}$: the first satellite at $\epsilon_c^{\text{Sat1}}$ corresponds to a single excitation where one electron is promoted from state $o$ to $v$. The second, $\epsilon_c^{\text{Sat2}}$, corresponds to a double excitation where both electrons in $o$ are excited to $v$. The intensity of the second satellite is significantly smaller because it couples only indirectly with the ground state configuration. Triple- and higher order excitations are not possible, as the model considers only valence shell excitations and the valence shell contains only two electrons. The positions and intensities of the satellites depend on the ratio $\rho/\Omega$. Increasing ratios enhance the intensities of the satellites and increase the separation between QP peak and satellite positions, see Figure~\ref{fig:model_system}b and c.
\par
In the $GW+C$ approximation, each order of $G_n^{1, 2, 3,\cdots}(\omega)$ creates exactly one satellite peak because there is only one excitation weight $\gamma$. 
The satellites occur at $\epsilon_c^{\mathrm{QP}} - n\Omega$ where $n\in \mathbb{N}$, and the intensities follow the Poisson distribution.
The intensity of each satellite is proportional to $\gamma^n$, thus intensities depend on the ratio $\rho/\Omega$, as for the exact solution. For $\rho/\Omega= 0.5$, only the first two satellites (corresponding to single and double excitations) carry considerable spectral weight, whereas four satellites with non-vanishing intensity are visible for $\rho/\Omega= 1.0$. The third and fourth satellite correspond to triple and quadruple excitations, which violate the Pauli principle because we can have only two electrons in state $v$, and only two electrons can be excited from $o$. In addition to this qualitatively wrong description, we find that the positions of the first two satellites deviates strongly from the exact solution for $\rho/\Omega=1.0$, whereas the agreement is better for $\rho/\Omega=0.5$.
\par
In $G_0W_0$ only a single satellite is generated, regardless of the coupling strength. 
The satellite appears at higher binding energies compared to the exact and $GW+C$ solution, consistent with the discussion in Section~\ref{subsec:gw_satellites}.
We emphasize that for this minimal model with only a single higher-order excitation, the error of $G_0W_0$ and $GW+C$ is of comparable magnitude.
For realistic systems with many electrons and a large virtual space, we expect an improvement of $GW+C$, since the number of physical higher-order excitations captured by $GW+C$ grows rapidly with the system size.
\par
To avoid spurious higher-order satellites, one could simply neglect the terms $G_n^{2,3...}$, as done in Ref.~\citenum{loos2024cumulant}.
However, a separation into lower-order terms $G_n^{1}$ and higher order terms $G_n^{2,3...}$ as in Eq.~\eqref{expanded_propagator} is not possible in our numerical implementation for large systems described in Section~\ref{sec:Implementation}. Instead, we derive conditions under which a cumulant approach like $GW+C$ is approximately valid, following the discussion of Cederbaum and co-workers in Ref.~\citenum{cederbaum1980many}.
For the model system, they formulated the following inequality
\begin{equation} 
  \label{core_cumulant_inequality}
  |\rho| \ll \left|\Omega\right| 
\end{equation}
If the inequality is kept, the excitation weights $\gamma$ are small and the intensity of higher order (non-physical) contributions quickly vanishes due to the Poisson distribution. This corresponds rather to a case as shown in Figure~\ref{fig:model_system}b.
\par
In cases where Eq.~\eqref{core_cumulant_inequality} is violated, 
spurious contributions from higher-order terms must be eliminated.
This can be achieved by employing non-linear variants of the cumulant expansion instead of the linear parametrization in Eq.~\eqref{cumulant}.\cite{mahan1982core, tzavala2020nonlinear}
Different paths like connections to real-time time-dependent DFT\cite{tzavala2020nonlinear} and real-time coupled cluster approximations\cite{vila_real-time_2020, vila2021equation, vila2022real} have been proposed and applied to model systems or small molecules. The combination of $GW$ with a non-linear cumulant approach is has not been achieved yet and is also not the goal of the present work. Here, we aim to demonstrate that $GW+C$ provides reliable results for molecular core-level satellites generalizing the inequality~\eqref{core_cumulant_inequality} for the many-electron case.
%%%%
\begin{table}[t]
    \fontsize{10}{12}\selectfont
    \centering
    \begin{tabular}{c c c c c}
    \toprule
         Basis Set & $\Omega^{\mathrm{Sat1}}$ & $\rho^{\mathrm{Sat1}}_\mathrm{C1s, C1s}$  & $\rho^{\mathrm{Sat1}}_\mathrm{C1s, 2}$  & $\rho^{\mathrm{Sat1}}_\mathrm{C1s, 3=4=5}$ \\
    \midrule
         cc-pVTZ& 13.91 & 0.3352 & 5.366$\cdot 10^{-7}$ & 0.0000 \\
         cc-pVQZ&  13.22& 0.2099 & 9.258$\cdot 10^{-7}$& 0.0000 \\
         cc-pV5Z&12.58 & 0.2453&6.413$\cdot 10^{-7}$& 0.0000 \\
    \bottomrule
    \end{tabular}
    \caption{Transition moments $\rho^\mathrm{Sat1}_{nm}$ and excitation energy $\Omega^\mathrm{Sat1}$ of the lowest shake-up satellite with a non-zero coupling in \ce{CH4} computed with the fully analytical $G_0W_0$ implementation in PySCF.\cite{sun2020recent} All values in~eV.}
    \label{tab:cumulant_decoupling}
\end{table}
\subsection{Assessing the validity of \textit{GW+C} for molecular shake-ups}
\label{subsec:correlation_satellites}
For a realistic system with many orbitals, first order satellites in $GW+C$ appear at frequencies
\begin{equation}
\label{gwc_satellite_positions}
    \omega = \epsilon_n^\mathrm{QP} + \epsilon_m - \epsilon_n - \Omega^\nu
\end{equation}
 in accordance with Eq.~\eqref{gwc_first_order}.
The inequality in Eq.~\eqref{core_cumulant_inequality} is then generalized to
\begin{equation} 
\label{general_cumulant_inequality}
  \left|\rho^\nu_{mn}\right| \ll \left|\epsilon_m - \epsilon_n - \Omega^\nu\right| 
\end{equation}

Equations~\eqref{gwc_satellite_positions} and \eqref{general_cumulant_inequality} cover two different types of satellites in the spectral function of a level $n$: Satellites generated by the diagonal elements ($m=n$) and satellites generated by the off-diagonal elements ($m\neq n$). 
For the first case, the DFT orbital energies in Eq.~\eqref{gwc_satellite_positions} vanish, and the satellites are placed relative to the QP peak at frequencies $\omega=\epsilon^\mathrm{QP}_n - \Omega^\nu$. 
These satellites correspond to shake-up states, i.e. charge neutral excitations coupled to the ionized (core) level $n$. We will show this explicitly in Section~\ref{subsec:decoupling}.
\par
We provide an example demonstrating that the inequality~\eqref{general_cumulant_inequality} is satisfied for the shake-up satellites. Table~\ref{tab:cumulant_decoupling} lists the transition energy, $\Omega^{\mathrm{Sat1}}$, and the corresponding transition moments, $\rho^{\mathrm{Sat1}}_{\text{C1s},m}$, of the lowest symmetry-allowed satellite in the C1s spectral function of an isolated \ce{CH4} molecule, calculated using different Gaussian basis sets. For the diagonal elements Eq.~\eqref{general_cumulant_inequality} simplifies to $|\rho^\nu_{\text{C1s},\text{C1s}}| \ll |\Omega^\nu|$. For the \ce{CH4} example, this inequality is clearly satisfied, as $\rho_\mathrm{C1s, C1s}^{\mathrm{Sat1}}$ is two orders of magnitude smaller than $\Omega^{\mathrm{Sat1}}$ across all basis sets. Since molecules have typically large optical gaps of several electronvolts, it is reasonable to assume that Eq.~\eqref{general_cumulant_inequality} is generally holds for molecular 1s levels. 
\par
Satellites due to the off-diagonal elements appear at frequencies $\omega = \epsilon_m - \Omega^\nu + \Delta H_n$. These satellites, previously termed \textit{correlation satellites},\cite{cederbaum1986correlation} couple the ionization of a level with energy $\epsilon_m$ to the core orbital $n$. 
Or in other words, a shake-up peak in the spectral function of a level $m$, for example the highest occupied molecular orbital (HOMO), shows also up as correlation satellite in the spectral function of the core level $n$. 
Generally, we expect the intensity of correlation satellites to be very small because the off-diagonal transition moments $\rho_{mn}^{\nu}$ are negligible. This expectation is supported by the data in Table~\ref{tab:cumulant_decoupling}, where $\rho_{\mathrm{C1s,},m}^{\mathrm{Sat1}}$ is found to be orders of magnitude smaller than $\rho_\mathrm{C1s, C1s}^{\mathrm{Sat1}}$ for the semicore state ($m = 2$) and vanishes entirely for the valence states ($m = 3$–5).
\par
Although the transition moments $\rho_{\mathrm{C1s}, m}^\nu$ are very small, Eq.~\eqref{general_cumulant_inequality} may still be violated because the right-hand side can be zero by chance if $\epsilon_m-\epsilon_n=\Omega^{\nu}$. If $m$ is a valence state and $n$ a 1s core-level this implies that $\Omega^{\nu} > 250$~eV because, for example, C1s, N1s and O1s core-levels are around -290~eV, -410~eV or -540~eV respectively, while valence excitations are typically in the order of tens of eV. Such large RPA excitation energies $\Omega^{\nu}$ are mostly due to transitions from valence states to very high lying virtual states. In a plane-wave basis-set framework, these high lying virtual states are correctly described as continuum,
and transitions to these states should appear as smooth offset in the spectral function. However, using a localized basis set, which is an computationally efficient choice for all-electron core level calculations, the high-level virtual states are discrete,\cite{van2014physical} which we will also demonstrate in Section~\ref{subsec:decoupling}. The valence transitions to the high-lying states then generates sharp peaks in the spectral functions, which might gain weight if Eq.~\eqref{general_cumulant_inequality} is not satisfied.

\subsection{Core-valence decoupling}
\label{subsec:decoupling_approximation}
The spurious contributions from correlation satellites can be removed by making use of the \textit{decoupling approximation} for core-levels $c$ in Eq.~\eqref{sigma_spectral}.\cite{zhou_dynamical_2015}
\begin{equation}
  \label{sigma_spectral_decoupled}
  \Sigma^{c, G_0W_0}_{c}(\omega) \approx \sum_{\nu} \, \frac{\rho^\nu_{cc}\rho^\nu_{cc}}{\omega - \epsilon_c - \Omega^\nu - \mathrm{i}\eta}
\end{equation}
In the decoupling approximation, only the contributions of core-level shake-up satellites are kept in the self-energy.
The approximation is well justified for deep core-levels, since density shifts to other levels are very small, i.e. $\rho_{cm}^\nu \approx 0$ (see Table~\ref{tab:cumulant_decoupling}).
The situation is slightly more complicated in the case of a core-level delocalized over several $N_\mathrm{Sym}$ symmetry-equivalent cores, e.g. in benzene.
In that case, all contributions from symmetry-related levels have to be considered.
\begin{equation}
  \label{sigma_spectral_decoupled_symmetric}
  \Sigma_{c}^{c, G_0W_0}(\omega) = \sum_{\nu}\sum_i^{N_\mathrm{Sym}} \, \frac{\rho^\nu_{ci}\rho^\nu_{ci}}{\omega - \epsilon_i - \Omega^\nu - \mathrm{i}\eta}
\end{equation}
Such a drastic simplification as the decoupling approximation comes at a price, specifically the QP shift is altered due to the missing relaxing valence orbital contributions.
However, we can calculate the satellite term $C^\mathrm{S}_n$ independently of the QP peak $G_{0, n}^\mathrm{QP}$, applying the decoupling approximation only to the satellite spectrum.
\par
The decoupling approximation follows a similar spirit as the CVS approximation, which is applied in many different flavors of wave-function methods, e.g. in ADC\cite{cederbaum1980many,barth1981many, wenzel2014calculating, wenzel2014calculating2} or CC\cite{coriani2015communication, coriani2016erratum} approaches, and exploits the low overlap between core- and valence region by systematically neglecting integrals involving products of core- and valence orbitals.\cite{norman2018simulating}
However, both approximations are not exactly equal, since we do not apply a similar approximation for the calculation of the RPA couplings and include integrals of the type $(cc|ca)$ in $\rho_{cc}^\nu$, which would be zero in the CVS approximation.
\section{Implementation}
\label{sec:technical_details}
\subsection{Numerical \textit{GW+C} scheme}
\label{sec:Implementation} 
%A flowchart of our implementation in FHI-aims is displayed in Figure~\ref{fig:implementation}. 
In our implementation, the working equations do not rely on the fully analytical expressions given in Section~\ref{subsec:gwc}. Instead, we employ the more commonly used numerical $GW+C$ scheme,\cite{aryasetiawan1996multiple, hedin1999correlation, guzzo2011valence,kas_cumulant_2014,zhou_dynamical_2015} widely adopted within the solid-state physics community.
In this numerical $GW+C$ approach, we employ the spectral representation of the self-energy.\cite{aryasetiawan1998gw,bruneval2005exchange,kas_cumulant_2014, zhou_dynamical_2015}
\begin{equation}
 \label{numerical_spectral_sigma}
    \Sigma_{n}^{G_0W_0}(\omega) = \Sigma_{n}^{x, G_0W_0} + \int \frac{\mathrm{d}\omega^\prime}{\pi} \frac{\left|\mathrm{Im}\Sigma^{c, G_0W_0}_{n}(\omega^\prime) \right| }{\omega - \omega^\prime + \mathrm{i}\eta}
\end{equation}
We obtain the cumulant $C_n(t)$ by inserting Eq.~\eqref{numerical_spectral_sigma} into Eq.~\eqref{cumulant_shifted_fourier}.
The exchange term $C_n^x(t)$ is given by Eq.~\eqref{cumulant_sigma_x} as before.
The correlation part $C_{n}^c(t)$ is obtained from a double-frequency integral
\begin{equation}
    \label{cumulant_double_integral}
    C_{n}^c(t) = \frac{\mathrm{i}}{2\pi^2}\int \mathrm{d}\omega^\prime\mathrm{d}\omega   \frac{e^{-\mathrm{i}\omega t}\left|\mathrm{Im}\Sigma^{c, G_0W_0}_{n}(\omega^\prime + \epsilon_n)\right| }{\left(\omega + \mathrm{i}\eta\right)^2\left(\omega - \omega^\prime + \mathrm{i}\eta\right)}
\end{equation}
Carrying out the integration over $\omega$ and relabeling $\omega^\prime$ as $\omega$ gives the Landau form of the cumulant, in analogy to Eq.~\eqref{landau_cumulant}.
\begin{equation}
  % TODO: Doublecheck signs for beta
    \label{landau_cumulant}
    C_{n}^c(t) = \frac{1}{\pi}\int \mathrm{d}\omega \, \frac{\left| \mathrm{Im} \Sigma_{n}^{c, G_0W_0}(\omega + \epsilon_n) \right|}{\omega^2} \left(e^{-\mathrm{i}\omega t} + \mathrm{i}\omega t - 1\right) 
\end{equation}\par
We split $C_n^c(t)$ into
\begin{equation}
    \label{landau_cumulant_numerical_part}
    %C_{n}^c(t) = -\mathrm{i}t\Sigma_{n}^{c, G_0W_0}(\epsilon_n) + \tilde{C}_{n}^c(t)
    C_{n}^c(t) = \hat{C}_{n}^c(t) + \tilde{C}_{n}^c(t)
\end{equation}
where
\begin{align}
 \label{landau_cumulant_truncated_Chat}
    \hat{C}_{n}^c(t) &= \frac{1}{\pi}\int \mathrm{d}\omega \, \frac{\left|\mathrm{Im} \Sigma_{n}^{c, G_0W_0}(\omega + \epsilon_n) \right| }{\omega^2} \left(i\omega t\right) \\\label{landau_cumulant_truncated_Chat_integrated}
    & = -\mathrm{i}t\Sigma_{n}^{c, G_0W_0}(\epsilon_n)
\end{align}
and
\begin{equation}
  % TODO: Doublecheck signs for beta
    \label{landau_cumulant_truncated}
    \tilde{C}_{n}^c(t) = \frac{1}{\pi}\int \mathrm{d}\omega \, \frac{\left|\mathrm{Im} \Sigma_{n}^{c, G_0W_0}(\omega + \epsilon_n)\right| }{\omega^2} \left(e^{-\mathrm{i}\omega t} - 1\right) 
\end{equation}
The cumulant Green's function $G_n^C(t)$ is derived by inserting Eq.~\eqref{eq:G0time} and Eq.~\eqref{cumulant_split} into Eq.~\eqref{cumulant_propagator}, using Eqs.~\eqref{landau_cumulant_numerical_part} and \eqref{cumulant_sigma_x} for $C_n^c(t)$ and 
$C_n^x$, respectively. Furthermore, we use Eq.~\eqref{landau_cumulant_truncated_Chat_integrated} for $\hat{C}_n^c(t)$ and absorb this contribution into the QP propagator $G_{0,n}^{\mathrm{QP}}$ defined in Eq.~\eqref{G_QPGWC}, yielding
\begin{equation}
 \label{cumulant_propagator_numerical_time}
    G_n^C(t) = G_{0, n}^\mathrm{QP}(t) e^{\tilde{C}^c_n(t)}
\end{equation}
In the numerical scheme, the renormalization and satellite terms, $Z_n$ and $\exp{\{C_{n}^\mathrm{S}(t)\}}$, are contained in $\exp\{\tilde{C}_{n}^c(t)\}$.\par
The frequency-dependent propagator is computed by making use of the convolution theorem.
\begin{equation}
 \label{cumulant_propagator_convolution}
    \begin{split}
    G_n^C(\omega) &= \int \mathrm{d}t G_{0, n}^\mathrm{QP}(t) e^{\tilde{C}^c_n(t)} e^{\mathrm{i}\omega t}\\
    &=\int \mathrm{d}\omega^\prime G_{0, n}^\mathrm{QP}(\omega + \omega^\prime)e^{\tilde{C}^c_n(\omega^\prime)}
    \end{split}
\end{equation}
$G_{0, n}^\mathrm{QP}(\omega)$ is known from Eq.~\eqref{eq:GQP}, whereas $e^{\tilde{C}^c_n(\omega)}$ has to be computed numerically from the Fourier transform.
\begin{equation}
 \label{cumulant_propagator_fourier}
    e^{\tilde{C}^c_n(\omega)} = \int \mathrm{d}t \, e^{\tilde{C}^c_n(t)} e^{\mathrm{i}\omega t}
\end{equation}
Our final quantity, the spectral function $A^{GW+C}_n(\omega)$, is obtained from $G_n^C(\omega)$ as in Eq.~\eqref{eq:AGWC}.\par
$A^{GW+C}_n$ can be approximated in terms of $\mathrm{Im} \, \Sigma^{c, G_{\Delta H}W_0}_n(\omega)=\mathrm{Im} \, \Sigma^{c, G_0W_0}_n(\omega-\Delta H_n)$ by making use of a Taylor expansion. To first order, the spectral function is approximated as
\begin{equation}
    \label{im_sigma_satellites}
    A_n^{GW+C}(\omega) \approx Z_n\left(\delta\left(\omega-\epsilon_n^\mathrm{QP}\right) + \frac{\left|\mathrm{Im}\,\Sigma^{c, G_{\Delta H}W_0}_n\left(\omega \right)\right|}{\pi \left(\omega - \epsilon^\mathrm{QP}_n\right)^2}\right)
\end{equation}
A detailed derivation of Eq.~\eqref{im_sigma_satellites} is provided in Section~S2 in the SI.
Following Eq.~\eqref{im_sigma_satellites}, every peak in $\text{Im}\Sigma^{c, G_{\Delta H}W_0}_n$ is directly related to a satellite peak in $A_n^{GW+C}(\omega)$.
\subsection{Workflow}
\label{subsec:workflow}
    \begin{figure}[!htb]
        \centering
        \resizebox{\columnwidth}{!}{%
            \begin{tikzpicture}[node distance=2cm]

%\node (start) [startstop] {Start};
\node (in1) [io] {DFT Starting Point};
\node (pro1) [process, below of=in1] {All-electron $G_{\Delta H}W_0$ with CD(-WAC) self-energy};
\node (dec1) [decision, below of=pro1,  yshift=-1cm] { Decoupling?};

\node (pro2a) [process, left of=dec1, xshift=-2.5cm] {$\mathrm{Im} \, W^c_{nn}(\left|\omega - \epsilon_n\right|)$};

\node (pro2b) [process, right of=dec1, xshift=2.5cm] {$\sum_i \mathrm{Im} \, W^c_{ni}(\left|\omega - \epsilon_i\right|)$};
\node (pro3) [process, below of=dec1,  yshift=-1cm] {$\tilde{C}^c_n(t)$};
\node (pro4) [process, below of=pro3] {$\tilde{C}^c_n(\omega)$};
\node (pro5) [process, below of=pro4] {$G_n(\omega)$};
\node (out1) [io, below of=pro5] {$A^{GW+C}_n(\omega)$};

%\draw [arrow] (start) -- (in1);
\draw [arrow] (in1) -- (pro1);
\draw [arrow] (pro1) -- (dec1);
\draw [arrow] (dec1) -- node[anchor=south] {\large Yes} (pro2a);
\draw [arrow] (dec1) -- node[anchor=south] {\large No} (pro2b);
\draw [arrow] (pro2b) |- node[anchor=north] {\large Integration  Eq. \eqref{landau_cumulant_truncated}} (pro3);
\draw [arrow] (pro2a) |- node[anchor=north] {\large Integration Eq. \eqref{landau_cumulant_truncated}} (pro3);
\draw [arrow] (pro3) -- node[anchor=east] {\large FFT Eq.~\eqref{cumulant_propagator_fourier}}(pro4);
\draw [arrow] (pro4) -- node[anchor=east] {\large Convolution Eq. \eqref{cumulant_propagator_convolution}}(pro5);
\draw [arrow] (pro5) -- (out1);

  \begin{pgfonlayer}{background}
    \filldraw [line width=4mm,join=round,red!10]
      (in1.north  -| pro2b.east)  rectangle (dec1.south  -| pro2a.west);
  \end{pgfonlayer}
  \begin{pgfonlayer}{background}
    \filldraw [line width=4mm,join=round,blue!10]
(out1.south  -| pro2b.east)  rectangle (pro3.north  -| pro2a.west);
  \end{pgfonlayer}
\end{tikzpicture}
        }
        \caption{Flowchart of the $GW+C$ workflow. The red-shaded part comprises a standard $G_{\Delta H}W_0$ calculation as implemented in the FHI-aims package. Subsequently, the newly implemented $GW+C$ procedure (blue-shaded) is carried out as a post-processing step.}
    \label{fig:implementation}
\end{figure}
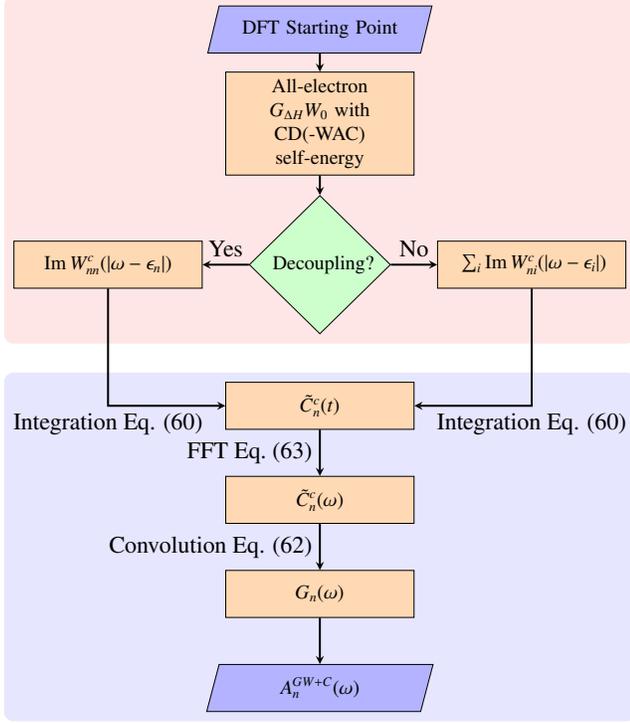
We implemented the $GW+C$ approach in the FHI-aims package.\cite{blum2009ab}
FHI-aims is an all-electron electronic structure code based on numeric atom-centered orbitals (NAOs) defined as
\begin{equation}
    \label{nao}
    \varphi_{i [l,m]}\left(\mathbf{r}\right)=\frac{u_i\left(r\right)}{r}Y_{l,m}(\Omega)
\end{equation}
The angular dependency is captured by the complex spherical harmonics $Y_{l,m}$, while the radial part $u_i$ is evaluated on a numerical grid and thus very flexible.
Standard quantum chemistry basis sets, e.g. Gaussian or Slater type orbitals, are a special case of Eq.~\eqref{nao} where $u_i(r)$ is a Gaussian or Slater function.
\par
Equations~\eqref{landau_cumulant_truncated}, \eqref{cumulant_propagator_convolution} and \eqref{cumulant_propagator_fourier} were implemented as post-processing step following a $GW$ calculation.
The workflow of our implementation is summarized in Figure~\ref{fig:implementation}.
For convenience, we use the $G_{\Delta H}W_0$ scheme instead of $G_0W_0$ as the starting point for the $GW+C$ calculation, as it directly provides the $GW+C$ QP energy $\epsilon_n^{\mathrm{QP}}$ defined in Eq.~\eqref{QPGWC}, which is needed to calculate $G_{n,0}^{\mathrm{QP}}$. We also avoid numerical artifacts, such as the non-convergence of the QP equation~\eqref{quasiparticle_equation} at the $G_0W_0@$PBE level, due to the multisolution behavior for core levels discussed in Section~\ref{subsec:corelevels}. In the $GW$ step, $\mathrm{Im}\,\Sigma^{c, G_{\Delta H}W_0}_n(\omega)$ is evaluated on a discrete, equidistant frequency grid using the CD or CD-WAC technique. If the decoupling approximation Eq.~\eqref{sigma_spectral_decoupled} is invoked, $\Sigma^{c, G_{\Delta H}W_0}_n(\omega)$ contains only the diagonal part of the screened interaction, as discussed in more detail ins Section~\ref{sec:contour_deformation}.
\par 
Next, we evaluate $\tilde{C}_n^c(t)$, using the 
Hedin-shifted self-energy matrix elements as input following Eq.~\eqref{hedin_sigma_spectral}.
The frequency integral in Eq.~\eqref{landau_cumulant_truncated} is carried out numerically for a set of $n_t$ discrete times, with time steps chosen such that we obtain the desired energy resolution after the final Fourier transform.
Due to the smooth nature of the imaginary part of the self-energy, optimized integration grids, which are commonly applied in $GW$,\cite{azizi2023time, azizi2024validation} are note necessary.
The Fourier transform from time-to-frequency space in Eq.~\eqref{cumulant_propagator_fourier} is performed using the fast Fourier transform (FFT) utility of the FFTW package for a set of $n_\mathrm{FFT}$ frequencies.\cite{frigo2005design}
Finally, we obtain the propagator in Eq.~\eqref{cumulant_propagator_convolution} in the desired frequency range by numerically integrating over the FFT frequencies $\omega^\prime$.
\par
The computational cost is dominated by the evaluation of the self-energy matrix elements $\Sigma_{n}^{c, G_{\Delta H}W_0}(\omega)$. The contribution of the cumulant evaluation to the total computational time is negligible. While the analytical $GW+C$ formulation in Section~\ref{subsec:gwc} requires an analytic evaluation of the self-energy, resulting in an unfavorable $O(N^6)$ scaling with system size $N$, our numerical implementation leverages the lower-scaling CD and CD-WAC algorithms for evaluating $\Sigma_{n}^{c, G_0W_0}(\omega)$.

\subsection{Self-energy evaluation}
\label{sec:contour_deformation}
For core-level calculations, we use the CD technique to perform the frequency integration of the self-energy.
The CD implementation in FHI-aims is based on the resolution-of-the-identity (RI) approximation\cite{ren2012resolution, vahtras1993integral} and is described in detail in our previous publication.\cite{golze2018core}
The CD technique allows the numerical exact evaluation of the self-energy at a given real frequency.
\par
In the CD formulation, the integral Eq.~\eqref{gw_sigma} for the correlation part is divided into two parts, one including an integral over the imaginary axis $I_n(\omega)$ and a residue part $R_n(\omega)$, which sums up the residues in $G_0$.
\begin{equation}
  \label{cd_terms}
  \Sigma^c_{n}(\omega) = -I_{n}(\omega) + R_{n}(\omega)
\end{equation}
The integral along the imaginary axis is defined as 
\begin{equation}
  \label{cd_imaginary}
  I_{n}(\omega) = \frac{1}{2\pi} \sum_m \int_{-\infty}^{\infty}\mathrm{d}\omega^\prime \, G_{0, m}(\omega + \mathrm{i}\omega^\prime + \mathrm{i}\eta)W_{mn}(\mathrm{i}\omega^\prime  +\mathrm{i}\eta)
\end{equation}
and the residue term is given by
\begin{equation}
  \label{cd_residue}
  R_{n}(\omega)=\sum_m f_m W_{mn}(\left|\epsilon_m-\omega\right| +\mathrm{i}\eta)
\end{equation}
The prefactors $f_m$ are defined as
\begin{equation}
  \label{cd_prefactors} 
  f_{m} = -\Theta(\epsilon_F-\epsilon_m)\Theta(\epsilon_m - \omega) + \Theta(\epsilon_m - \epsilon_F)\Theta(\omega - \epsilon_m)
\end{equation}
For a frequency in the core-level range, the sum in Eq.~\eqref{cd_residue} runs over all occupied levels. Since the calculation of $W_{mn}$ at a specific frequency scales $O(N^4)$ with respect to system size $N$, this leads to an overall scaling of $O(N^5)$ for $R_n$. The evaluation of $I_n$ is $O(N^4)$; for a detailed discussion of the scaling we refer the reader to Ref.~\citenum{golze2018core}.
\par
The central idea of the CD-WAC approach~\cite{duchemin2020robust,panades2023accelerating} is to reduce the scaling of the $R_n$ term to $N^4$ by approximating the screened interaction $W$ using analytical continuation (AC) techniques, specifically a Pad\'e approximation Eq.~\eqref{pade_approximation}.
\begin{equation}
    \label{pade_approximation}
    W^c_{mn}(\omega) \approx \cfrac{a_0}{1 + \cfrac{a_1(\omega - \tilde{\omega}_1)}{1+\cfrac{a_2(\omega - \tilde{\omega}_2)}{1+\dots}}}
\end{equation}
To obtain the parameters $a_i$, we compute the matrix elements $W^c_{mn}(\tilde{\omega})$ for a fixed number of reference frequencies $\tilde{\omega}$.
The reference frequencies $\tilde{\omega}$ are taken along the real- and the imaginary axis.
The points along the imaginary axis are reused from the numerical integration in Eq.~\eqref{cd_imaginary}.
The real-valued frequencies are picked in the core / valence region, and the range is determined by the numerical parameters $\omega_\mathrm{C}^\mathrm{min/max}$ for the core region and $\Delta_\mathrm{lower/upper}$ for the valence region.
The number of real-frequency reference points in the core- and valence region ($N_\mathrm{C}^\mathrm{WAC}$ / $N_\mathrm{V}^\mathrm{WAC}$) determines the accuracy of the fit.
Sensible choices for these parameters were derived in Ref.~\citenum{panades2023accelerating}.\par
After the Pad\'{e} fit, the computation of $W_{mn}(\omega)$ requires only the evaluation of an analytic function, reducing the overall scaling to $N^4$ for core-levels. In Ref.~\citenum{panades2023accelerating}, we demonstrated the scalability of the CD-WAC implementation, achieving a tenfold speed-up in the QP calculations for the largest system, consisting of 116 atoms, compared to CD.
The reduction in computational cost is particularly significant when calculating spectral functions because this requires evaluating the self-energy at several thousand frequency points, compared to only 10-15 points when iterating the QP equation~\eqref{quasiparticle_equation}. The CD-WAC speed-up with respect to CD is in the range of a factor of 100 to 1000, depending on the chosen frequency range and resolution of $A^{GW+C}_n(\omega)$. Practically, CD-WAC produces spectral functions with negligible computational overhead after the QP evaluation.
\par
In CD(-WAC), the integral in $I_{n}(\omega)$ is real for $\eta \rightarrow 0$ due to the symmetry relation $W_{mn}(\mathrm{i}\omega)=W_{mn}^*(-\mathrm{i}\omega)$.\cite{bruneval2005exchange}
Therefore, the imaginary part of the self energy, which is required for the calculation of the cumulant in Eq.~\eqref{landau_cumulant_truncated}, depends only on $R_{n}(\omega)$, and we can rewrite $\mathrm{Im}\,\Sigma^{c, G_0W_0}_{n}(\omega)$ as
\begin{equation}
  \label{cd_sigma_imaginary}
  \mathrm{Im}\,\Sigma^{c, G_0W_0}_{n}(\omega) = \sum_m f_m \, \mathrm{Im}\, W^c_{nm}(\left|\epsilon_m - \omega \right| +\mathrm{i}\eta)
\end{equation}
By inspecting Eq.~\eqref{W_spectral} and Eq.~\eqref{sigma_spectral_decoupled}, we note that the decoupling approximation Eq.~\eqref{sigma_spectral_decoupled} can be enforced by approximating 
\begin{equation}
  \label{cd_sigma_imaginary_decoupled}
  \mathrm{Im}\Sigma^{c, G_0W_0}_{n}(\omega) \approx f_n\, \mathrm{Im} W^c_{nn}(\left| \epsilon_n - \omega \right| +\mathrm{i}\eta)
\end{equation}
For symmetry related core-levels, all contributions from equivalent core-levels have to be included as in Eq.~\eqref{sigma_spectral_decoupled_symmetric}.

 \section{Computational Details}
 \label{sec:ComputationalDetails}
We performed all-electron $G_{\Delta H}W_0$ and $GW+C$ calculations for the CORE65 benchmark set\cite{golze2020accurate} as well as for the acene series $C_{4n + 2}H_{2n+4}$ ($n$=1-5) using the FHI-aims program package.\cite{blum2009ab,ren2012resolution}
The CORE65 benchmark set includes 65 1s core-level binding energies for 32 organic molecules containing the elements H, C, N, O and F.\cite{golze2020accurate}
The geometries for the benchmark set are available in the original publication.\cite{golze2020accurate}
We generated the acene structures for $n>1$ by performing geometry optimizations at the DFT level using NAOs of \textit{tier~2} quality (FHI-aims-2020 default). We employed the PBE exchange-correlation functional\cite{perdew1996generalized} including van-der Waals interactions via the Tkatchenko-Scheffler dispersion correction~\cite{tkatchenko2009accurate} and scalar-relativistic effects using the zeroth order 
regular approximation (ZORA).\cite{blum2009ab}
\par
We used the PBE functional as starting point for the $G_{\Delta H}W_0$ calculations and the CD(-WAC) technique for the frequency integration.
The integration along the imaginary axis (Eq.~\eqref{cd_imaginary}) was performed using a 200-points modified Gauss-Legendre grid.
For the CD-WAC calculation, we calculated $W_{mn}^c$ at 400 reference frequency points, consisting of 200 imaginary frequencies from the modified Gauss-Legendre grid and another 200 real-valued frequencies points, with $N_\mathrm{C}^\mathrm{WAC} = N_\mathrm{V}^\mathrm{WAC}=100$. The frequency ranges for the real points are $\omega_\mathrm{C}^\mathrm{min}=0$~eV, $\omega_\mathrm{C}^\mathrm{max}=40$~eV for the core region and $\Delta_\mathrm{lower}=8\,\%$, $\Delta_\mathrm{upper}=1\,\%$ for the valence region (see Ref.~\citenum{panades2023accelerating}).
\par
We used several common Gaussian- and NAO-type all-electron basis sets for convergence studies: 
The Gaussian basis sets comprise Dunning's correlation-consistent basis sets (cc-pV$X$Z\cite{dunning1989gaussian}, $X$=3-6), with additional diffuse functions (aug-cc-pV$X$Z\cite{kendall1992electron}, $X$=3-6) and with core-optimized (C) functions (aug-cc-pCV$X$Z, $X$=3-6) as well as the core-rich basis sets of the ccX-$X$Z family ($X$=3-5).\cite{ambroise2021probing, mejia2022basis}
For the NAO basis sets, we employed the FHI-aims-2020 \textit{tier~2} (T2) basis set, augmented with two additional diffuse (+aug2) functions obtained from the ($l=0, 1$) augmentation functions of the aug-cc-pV5Z basis set.\cite{liu2020all}
Furthermore, a set of highly localized Slater functions (STO$X$) was added to the NAO basis sets, abbreviated as T2+aug2+STO$X$ ($X$=1-5).\cite{yao2022all}
\par
As a measure for the basis set convergence, we assess the splitting $\Delta^\mathrm{Sat1-QP}$ of the lowest lying satellite (with a non-zero intensity) at $\omega^\mathrm{Sat1}$ and the QP peak.
\begin{equation}
  \label{sat_splitting}
  \Delta^\mathrm{Sat1-QP} = \omega^\mathrm{Sat1} - \epsilon_\mathrm{1s}^\mathrm{QP}
\end{equation}
For every basis set, we calculated the absolute error with respect to the largest basis set utilized here (aug-cc-pCV6Z) and averaged over all 65 levels in the CORE65 benchmark set to obtain to the mean average error (MAE):
\begin{equation}
  \label{rel_splitting}
  \bar{\Delta}_\mathrm{Basis}^\mathrm{MAE} = \frac{1}{65}\sum_i^\mathrm{CORE65}\left|\Delta^\mathrm{Sat1-QP}_{i, \mathrm{aug-cc-pCV6Z}} - \Delta^\mathrm{Sat1-QP}_{i, \mathrm{Basis}}\right|
\end{equation}
The position of the lowest satellite is determined based on a intensity criterion outlined in Section~S3 in the SI.
\par
For $GW+C$ calculations, we used $n_\mathrm{t}=n_\mathrm{FFT}=2^{18}$ and a final energy resolution of 0.01~eV. 
An imaginary broadening parameter $\eta=0.1 \, \mathrm{eV}$ is used throughout the $GW+C$ calculations if not stated otherwise.
The input and output files of all the FHI-aims calculations are available in the NOMAD database.\cite{nomad_reference}
\par
Reference RPA, time-dependent DFT (TDDFT) and analytical transition moments calculations were carried out in PySCF.\cite{sun2015libcint, sun2018pyscf,sun2020recent}
\section{Results and Discussion}
\label{sec:results}
We start with assessing the validity of the decoupling approximation, introduced and motivated in Section~\ref{subsec:decoupling_approximation}, and proceed with the analysis of important technical settings like the frequency integration technique, the basis set dependence and the choice of the DFT starting point.
We also compare the $GW+C$ spectral functions of small molecules with both experimental data and $G_{\Delta H} W_0$ spectral functions.
Finally, we showcase a possible application of our approach for the interpretation of experimental data by studying the acene series from benzene to pentacene.

\begin{figure*}[ht]
  \centering
  \includegraphics[width=\linewidth]{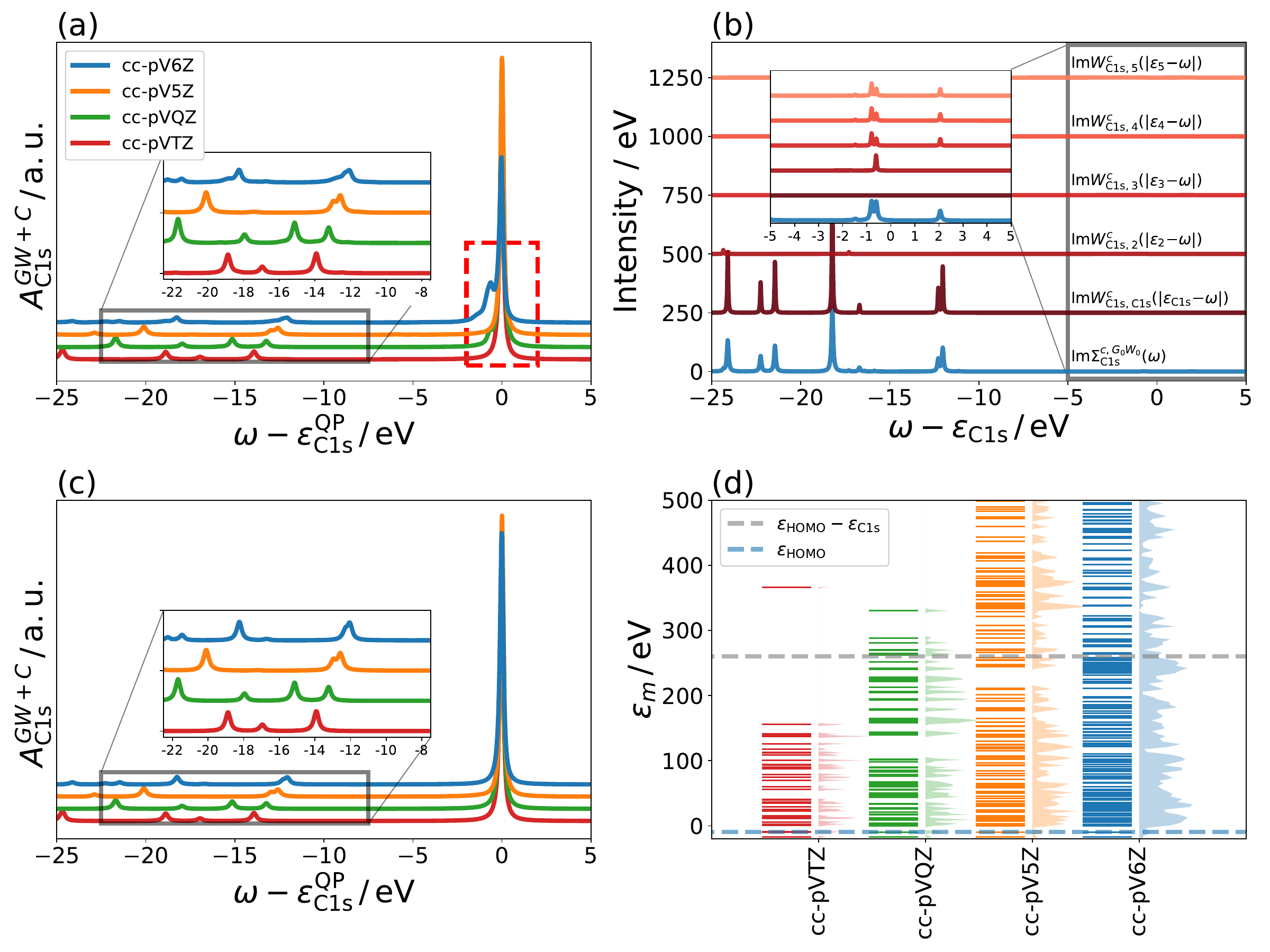}
  \caption{Analysis of the decoupling approximation for the $\mathrm{C1s}$ core-level in \ce{CH4}. (a): 
$GW+C$ spectral function of the satellite region using the full self-energy as in Eq.~\eqref{cd_sigma_imaginary}.
(b): Contributions to $\mathrm{Im}\Sigma_\mathrm{C1s}^{c, G_0W_0}(\omega)$ from diagonal- and off-diagonal elements of the screened interaction in the cc-pV6Z basis.
Low-intensity features close to the QP peak are highlighted in the inset, demonstrating that such spurious features arise from the off-diagonal elements of the screened interaction.
In the $GW+C$ spectral function, these satellites gain intensity following Eq.~\eqref{general_cumulant_inequality}.
 (c): Spectral functions obtained after applying the decoupling approximation (Eq.~\eqref{cd_sigma_imaginary_decoupled}), yielding a satellite spectrum free of spurious correlation satellites. (d): Energy distribution of the DFT orbital energies $\epsilon_m$ for the occupied valence states and virtual states, showing their exact positions as horizontal lines and their Gaussian-broadened distribution as a shaded plot.   
} 
  \label{fig:cd_decoupling}
\end{figure*}
\subsection{Decoupling approximation}
\label{subsec:decoupling}
In Figure~\ref{fig:cd_decoupling}a, the spectral function $A_\mathrm{C1s}^{GW+C}$ is displayed for the cc-pV$X$Z ($X$=3-6) basis set family without using the decoupling approximation. For convenience, we center the spectral functions at the C1s QP peak. 
For all basis sets, we observe the occurrence of a series of shake-up satellites between $-12$ and $-20$~eV, as shown in the inset.
However, when employing the cc-pVQZ and cc-pV6Z basis sets, additional signals appear near the QP peak, highlighted by the red dashed box.
These peaks carry considerable spectral weight, but have no counterpart in the cc-pV5Z and cc-pVTZ basis sets.
For the cc-pV6Z basis set, a very intense peak appears roughly $-1$~eV from the QP peak, at least one order of magnitude more intense compared to the satellites in the inset.
Due to the multiplicative nature of the cumulant Eq.~\eqref{cumulant_taylor}, this peak creates a set of replicas, i.e. higher order satellites, on top of every peak in the spectrum, causing additional broadening and peaks in the spectrum.
\par
To analyze the origin of the satellites close to the QP peak (red dashed box in Figure~\ref{fig:cd_decoupling}a), we first note that satellites in $A_n^{GW+C}$ can be either shake-up satellites or correlation satellites, as discussed in Section~\ref{subsec:correlation_satellites}.
Shake-up satellites appearing at $A_n^{GW+C}(\omega\approx \epsilon^\mathrm{QP}_n)$ require a very small charge-neutral excitation energy of $\Omega^\nu \approx 0$, while correlation satellites coupling valence- and core-levels need to have very large excitation energies of $\Omega^\nu > 250$~eV to show up in the C1s core region.
For the satellite in the cc-pV6Z spectral function at $\omega \approx \epsilon_\mathrm{C1s}^\mathrm{QP}$, a shake-up satellite would require an RPA excitation with $\Omega^\nu \approx 1\,\mathrm{eV}$.
For \ce{CH4}, such an excitation does not exist, since the lowest RPA excitation energy in \ce{CH4} is above 10~eV.
In turn, a correlation satellite, coupling e.g. HOMO and C1s level, requires an RPA excitation energy of roughly $\Omega^\nu \approx 260 \, \mathrm{eV}$ to appear at the same frequency in the spectral function, i.e. at $\omega = \epsilon_{\text{C1s}}^{\mathrm{QP}}+\epsilon_{\text{HOMO}}-\epsilon_{\text{C1s}}-\Omega^{\nu} \approx \epsilon_{\text{C1s}}^{\mathrm{QP}}$ (Eqs.~\eqref{cumulant_deltas} and \eqref{gwc_first_order}), where $\epsilon_{\text{C1s}}^{\mathrm{QP}}=-290$~eV, $\epsilon_{\text{HOMO}}=-9.4$~eV and $\epsilon_{\text{C1s}}=-268.5$~eV for that particular example.
\par
To assess the likelihood of such high-energy RPA excitations, we display the energy distribution of the virtual states with increasing basis set size in Figure~\ref{fig:cd_decoupling}d.
The RPA excitations $\Omega^\nu$ are close to the orbital differences $\epsilon_a - \epsilon_i$, and therefore the DFT eigenvalues can indicate the occurrence of RPA excitations with matching energy.
Near the Fermi level ($\epsilon_m \approx -0-20$~eV), the number of virtual states increases rapidly with the basis set size, and the number of high-energy states extending to several hundred electronvolts also grows from cc-pVTZ to cc-pV6Z. 
However, the density of states at high energies does not become continuous; instead, discrete levels with large gaps between them are observed.
\par
For the small cc-pVTZ basis set, there are no states in the 250–300~$\mathrm{eV}$ range. Consequently, $\Omega^\nu$ around 260~eV do not exist and only a pure shake-up spectrum is observed in Figure~\ref{fig:cd_decoupling}a. In contrast, for the cc-pVQZ, cc-pV5Z, and cc-pV6Z basis sets, states are present in the 250–300~$\mathrm{eV}$ range, and correlation satellites do appear in the spectral function, as shown in Figure~\ref{fig:cd_decoupling}a. The number of states in this energy range increases with larger basis sets, making the occurrence of correlation satellites more likely for the largest basis set.
\par
For a deeper analysis and to justify the decoupling approximation, we trace the correlation satellite contributions in our numerical $GW+C$ scheme by plotting $\mathrm{Im} \, \Sigma_\mathrm{C1s}^{c}(\omega)$ for the cc-pV6Z basis set in Figure~\ref{fig:cd_decoupling}b. Following Eq.~\eqref{im_sigma_satellites}, the peaks in $\mathrm{Im} \, \Sigma_n^{c}(\omega)$ directly show up with modulated intensity in the $GW+C$ spectral function as satellites. Furthermore, we plot the individual contributions to $\mathrm{Im} \, \Sigma_n^{c}(\omega)$ in Figure~\ref{fig:cd_decoupling}b, i.e., the matrix elements $\mathrm{Im}\, W^c_{\mathrm{C1s}, m}(\left|\epsilon_m - \omega\right|)$ (see Eq.~\eqref{cd_sigma_imaginary}). Methane has five occupied states and we have consequently five contributions because the sum over $m$ in Eq.~\eqref{cd_sigma_imaginary} includes all occupied levels for the frequency range under consideration ($\omega < \epsilon_{\text{C1s}}$).
\par
In Figure~\ref{fig:cd_decoupling}b, the most intense peaks in $\mathrm{Im} \, \Sigma_\mathrm{C1s}^{c}$ appear in the region between $-25$ to $-12$ $\mathrm{eV}$, originating solely from the diagonal element $\mathrm{Im}\, W^c_\mathrm{C1s, C1s}(\left|\epsilon_\mathrm{C1s} - \omega \right|)$. For $\omega<\epsilon_{\text{C1s}}$, peaks in $\mathrm{Im}\, W^c_\mathrm{C1s, C1s}(\left|\epsilon_\mathrm{C1s} - \omega \right|)$ are due to $\Omega^{\nu}=12-25$~eV, as evident from Eq.~\eqref{W_spectral} (poles in the real part correspond to peaks in the imaginary part). These are transitions from valence states to low-energy virtual states, generating the actual shake-up satellites. In addition, we observe small peaks for the off-diagonal elements $\mathrm{Im}\, W^c_\mathrm{C1s, 2-5}(\left|\epsilon_\mathrm{2-5} - \omega \right|)$ between $-5$ to 5~eV, see inset in Figure~\ref{fig:cd_decoupling}b. Following Eq.~\eqref{W_spectral}, these peaks correspond to $\Omega^{\nu} > 250$~eV because $\epsilon_{2-5}$ ranges from $-17$ to $-9.4$~eV. This demonstrates that the off-diagonal elements $\mathrm{Im}\, W^c_\mathrm{C1s, 2-5}(\left|\epsilon_\mathrm{2-5} - \omega \right|)$ are responsible for the correlation satellites. However, the intensity of the off-diagonal contributions is orders of magnitude smaller compared to those from the diagonal elements. Nevertheless, the off-diagonal contributions in $\mathrm{Im} \, \Sigma_\mathrm{C1s}^{c}$ gain artificial spectral weight in $GW+C$ because they appear very close to the QP peak. This follows directly from inspecting the second term in Eq.~\eqref{im_sigma_satellites}, corresponding to violation of the inequality ~\eqref{general_cumulant_inequality}.
\par
The correlation satellites appear at random positions and are likely to occur in large basis sets, which are needed to converge the calculations, see Section~\ref{sec:basis_convergence}. Therefore, it is crucial to remove spurious correlation satellites from $\mathrm{Im} \, \Sigma_\mathrm{C1s}^{c}$ to prevent a breakdown of the $GW+C$ approximation. To achieve this, we apply the decoupling approximation, as outlined in Section~\ref{subsec:decoupling_approximation}. Following Eq.~\eqref{cd_sigma_imaginary_decoupled}, we neglect all off-diagonal elements $\mathrm{Im}\, W^c_\mathrm{C1s, 2-5}(\left|\epsilon_\mathrm{2-5} - \omega \right|)$, thereby effectively removing all correlation satellites.
As demonstrated in Figure~\ref{fig:cd_decoupling}c, the resulting spectrum is "clean" and the shake-up spectrum is equally well captured as in Figure~\ref{fig:cd_decoupling}a.

\subsection{CD-WAC}
\label{sec:cdwac}
\begin{figure}[!t]
  \centering
  \includegraphics[width=\linewidth]{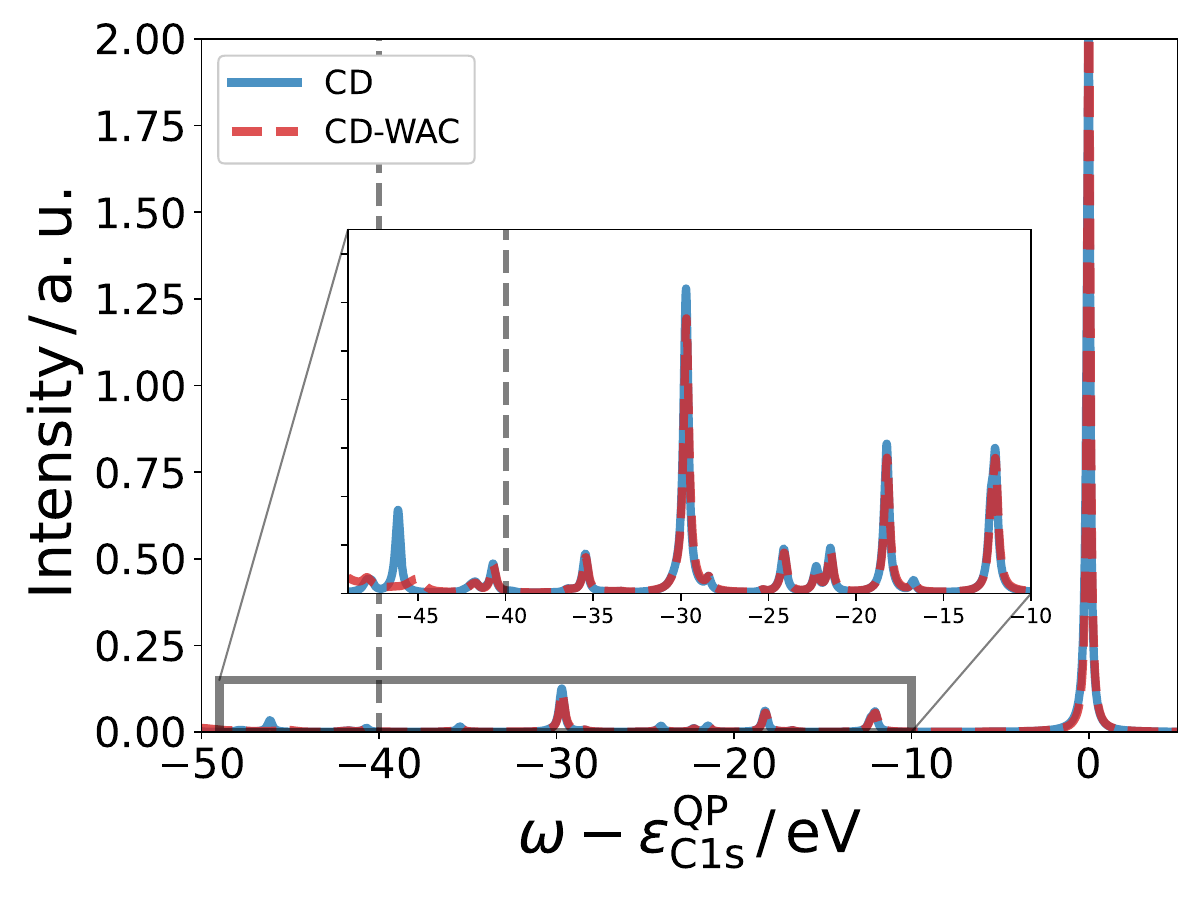}
  \caption{Comparison of $A_\mathrm{C1s}^{GW+C}$ of \ce{CH4} (cc-pV6Z basis, with decoupling approximation) computed with CD and CD-WAC self-energies. The fitting threshold $\omega_\mathrm{C}^\mathrm{max}$ is indicated by a grey line.}
  \label{fig:cdwac_comparison}
\end{figure}
We showed in Ref.~\citenum{panades2023accelerating} that CD-WAC reproduces QP energies within 4~meV of the CD reference.
Here, we briefly comment on the accuracy of the CD-WAC approximation for spectral functions.
\par
In Figure~\ref{fig:cdwac_comparison}, the $GW+C$ spectral functions based on CD- and CD-WAC self-energies are compared for \ce{CH4}, employing the decoupling approximation. 
Both $GW+C$ spectral functions match well, with satellite positions and intensities in excellent agreement to each other.
Beyond the fitting threshold $\omega_\mathrm{C}^\mathrm{max}$ (i.e., more than 40~eV from the QP peak, as indicated by the dashed grey line), the accuracy of the CD-WAC fit begins to deteriorate, and the satellites are slightly shifted relative to the CD reference.
In this work, we are primarily interested in low-lying satellites within a few~eV of the QP peak, which is accurately described by the CD-WAC fit.
Moreover, the satellites beyond $-40$~eV of the QP energy are mostly excitations to the continuum, which appear as spurious peaks in a localized basis set as discussed in Section~\ref{subsec:correlation_satellites}.
\par
We note that CD-WAC fits also the off-diagonal elements $W^c_{\text{C1s},m}$ well and reproduces therefore also the correlation satellites (see Section~S4 in SI). Consequently, the decoupling approximation must be also employed for $GW+C$ calculations based on a CD-WAC self-energy.
\subsection{Basis Set convergence}

\label{sec:basis_convergence}
\begin{figure}[!t]
  \centering
  \includegraphics[width=\linewidth]{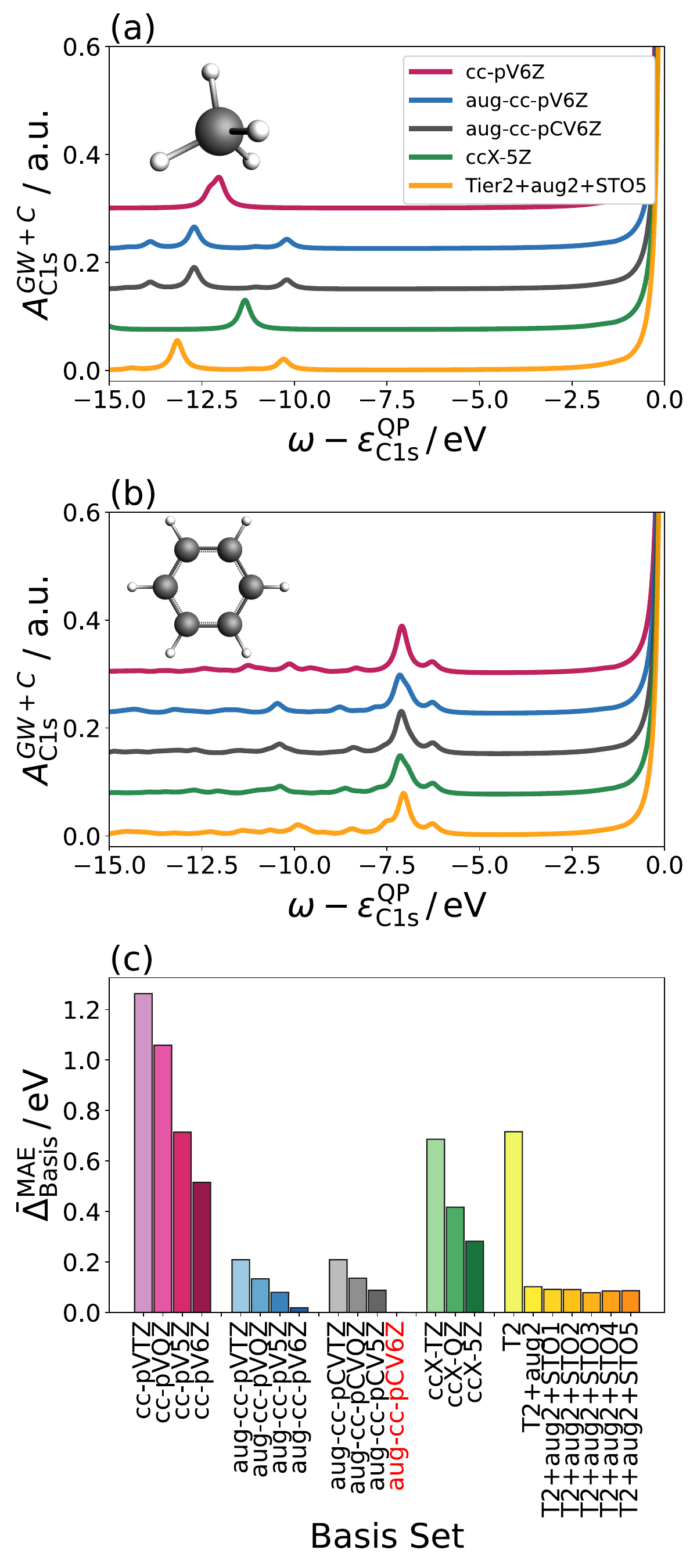}
  \caption{Basis set convergence of the lowest-energy satellite for the CORE65 benchmark set relative to the largest basis set (aug-cc-pCV6Z, marked in red). Figure (a, b) show the satellite spectrum of the \ce{CH4} and C$_6$H$_6$ C1s level for the largest basis set of each family. Figure (c) displays $\bar{\Delta}_\mathrm{Basis}^\mathrm{MAE}$ relative to the aug-cc-pCV6Z basis set as specified in Eq.~\eqref{rel_splitting}.}
  \label{fig:core65_basis_comparison}
\end{figure}

We investigate the basis set dependence of the satellite spectrum using the CORE65 benchmark set, employing four different GTO basis set families and customized NAO basis sets as detailed in Section~\ref{sec:ComputationalDetails}.
\par
The effect of different basis sets is displayed in Figure~\ref{fig:core65_basis_comparison}a and b for methane (\ce{CH4}) and benzene (\ce{C6H6}) using the largest basis set of each family.
For \ce{C6H6} in Figure~\ref{fig:core65_basis_comparison}b, all basis sets yield nearly identical satellite spectra, with the onset of the satellite region at $-6$~eV and a dominating peak at $-7$~eV.
For \ce{CH4} in Figure~\ref{fig:core65_basis_comparison}a, the basis sets with diffuse (aug-) functions produce similar spectra, with a first satellite appearing at ca. $-10$~eV and a second, more intense satellite around $-13$~eV.
However, the cc-pV6Z and ccX-5Z spectra lack the peak near $-10$~eV and instead show more intense satellite peaks around $-12$~eV and $-11$~eV. This suggests that the basis set dependence varies significantly across different systems.
\par
For a more systematic assessment, we computed $A_{1s}^{GW+C}$ for the 65 excitations in the CORE65 benchmark set and calculated $\Delta^\mathrm{Sat1-QP}$ ( Eq.~\eqref{sat_splitting}) for each spectral function. The MAE of $\Delta^\mathrm{Sat1-QP}$ with respect to the aug-cc-pCV6Z result, see definition in Eq.~\eqref{rel_splitting}, is shown in Figure~\ref{fig:core65_basis_comparison} for all 22 basis sets.
For completeness, we report the MAEs of the 1s QP excitations with respect to experiment and with respect to the aug-cc-pCV6Z result for each basis set in Section~S5 of the SI. 
\par
In Figure~\ref{fig:core65_basis_comparison}c, we observe that for the correlation-consistent (cc) Gaussian basis set families, $\bar{\Delta}_\mathrm{Basis}^\mathrm{MAE}$ systematically decreases with increasing basis set size, reducing the error by a factor of 2–3 when moving from triple- to sextuple-$\zeta$ quality.
The inclusion of diffuse (aug) functions has the most drastic effect, decreasing the error by a factor of roughly 6.
As an example, comparing the cc-pVTZ and aug-cc-pVTZ basis sets, $\bar{\Delta}_\mathrm{Basis}^\mathrm{MAE}$ is reduced from ca.~1.2~eV to 0.2~eV upon inclusion of diffuse functions. Similarly for the NAOs, $\bar{\Delta}_\mathrm{Basis}^\mathrm{MAE}$ decreases by 0.62~eV comparing T2 to T2+aug2. While adding augmentation functions significantly impacts the satellite positions, the QP positions are barely affected (see SI Table~S2). This suggests that augmentation functions are crucial for charge-neutral excitations but not for charged excitations. 
This observation aligns with previous TDDFT\cite{bruneval2015systematic} and BSE studies,\cite{bruneval2015systematic, liu2020all, yao2022all} which found that adding augmentation functions to localized basis sets is essential for improving the description of the virtual space.\par
We note that the unaugmented ccX-$X$Z basis sets yield smaller $\bar{\Delta}_\mathrm{Basis}^\mathrm{MAE}$ values than the cc-pV$X$Z basis sets because they are larger by a factor of ca. $1.5-2.$ However, the performance of the ccX-$X$Z basis sets is significantly worse compared to any of the (smaller) augmented basis sets, underpinning the importance of very diffuse functions in the basis set.\par
The addition of core-optimized (C) functions in the aug-cc-pCV\textit{X}Z basis sets does not improve $\bar{\Delta}_\mathrm{Basis}^\mathrm{MAE}$ compared to the aug-cc-pV\textit{X}Z family. A similar trend is observed for the NAO basis sets T2+aug+STO\textit{X}: the inclusion of additional steep STO functions hardly affects $\bar{\Delta}_\mathrm{Basis}^\mathrm{MAE}$.
This highlights that even for core-level satellites, satellite features depend only on the quality of the basis set in the valence space. However, we stress that core-optimized functions or additional STOs improve the accuracy of the 1s QP energies tremendously, decreasing the MAE with respect to the aug-cc-pCV6Z basis set from 1.1~eV to 0.1~eV going from T2+aug2 to T2+aug2+STO5 (see Table~S2 in the SI). 
\par
For a balanced description of the full spectrum, including QPs and satellite features, we recommend using the aug-cc-pCV5Z basis set or T2+aug2+STO2 basis set, which both reliably reproduce satellite features within 0.1~eV of the aug-cc-pCV6Z reference.
The NAO basis set is computationally significantly more efficient. 
As an example, for benzene, the T2+aug2+STO2 basis uses 468 basis functions, while the same calculation with an aug-cc-pCV5Z basis set has 1566 basis functions in total, increasing the computational cost by at least an order of magnitude.
Therefore, will use the T2+aug2+STO2 basis set, unless otherwise noted.

\subsection{Starting point dependence}
\label{subsec:starting_point}

\begin{table}[]
   \fontsize{10}{12}\selectfont
    \centering
    \begin{tabular}{c c c c c}
    \toprule
      Functional& $\Delta_{\text{gap}}^{\text{DFT}}$& $\Omega^1$ & $\Omega^1_\mathrm{TDDFT}$& $\Delta\Omega^1$\\
    \midrule
        LDA &9.70 & 10.12& 9.91 & -0.21\\   
        PBE & 10.21 & 10.64& 10.43 & -0.21  \\
        PBE0 &12.52&12.92 & 11.01& -1.91 \\
        PBEh($\alpha$=0.45) &14.35& 14.73 &11.45 & -3.28\\
        HF &18.67 & 19.00 &12.01 & -6.99\\
   \bottomrule
    \end{tabular}
   \caption{Fundamental DFT gap, $\Delta_{\text{gap}}^{\text{DFT}}=\epsilon_\mathrm{LUMO} - \epsilon_\mathrm{HOMO}$, lowest RPA ($\Omega^1$) and TDDFT ($\Omega^1_\mathrm{TDDFT}$) excitation energies and their difference $\Delta\Omega^1=\Omega^1_\mathrm{TDDFT} - \Omega^1$ for \ce{CH4} using a cc-pVTZ basis set computed with PySCF.\cite{sun2020recent} All values are in $\mathrm{eV}$.}
    \label{tab:starting_point}
\end{table}

In $GW+C$, the positions of the satellites relative to the QP peak are determined by the RPA excitation energies $\Omega^\nu$, obtained by solving Eq.~\eqref{rpa_casida}.
The starting point enters the RPA equations through the orbital energy differences in the diagonal elements $\epsilon_a - \epsilon_i$, while the off-diagonal (or coupling) elements in Eq.~\eqref{rpa_matrices} are rather independent of the DFT functional as they only contain the direct Coulomb interactions $(ia|jb)$.
Here, we aim to assess what constitutes a reasonable starting points for RPA excitations.
\par
To support the discussion, we list the lowest RPA excitation, $\Omega^1$, and the DFT gap between the HOMO and the lowest uncoccupied molecular orbital (LUMO) for \ce{CH4} in Table~\ref{tab:starting_point}, calculated using five different functionals. We include two DFT functionals without exact exchange, the local density approximation (LDA) and PBE, as well as two hybrid functionals, PBE0~\cite{adamo1999toward, ernzerhof1999assessment} and PBEh($\alpha=0.45$).\cite{atalla2013hybrid,golze2020accurate} The latter is identical to PBE0 but incorporates 45~\% exact exchange instead of 25~\%. Additionally, HF is included as the limiting case with full exact exchange.
As reference, we report the lowest TDDFT excitation, $\Omega^1_\mathrm{TDDFT}$, which is known to provide excitation energies within 0.3~eV of experimental values for TDDFT@PBE0.\cite{leang2012benchmarking, laurent2013td}
\par
Across all functionals, the RPA excitation energies $\Omega^1$ closely match the HOMO-LUMO gap, indicating that the Coulomb coupling elements in Eq.~\eqref{rpa_matrices} are small.
Going from LDA to HF, $\Omega^1$ increases gradually with the amount of exact exchange by 9~eV.
This enormous starting point dependence is mitigated in TDDFT, where exchange effects are added to Eq.~\eqref{rpa_matrices} by including the exchange-correlation kernel $f_{xc}$.
Comparing LDA to HF, $\Omega^1_\mathrm{TDDFT}$ increases by only $\approx 2$~eV.
Looking at the difference between TDDFT and RPA results $\Delta\Omega^1=\Omega^1_\mathrm{TDDFT}-\Omega^1$, LDA and PBE starting points produce RPA results close to TDDFT, with $\Delta \Omega^1=-0.21$. 
In contrast, for hybrid functionals and HF, the differences are substantial, with $\Delta \Omega^1$ gradually increasing from $-2$ to $-7$~eV. This shows that only DFT starting points without exact exchange, like PBE, yield RPA excitation energies of similar quality as TDDFT. This transfers also to the $GW+C$ satellite spectra as shown in Figure~S5 and S6 (SI). The spectral functions computed with hybrid starting points show non-physical satellite spectra, lacking any resemblance with experiment.
\par
These observations based on Table~\ref{tab:starting_point} can be generalized by looking at the physical meaning of orbital energy differences, like the HOMO-LUMO gap, in DFT based on non-hybrid functionals compared to HF.
As discussed in detail in Ref.~\citenum{baerends2013kohn}, the physical interpretation in both cases is different as consequence of different approximate mean-field potentials for virtual orbitals in HF and DFT theory.
The HF mean-field potential of the virtual orbitals lacks an exchange hole, and thus the virtual orbitals are defined with respect to the field of $N$ electrons, in difference to $N-1$ electrons for the occupied orbitals.\cite{baerends2013kohn}
Consequently, HF virtual orbital energies approximate electron affinities in a reverted Koopman's theorem\cite{baerends2013kohn, van2014physical} and orbital energy differences are no reasonable approximation to charge neutral excitations.\cite{gritsenko2004asymptotic}\par
In contrast, in DFT, virtual levels are computed for an (approximate) $N-1$ electron potential which an excited electron would experience.
Hence, DFT orbital energy differences resemble charge-neutral excitations, and RPA@PBE and RPA@LDA excitation energies are often a good approximation compared to TDDFT.\cite{baerends2013kohn, van2014physical}
Hybrid functionals interpolate between those limiting cases, and by increasing the amount of exact exchange the physical interpretation of the orbital energy differences changes towards the interpretation as fundamental gap.
In TDDFT, the exchange contributions in $f_{xc}$ counteract this effect, whereas RPA contains only Coulomb couplings which cannot adapt to this change.\cite{schambeck2024solving}
\par
As the excitation energies of RPA@PBE resemble the TDDFT@PBE results, we can estimate the expected accuracy of $GW+C$ satellites by comparison with previous TDDFT benchmarks:
For TDDFT@PBE, the best results were obtained for $\pi-\pi^*$ excitations, with an MAE of 0.3~eV compared to experimental values\cite{leang2012benchmarking}, while other valence excitations (MAE=0.6~eV) and Rydberg excitations (MAE=0.8~eV) were predicted with much lower accuracy.\cite{leang2012benchmarking, laurent2013td}
Therefore, we expect $GW+C$@PBE to work well for organic molecules with conjugated $\pi$ systems where the dominant satellite features are due to $\pi-\pi^*$ transitions.
Although hybrid functionals are known to provide improved TDDFT excitation energies with a broad range of applicability\cite{leang2012benchmarking}, their use in $GW+C$ would demand to go beyond RPA by including additional terms in Eq.~\eqref{rpa_casida}, which is effectively a vertex correction to the polarizability in $GW$.\cite{maggio2017gw, lewis2019vertex}

\begin{figure*}[t!]
  \centering
  \includegraphics[width=0.95\linewidth]{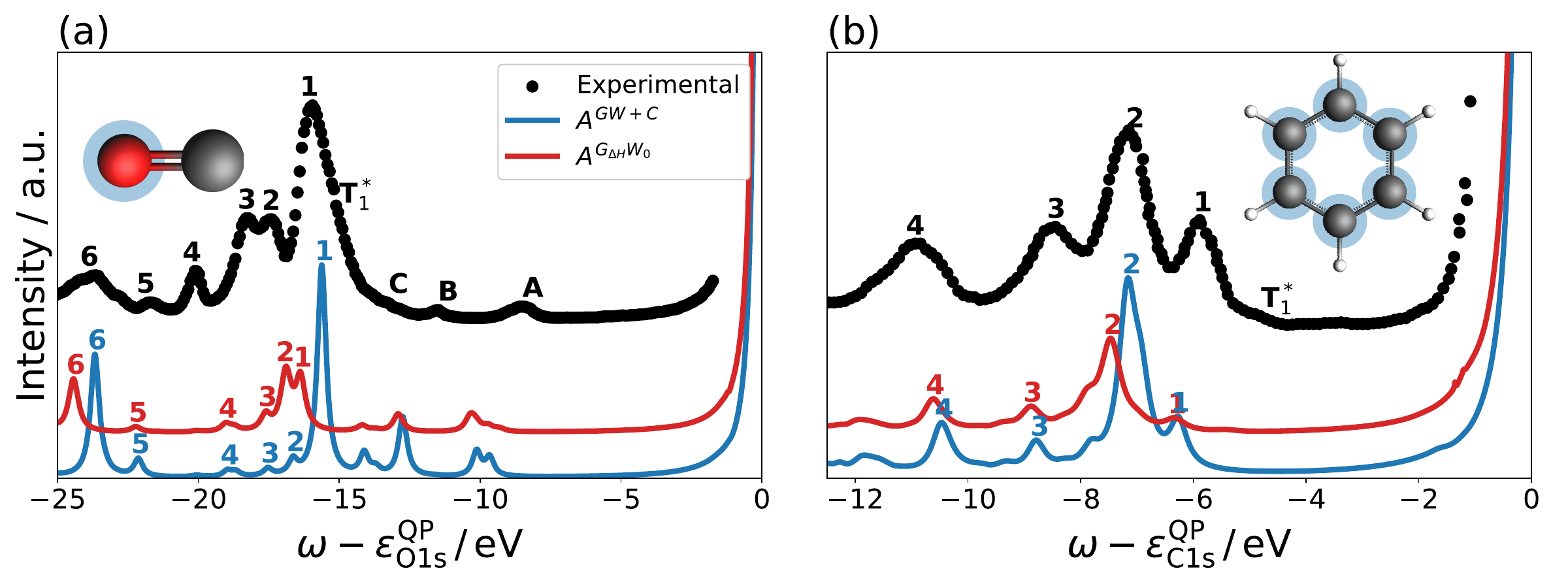}
  \caption{Comparison of the $G_{\Delta H}W_0$ and $GW+C$ spectral function (coloured) to the gas-phase experiment (black) for carbon monoxide\cite{schirmer1987high} and benzene\cite{nordfors1988experimental}.
The peak assignment follows approximately the experiment, with numbers labeling singlet shake-up satellites with relevant intensity, while letters denote either inelastic losses (A, B, C) or triplet excitations ($T_1^*$).
  }
  \label{fig:experimental_panel}
\end{figure*}

\subsection{Shake-up satellites in molecules}
\label{sec:gw_comparison}
In this work, we restrict the discussion to satellites associated with 1s excitations in molecules containing $\pi$-electrons.
In Figure~\ref{fig:experimental_panel}, we present our results for the O1s level of carbon monoxide (CO) as a polarized and unconjugated system, and the C1s level in benzene as an aromatic system, together with experimental gas-phase XPS data.\cite{schirmer1987high, nordfors1988experimental}
Both CO\cite{angonoa1987theoretical, fronzoni1999theoretical, ehara2006c1s} and benzene\cite{nordfors1988experimental, lunell1978theoretical, lunell1989semiempirical, brena2005functional} have been the subject of several theoretical studies on core-level shake-up satellites, which we use here as reference.
In addition, we provide spectral functions at the $G_{\Delta H}W_0$ level to identify the effect of the vertex corrections in $GW+C$.
Based on the findings in Section~\ref{subsec:decoupling} and \ref{sec:basis_convergence}, we use a PBE starting point with a T2+aug2+STO2 basis set and employ the decoupling approximation in combination with the CD-WAC algorithm.
We point out that we compare the satellite region directly to the experiment by aligning the QP peaks and do not require a scissor shift for the satellite region.
\subsubsection{Comparison to experiment}
The experimental spectra are shown in black in Figure~\ref{fig:experimental_panel}.
The peaks are assigned as singlet (1-6) and triplet ($T_1^*$) shake-up satellites as well as inelastic losses (A-C), following the assignment in Refs.~\citenum{schirmer1987high} and \citenum{nordfors1988experimental}.
We will focus on satellites due to singlet excitations, as $GW+C$ excludes other types of excitations.
\par
For CO, the O1s shake-up spectrum has six singlet shake-up satellites in the low-energy region.
The spectrum is dominated by the intense peak 1, which has been assigned to an admixture of $\pi-\pi^*$ and $\sigma-\sigma^*$ excitations by ADC\cite{angonoa1987theoretical}, QDPTCI\cite{fronzoni1999theoretical} and SAC-CI\cite{ehara2006c1s} calculations.
For the higher lying satellite peaks 2-5, the $\pi-\pi^*$ component decreases and is replaced by additional $\sigma-\sigma^*$ excitations, while peak 6 has been determined to be predominantly of Rydberg character.\cite{fronzoni1999theoretical}
Besides the singlet excitations, several additional signals appear closer to the QP peak which are assigned to inelastic losses (A-C) and a triplet excitation $T_1^*$.\cite{angonoa1987theoretical, ehara2006c1s}
\par
In Figure~\ref{fig:experimental_panel}a, $A_\mathrm{O1s}^{GW+C}$ shows in general good agreement compared to the experimental spectrum.
The integrated intensity of the satellite region is predicted as 16.9~\% of the QP peak, in excellent agreement with the experimental value of 18.9~\%.
The position of the intense peak~1 is predicted within $0.25 \, \mathrm{eV}$ of the experimental peak and correctly predicted to be the by far most intense shake-up state in the spectrum.
For the peaks 2-5, the agreement of $A^{GW+C}_\mathrm{O1s}$ is less satisfying, since the intensities are very low compared to the experiment, although the satellites appear within $0.5 \, \mathrm{eV}$ of the reference.
Peak 6 is again well predicted within $0.1 \, \mathrm{eV}$ of the experimental counterpart, albeit the agreement might be somewhat fortunate as it previously has been assigned to a Rydberg excitation.
Between $-10$ and $-15$~eV, several small peaks appear in $A^{GW+C}_\mathrm{O1s}$, which approximately correspond to the signals (A-C) in the experimental spectrum.
These peaks have previously been attributed to inelastic losses due to their pressure dependence,\cite{schirmer1987high} but they may also overlap with smaller singlet excitations.
\par
For the benzene C1s shake-up spectrum in Figure~\ref{fig:experimental_panel}, previous calculations assigned all peaks to singlet $\pi-\pi^*$ shake-up processes.
The shake-up transitions 1 and 2 have been assigned to HOMO-LUMO transitions, which split into several levels due to symmetry reduction during the ionization process.\cite{nordfors1988experimental}
Peak 3 and 4 are associated with a higher order $\pi-\pi^*$ transitions, although semi-empirical configuration interaction calculations suggest substantial $\sigma-\sigma^*$ admixtures for peak 4.\cite{lunell1989semiempirical}
As for CO, a low-intensity triplet shake-up excitation is observed as shoulder of peak~1.
\par
The $GW+C$ spectral function agrees overall very well with the experimental spectrum for benzene, and the integrated relative intensity of the satellites matches the experimental value of 15~\% exactly.
Peak~1 appears slightly too far from the QP peak in $GW+C$, but is still within $0.3 \, \mathrm{eV}$ of the experimental reference.
Peak~2 is calculated to be within 0.1~eV of the experimental signal and is correctly predicted as the most intense feature in the satellite spectrum.
The $GW+C$ spectral function exhibits several sharp features between $-8$ and $-10$~eV, therefore the assignment of the broad peak 3 is ambiguous.
We assign peak 3 to the shake-up satellite with the highest intensity in this region, which is within $0.25 \, \mathrm{eV}$ of the experimental signal.
For peak 4, the assignment can again be done unambiguously.
The calculated peak in the $GW+C$ spectrum matches the experiment up to $0.5 \, \mathrm{eV}$, which is a slightly higher deviation compared to the satellites~1-3.

\subsubsection{Comparison with $\boldsymbol{G_{\Delta H}W_0}$}
Comparing the $G_{\Delta H}W_0$ and $GW+C$ spectral functions, we observe strong differences in intensity and position of the satellite features.
In general, in $G_{\Delta H}W_0$, the spectral weight of the satellite region is only half of both experiment and $GW+C$, carrying only 9.2~\% and 8.8~\% of the QP peak intensity for CO and benzene respectively.
For CO, as an example, $G_{\Delta H}W_0$ predicts peaks~1 and 2 to have similar magnitudes, which is in striking contrast to the experimental and $GW+C$ results.
However, the changes relative to $GW+C$ are not systematic; some satellites, such as peak~2 in CO, are enhanced, while others, like peak~1 in CO, are suppressed. \par
Additionally, the satellites in $A^{G_{\Delta H}W_0}$ are shifted to higher binding energies compared to their $GW+C$ counterparts.
This effect is most pronounced for the intense peaks. For example, peaks~1 and 6 in CO shift by nearly 1~eV, while peak~2 in benzene shifts by approximately 0.4~eV. In contrast, less intense satellites show virtually no shift; for instance, peak~5 in CO appears almost at the same position in $G_{\Delta H}W_0$ and $GW+C$.
This is in line with the discussion in Section~\ref{subsec:gw_satellites} and \ref{subsec:validity_gwc}, because satellites in all kinds of $GW$ flavors appear at higher energies compared to the poles in the self-energy, while in $GW+C$ the first-order satellites appear exactly at the poles of the $G_{\Delta H}W_0$ self-energy, as evident by inspecting Eqs.~\eqref{gwc_first_order} and \eqref{hedin_sigma_spectral}.
Therefore, we stress that the vertex corrections provided by $GW+C$ approach are crucial for the interpretation of intense satellite features, within the boundaries derived in Section~\ref{subsec:validity_gwc}.

\subsection{Acene Series}
\label{sec:acenes}
\begin{figure}[!t]
  \centering
  \includegraphics[width=\linewidth]{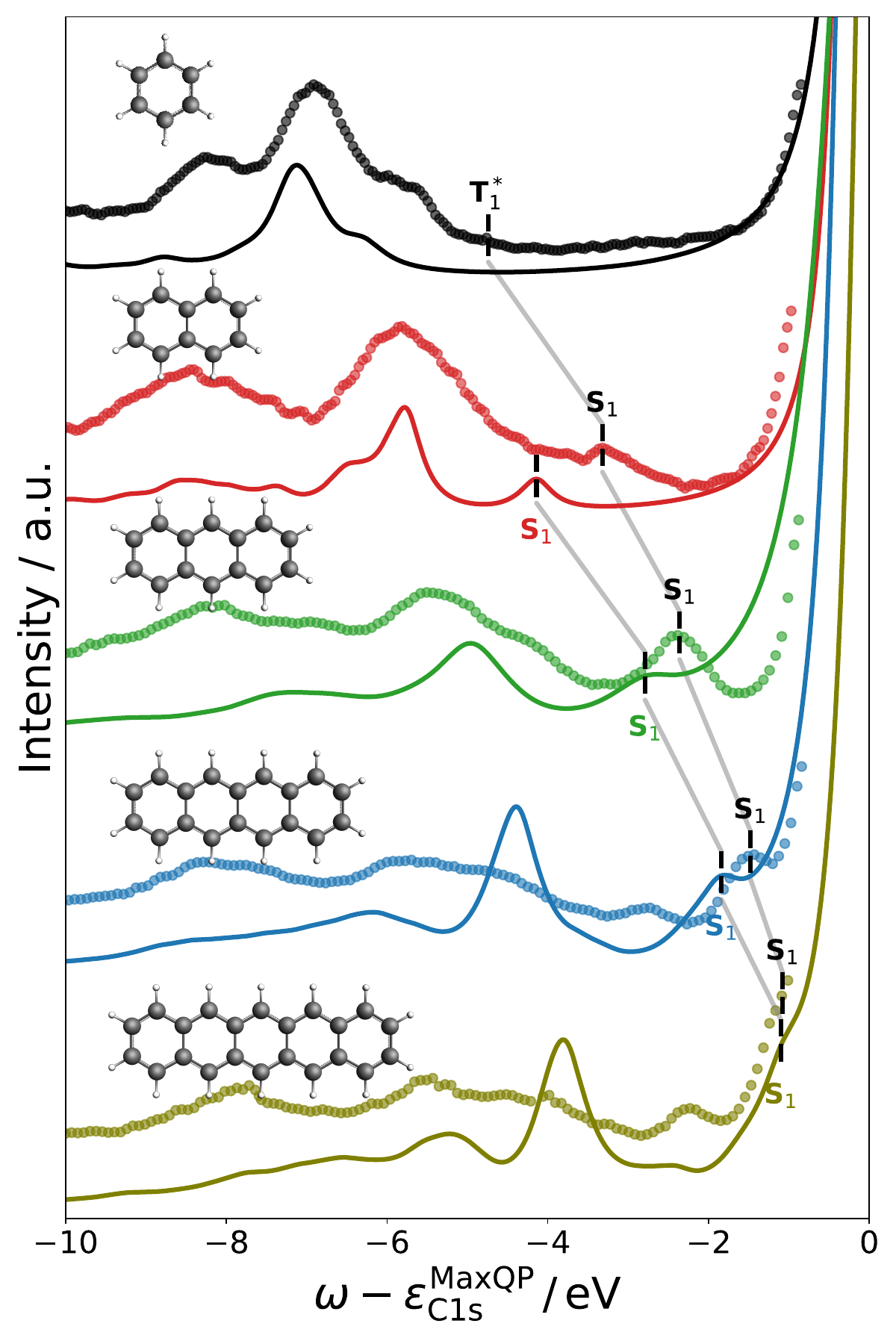}
  \caption{C1s spectral functions $A_\mathrm{C1s}^{GW+C}$ (solid line) of the acene series from benzene to pentacene compared to experimental XPS (dots) measured for multilayer films on Ag(111).\cite{rocco2008electronic} The satellite spectrum is shown relative to the maximum of the main C1s excitation.
  For each system, the position of the lowest singlet shake-up satellite ($S_1$) is marked with a dashed line in the experiment (black) and the $GW+C$ prediction (coloured). For benzene, only the triplet component ($T_1$) is observable.}
  \label{fig:acenes}
\end{figure}

\begin{figure}[!b]
  \centering
  \includegraphics[width=\linewidth]{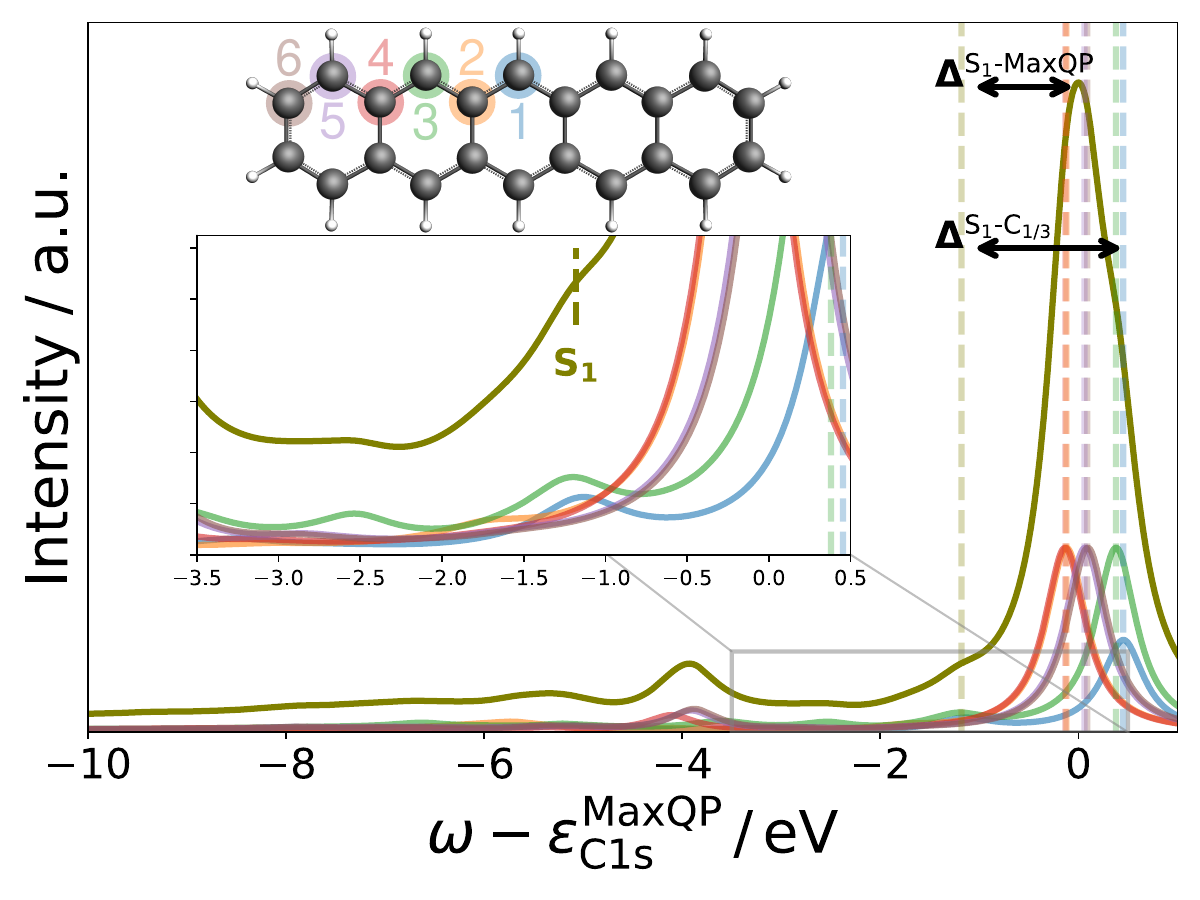}
  \caption{Contributions from all six inequivalent core-levels to the $GW+C$ spectral function of pentacene. The inset highlights the satellite spectra of each individual core-level in the region of the $S_1$ satellite.}
 \label{fig:pentacene}
\end{figure}

\begin{table*}[t]
    \fontsize{10}{12}\selectfont
    \centering
    \begin{tabular}{c c c c c}
    \toprule
         & Naphthalene & Anthracene & Tetracene & Pentacene  \\
    \midrule
         Optical Gap & 3.90\cite{colson1968direct} & 3.12\cite{glockner1969fluoreszenzspektrum,prikhotko1969spectral} & 2.37\cite{prikhotko1969spectral,mizuno1989exciton} & 1.85\cite{prikhotko1969spectral,jentzsch1998efficiency} \\[5pt]
          $\Delta^\mathrm{S_1-MaxQP}_{\mathrm{Exp}}$\cite{rocco2008electronic}& 3.32 & 2.36 & 1.48 & 1.08 \\[5pt]
          $\Delta^\mathrm{S_1-MaxQP}_{GW+C}$&4.14&2.79&1.86 &1.18 \\[5pt]
          $\Delta^\mathrm{S_1-C_x}_{GW+C}$&4.15&2.97&2.16&1.60\\[5pt]
          $\epsilon^\mathrm{MaxQP}_\mathrm{C1s} - \epsilon^\mathrm{QP}_\mathrm{C_x}$ &0.01 & 0.18 & 0.30 & 0.42 \\ [5pt]       
    \bottomrule
    \end{tabular}
    \caption{Experimental (Exp) and theoretical ($GW+C$) QP-satellite splitting $\Delta^\mathrm{S_1-MaxQP}$ between the $S_1$ satellite and the maximum of the C1s ionization peak $\epsilon^\mathrm{MaxQP}_\mathrm{C1s}$ for the acene series compared to the experimental optical gap and the calculated actual QP-satellite splitting $\Delta^\mathrm{S_1-C_x}_{GW+C}$ between the satellite and the generating core-level with the QP energy $\epsilon^\mathrm{QP}_\mathrm{Cx}$. 
    All values in $\mathrm{eV}$.}
    \label{tab:acenes_excitation}
\end{table*}
We use the acene series with the general formula $\text{C}_{4n + 2}\text{H}_{2n+4}$ to demonstrate both the scalability the applicability of our $GW+C$ implementation for the interpretation of shake-up features in molecules.
In Figure~\ref{fig:acenes}, we compare $A^{GW+C}_\mathrm{C1s}(\omega)$ with solid-state experimental XPS data from Ref.~\citenum{rocco2008electronic} for benzene (\ce{C6H6}) to pentacene (\ce{C22H14}).
As before, the computed and measured spectra are aligned at the maximum of the QP peaks at $\epsilon_\mathrm{C1s}^\mathrm{MaxQP}$.
For systems with more than one inequivalent core-level, i.e. all acenes but benzene, $\epsilon_\mathrm{C1s}^\mathrm{MaxQP}$ is a superposition of several C1s peaks.
Due to weak intermolecular interactions in molecular aggregates, the solid-state shake-up spectra are expected to be similar to the gas-phase spectra. However, in solid-state systems, additional effects such as extrinsic losses and vibrational broadening may modulate the intensity and peak shape. A comparison of the solid-state spectrum of benzene in Figure~\ref{fig:acenes} with the gas-phase data in Figure~\ref{fig:experimental_panel} confirms that the primary difference is the increased peak broadening in the solid-state case.
\par
The first satellite has been assigned to the lowest singlet ($S_1$) excitation, which is dominated by a $\pi-\pi^*$ (HOMO-LUMO) transition.
The $S_1$ satellite appears in the spectra of all acenes besides benzene, where the respective peak is excluded for symmetry reasons and only the triplet component is visible.\cite{rocco2008electronic}
We define $\Delta^\mathrm{S_1-MaxQP}$ as separation between $S_1$ and the maximum of the main line at $\epsilon^\mathrm{MaxQP}_\mathrm{C1s}$ and list the experimental  ($\Delta^\mathrm{S_1-MaxQP}_\mathrm{Exp}$) and theoretical ($\Delta^\mathrm{S_1-MaxQP}_\mathrm{GW+C}$) values in Table~\ref{tab:acenes_excitation}.
With increasing chain length, we observe that $\Delta^\mathrm{S_1-MaxQP}_\mathrm{Exp}$ decreases by more than $2 \, \mathrm{eV}$ going from naphthalene to pentacene, a trend which is is well reproduced by $GW+C$. The deviation of $\Delta^\mathrm{S_1-MaxQP}_\mathrm{Exp}$and $\Delta^\mathrm{S_1-MaxQP}_\mathrm{GW+C}$ is in the range of $0.8-0.1$~eV, gradually decreasing with the chain length. \par
Beyond the $S_1$ satellite, several broad, overlapping peaks appear between $-4$ and $-8$~eV.
For benzene, naphthalene and anthracene, the most intense satellite is correctly predicted by $GW+C$ and within $0.1-0.4$~eV of the experiment.
In the case of tetracene and pentacene, the position of the intense peak between $-4$ and $-5$~eV matches the reported experimental peaks equally well.
However, the intensity of the satellite is overestimated in the $GW+C$ spectral function, which might be explained by extrinsic effects like vibrations in the experiment.
\par
In Ref.~\citenum{rocco2008electronic}, it was expected that the lowest satellite excitation, i.e. $\Delta^\mathrm{S_1-MaxQP}_\mathrm{Exp}$, should coincide with the optical gap (see Table~\ref{tab:acenes_excitation}).
However, experimentally it was found that $\Delta^\mathrm{S_1-MaxQP}_\mathrm{Exp}$ is substantially lower by $0.6-0.9$~eV compared to the experimental optical gap.
This puzzling effect was interpreted in terms of a reorganization of the valence shell upon ionization, which seems to become more important with increasing chain length.
For pentacene, this effect was estimated to be as large as $0.8 \, \mathrm{eV}$, which accounts for roughly $50 \, \%$ of the optical gap.
However, here we stress that for systems with more than one C1s core-level, the interpretation of $\Delta^\mathrm{S_1-MaxQP}_\mathrm{Exp}$ as excitation energy is not accurate.
Since both the satellite peak and the QP peak are aggregated from the sum of all nonequivalent C1s levels to the spectral functions, the direct comparison of both makes the \textit{intrinsic assumption that all core-levels contribute equally to the satellite spectrum}.
In the following, we will demonstrate for pentacene that this is not the case.
\par
In Figure~\ref{fig:pentacene}, we resolve the six individual contributions C$_{1-6}$ to the C1s spectral function of pentacene.
For each individual core-level $\mathrm{C_x}$, the distance $\Delta^\mathrm{S_1-C_x}$ equals $\Omega^\mathrm{1}$, which is the optical gap in the RPA approximation.
By looking at the contributions from the individual core-levels \ce{C1}-\ce{C6}, we observe that most of the intensity of the $S_1$ satellite is contributed by the C1s levels \ce{C1} and \ce{C3}.
The other core-levels (\ce{C2}, \ce{C4}-\ce{C6}) couple only weakly to the $S_1$ excitation, and therefore the position of the $S_1$ satellite in the aggregated spectral function has to be interpreted relative to the \ce{C1} and \ce{C3} QP peaks at $\epsilon_\mathrm{C_{1/3}}^\mathrm{QP}$ and not to the maximum at $\epsilon_\mathrm{C1s}^\mathrm{MaxQP}$.
%The position of $\epsilon_\mathrm{C1s}^\mathrm{MaxQP}$ is determined by the majority of the QP peaks.
Since the \ce{C1} and \ce{C3} peaks at $\epsilon_\mathrm{C_{1/3}}^\mathrm{QP}$ appear at lower binding energies than the maximum of the QP, the $S_1$ satellite appears shifted by $\epsilon^\mathrm{MaxQP}_\mathrm{C1s}-\epsilon^\mathrm{QP}_\mathrm{C_{1/3}}\approx$~0.42~eV.
By adding this shift to $\Delta^\mathrm{S_1-MaxQP}_\mathrm{Exp}$, the difference between the optical gap and the satellite excitation energy is decreased by more than half to only $0.35 \, \mathrm{eV}$.
\par
This example highlights that in systems with several inequivalent core-level the exact knowledge of the generating level(s) is mandatory for the interpretation of shake-up satellites.
This kind of knowledge is provided by $GW+C$, and can be directly applied to interpret both gas-phase and solid-state XPS spectra of molecules.
\par
We point out that the system size presented here is still far from the actual computational limit of our method.
For pentacene, the largest system in this study with 22 core states and 1600 basis functions in total, the computation took ca. 5600~CPU hours.
Based on the results presented in the Ref. \citenum{panades2023accelerating}, we expect our $GW+C$ implementation to be applicable to systems with up to 100 atoms and even beyond, depending on the number of core-levels and the basis set.
\section{Conclusion}
\label{sec:Conclusion}
In this work, we derived a scalable $GW+C$ implementation for the prediction of molecular core-level satellites.
Building on Ref.~\citenum{panades2023accelerating}, we combined $GW+C$ with the CD-WAC approach in an efficient all-electron NAO framework, enabling the calculation of core-level spectral functions with $N^4$ scaling for systems with more than 100~atoms.
We derived several key recommendations for calculating core-level spectral functions with $GW+C$:
i) The decoupling approximation is essential for localized basis sets;
ii) The T2+aug2 basis set is suitable for accurate satellite properties, with additional core-level STOs improving QP energies;
iii) A PBE (GGA) starting point ensures reliable satellite properties.
We tested our computational framework for CO and benzene and the acene series up to pentacene, yielding an agreement of roughly 0.5~eV for the dominant satellite features and correctly predicting qualitative trends.
For pentacene, we demonstrated how $GW+C$ spectral functions can be a valuable tool for the interpretation of experimental data shake-up spectra, paving the way for future applications.
\par
Our ongoing and future work includes incorporating additional vertex corrections, e.g.,
%by using TDDFT instead of RPA in the computation of $W$ and 
by making use of recent developments connecting $GW$ and coupled cluster theory.\cite{vila_real-time_2020,tolle2023exact, lange2018relation, quintero2022connections} We also aim to develop a non-linear extension of the $GW+C$ method to avoid non-physical contributions in the spectral function in all cases.

\begin{acknowledgement}
\fontsize{10}{12}\selectfont

D.G. acknowledges funding by the Emmy Noether Programme of the German Research Foundation (project number 453275048). The authors acknowledge funding by the German Research Foundation (GRK2861–491865171).
We gratefully acknowledge the computing time provided on the high-performance computer Noctua 2 at the NHR Center PC2 and at the J\"ulich Supercomputer Center.
\end{acknowledgement}

\begin{suppinfo}
 The supplementary information is available free of charge.
 We include a discussion of the $G_{\Delta H}W_0$ quasiparticle peak and a detailed derivation of Eq.~\eqref{im_sigma_satellites}.
 Furthermore, we provide a comparison of the CD-WAC fit for diagonal and off-diagonal matrix elements.
 We also include a technical description of our satellite selection criterion used for the calculation of Eq.~\eqref{sat_splitting}. Tables~S1 and S2 include statistics for the satellite- and quasiparticle energies for the CORE65 benchmark set. In addition, we provide $GW+C$ spectral functions based on starting points with various amount of exact exchange for \ce{CH4} and CO.
 $G_{\Delta H}W_0$ core-level binding energies are provided for the CORE65 benchmark set for all 22 basis sets in the accompanying text file.
\end{suppinfo}

\bibliography{references}

\providecommand{\latin}[1]{#1}
\makeatletter
\providecommand{\doi}
  {\begingroup\let\do\@makeother\dospecials
  \catcode`\{=1 \catcode`\}=2 \doi@aux}
\providecommand{\doi@aux}[1]{\endgroup\texttt{#1}}
\makeatother
\providecommand*\mcitethebibliography{\thebibliography}
\csname @ifundefined\endcsname{endmcitethebibliography}  {\let\endmcitethebibliography\endthebibliography}{}
\begin{mcitethebibliography}{132}
\providecommand*\natexlab[1]{#1}
\providecommand*\mciteSetBstSublistMode[1]{}
\providecommand*\mciteSetBstMaxWidthForm[2]{}
\providecommand*\mciteBstWouldAddEndPuncttrue
  {\def\EndOfBibitem{\unskip.}}
\providecommand*\mciteBstWouldAddEndPunctfalse
  {\let\EndOfBibitem\relax}
\providecommand*\mciteSetBstMidEndSepPunct[3]{}
\providecommand*\mciteSetBstSublistLabelBeginEnd[3]{}
\providecommand*\EndOfBibitem{}
\mciteSetBstSublistMode{f}
\mciteSetBstMaxWidthForm{subitem}{(\alph{mcitesubitemcount})}
\mciteSetBstSublistLabelBeginEnd
  {\mcitemaxwidthsubitemform\space}
  {\relax}
  {\relax}

\bibitem[Fadley(2010)]{fadley2010x}
Fadley,~C.~S. X-ray photoelectron spectroscopy: {Progress} and perspectives. \emph{J. Electron Spectros. Relat. Phenomena} \textbf{2010}, \emph{178}, 2--32\relax
\mciteBstWouldAddEndPuncttrue
\mciteSetBstMidEndSepPunct{\mcitedefaultmidpunct}
{\mcitedefaultendpunct}{\mcitedefaultseppunct}\relax
\EndOfBibitem
\bibitem[Bagus \latin{et~al.}(2013)Bagus, Ilton, and Nelin]{bagus2013interpretation}
Bagus,~P.~S.; Ilton,~E.~S.; Nelin,~C.~J. The interpretation of {XPS} spectra: {Insights} into materials properties. \emph{Surf. Sci. Rep.} \textbf{2013}, \emph{68}, 273--304\relax
\mciteBstWouldAddEndPuncttrue
\mciteSetBstMidEndSepPunct{\mcitedefaultmidpunct}
{\mcitedefaultendpunct}{\mcitedefaultseppunct}\relax
\EndOfBibitem
\bibitem[Bagus \latin{et~al.}(2018)Bagus, Ilton, and Nelin]{bagus2018extracting}
Bagus,~P.~S.; Ilton,~E.; Nelin,~C.~J. Extracting chemical information from {XPS} spectra: a perspective. \emph{Catal. Lett.} \textbf{2018}, \emph{148}, 1785--1802\relax
\mciteBstWouldAddEndPuncttrue
\mciteSetBstMidEndSepPunct{\mcitedefaultmidpunct}
{\mcitedefaultendpunct}{\mcitedefaultseppunct}\relax
\EndOfBibitem
\bibitem[Bernadotte \latin{et~al.}(2013)Bernadotte, Evers, and Jacob]{bernadotte2013plasmons}
Bernadotte,~S.; Evers,~F.; Jacob,~C.~R. Plasmons in molecules. \emph{J. Phys. Chem. C} \textbf{2013}, \emph{117}, 1863--1878\relax
\mciteBstWouldAddEndPuncttrue
\mciteSetBstMidEndSepPunct{\mcitedefaultmidpunct}
{\mcitedefaultendpunct}{\mcitedefaultseppunct}\relax
\EndOfBibitem
\bibitem[Van~Veenendaal and Sawatzky(1993)Van~Veenendaal, and Sawatzky]{van1993nonlocal}
Van~Veenendaal,~M.; Sawatzky,~G. Nonlocal screening effects in 2p x-ray photoemission spectroscopy core-level line shapes of transition metal compounds. \emph{Phys. Rev. Lett.} \textbf{1993}, \emph{70}, 2459\relax
\mciteBstWouldAddEndPuncttrue
\mciteSetBstMidEndSepPunct{\mcitedefaultmidpunct}
{\mcitedefaultendpunct}{\mcitedefaultseppunct}\relax
\EndOfBibitem
\bibitem[De~Groot and Kotani(2008)De~Groot, and Kotani]{de2008core}
De~Groot,~F.; Kotani,~A. \emph{Core level spectroscopy of solids}; CRC press, 2008\relax
\mciteBstWouldAddEndPuncttrue
\mciteSetBstMidEndSepPunct{\mcitedefaultmidpunct}
{\mcitedefaultendpunct}{\mcitedefaultseppunct}\relax
\EndOfBibitem
\bibitem[Martin and Shirley(1976)Martin, and Shirley]{martin1976theory}
Martin,~R.; Shirley,~D. Theory of core-level photoemission correlation state spectra. \emph{J. Chem. Phys.} \textbf{1976}, \emph{64}, 3685--3689\relax
\mciteBstWouldAddEndPuncttrue
\mciteSetBstMidEndSepPunct{\mcitedefaultmidpunct}
{\mcitedefaultendpunct}{\mcitedefaultseppunct}\relax
\EndOfBibitem
\bibitem[Cederbaum(1974)]{cederbaum1974application}
Cederbaum,~L. Application of {Green}'s functions to excitations accompanying photoionization in atoms and molecules. \emph{Mol. Phys.} \textbf{1974}, \emph{28}, 479--493\relax
\mciteBstWouldAddEndPuncttrue
\mciteSetBstMidEndSepPunct{\mcitedefaultmidpunct}
{\mcitedefaultendpunct}{\mcitedefaultseppunct}\relax
\EndOfBibitem
\bibitem[Cederbaum and Domcke(1977)Cederbaum, and Domcke]{cederbaum1977theoretical}
Cederbaum,~L.; Domcke,~W. Theoretical aspects of ionization potentials and photoelectron spectroscopy: {A} {Green}’s function approach. \emph{Adv. Chem. Phys.} \textbf{1977}, \emph{36}, 205--344\relax
\mciteBstWouldAddEndPuncttrue
\mciteSetBstMidEndSepPunct{\mcitedefaultmidpunct}
{\mcitedefaultendpunct}{\mcitedefaultseppunct}\relax
\EndOfBibitem
\bibitem[Cederbaum \latin{et~al.}(1986)Cederbaum, Domcke, Schirmer, and Niessen]{cederbaum1986correlation}
Cederbaum,~L.; Domcke,~W.; Schirmer,~J.; Niessen,~W.~v. Correlation effects in the ionization of molecules: {Breakdown} of the molecular orbital picture. \emph{Adv. Chem. Phys.} \textbf{1986}, 115--159\relax
\mciteBstWouldAddEndPuncttrue
\mciteSetBstMidEndSepPunct{\mcitedefaultmidpunct}
{\mcitedefaultendpunct}{\mcitedefaultseppunct}\relax
\EndOfBibitem
\bibitem[Norman and Dreuw(2018)Norman, and Dreuw]{norman2018simulating}
Norman,~P.; Dreuw,~A. Simulating {X}-ray spectroscopies and calculating core-excited states of molecules. \emph{Chem. Rev.} \textbf{2018}, \emph{118}, 7208--7248\relax
\mciteBstWouldAddEndPuncttrue
\mciteSetBstMidEndSepPunct{\mcitedefaultmidpunct}
{\mcitedefaultendpunct}{\mcitedefaultseppunct}\relax
\EndOfBibitem
\bibitem[Schirmer and Cederbaum(1978)Schirmer, and Cederbaum]{schirmer1978two}
Schirmer,~J.; Cederbaum,~L. The two-particle-hole {Tamm}-{Dancoff} approximation (2ph-{TDA}) equations for closed-shell atoms and molecules. \emph{J. Phys. B} \textbf{1978}, \emph{11}, 1889\relax
\mciteBstWouldAddEndPuncttrue
\mciteSetBstMidEndSepPunct{\mcitedefaultmidpunct}
{\mcitedefaultendpunct}{\mcitedefaultseppunct}\relax
\EndOfBibitem
\bibitem[Cederbaum \latin{et~al.}(1980)Cederbaum, Domcke, and Schirmer]{cederbaum1980many}
Cederbaum,~L.~S.; Domcke,~W.; Schirmer,~J. Many-body theory of core holes. \emph{Phys. Rev. A} \textbf{1980}, \emph{22}, 206\relax
\mciteBstWouldAddEndPuncttrue
\mciteSetBstMidEndSepPunct{\mcitedefaultmidpunct}
{\mcitedefaultendpunct}{\mcitedefaultseppunct}\relax
\EndOfBibitem
\bibitem[von Niessen \latin{et~al.}(1984)von Niessen, Schirmer, and Cederbaum]{von1984computational}
von Niessen,~W.; Schirmer,~J.; Cederbaum,~L.~S. Computational methods for the one-particle green's function. \emph{Comput. Phys. Rep.} \textbf{1984}, \emph{1}, 57--125\relax
\mciteBstWouldAddEndPuncttrue
\mciteSetBstMidEndSepPunct{\mcitedefaultmidpunct}
{\mcitedefaultendpunct}{\mcitedefaultseppunct}\relax
\EndOfBibitem
\bibitem[Brisk and Baker(1975)Brisk, and Baker]{brisk1975shake}
Brisk,~M.~A.; Baker,~A. Shake-up satellites in {X}-ray photoelectron spectroscopy. \emph{J. Electron Spectros. Relat. Phenomena} \textbf{1975}, \emph{7}, 197--213\relax
\mciteBstWouldAddEndPuncttrue
\mciteSetBstMidEndSepPunct{\mcitedefaultmidpunct}
{\mcitedefaultendpunct}{\mcitedefaultseppunct}\relax
\EndOfBibitem
\bibitem[Arneberg \latin{et~al.}(1982)Arneberg, Müller, and Manne]{arneberg1982configuration}
Arneberg,~R.; Müller,~J.; Manne,~R. Configuration interaction calculations of satellite structure in photoelectron spectra of {H2O}. \emph{Chem. Phys.} \textbf{1982}, \emph{64}, 249--258\relax
\mciteBstWouldAddEndPuncttrue
\mciteSetBstMidEndSepPunct{\mcitedefaultmidpunct}
{\mcitedefaultendpunct}{\mcitedefaultseppunct}\relax
\EndOfBibitem
\bibitem[Lunell \latin{et~al.}(1989)Lunell, Keane, and Svensson]{lunell1989semiempirical}
Lunell,~S.; Keane,~M.; Svensson,~S. Semiempirical configuration interaction calculations of shake-up satellites in formaldehyde, benzene, and benzaldehyde. \emph{J. Chem. Phys.} \textbf{1989}, \emph{90}, 4341--4350\relax
\mciteBstWouldAddEndPuncttrue
\mciteSetBstMidEndSepPunct{\mcitedefaultmidpunct}
{\mcitedefaultendpunct}{\mcitedefaultseppunct}\relax
\EndOfBibitem
\bibitem[Chatterjee and Sokolov(2019)Chatterjee, and Sokolov]{chatterjee2019second}
Chatterjee,~K.; Sokolov,~A.~Y. Second-order multireference algebraic diagrammatic construction theory for photoelectron spectra of strongly correlated systems. \emph{J. Chem. Theory Comput.} \textbf{2019}, \emph{15}, 5908--5924\relax
\mciteBstWouldAddEndPuncttrue
\mciteSetBstMidEndSepPunct{\mcitedefaultmidpunct}
{\mcitedefaultendpunct}{\mcitedefaultseppunct}\relax
\EndOfBibitem
\bibitem[Marie and Loos(2024)Marie, and Loos]{marie2024reference}
Marie,~A.; Loos,~P.-F. Reference {Energies} for {Valence} {Ionizations} and {Satellite} {Transitions}. \emph{J. Chem. Theory Comput.} \textbf{2024}, \relax
\mciteBstWouldAddEndPunctfalse
\mciteSetBstMidEndSepPunct{\mcitedefaultmidpunct}
{}{\mcitedefaultseppunct}\relax
\EndOfBibitem
\bibitem[Nakatsuji(1991)]{nakatsuji1991description}
Nakatsuji,~H. Description of two-and many-electron processes by the {SAC}-{CI} method. \emph{Chem. Phys. Lett.} \textbf{1991}, \emph{177}, 331--337\relax
\mciteBstWouldAddEndPuncttrue
\mciteSetBstMidEndSepPunct{\mcitedefaultmidpunct}
{\mcitedefaultendpunct}{\mcitedefaultseppunct}\relax
\EndOfBibitem
\bibitem[Kuramoto \latin{et~al.}(2005)Kuramoto, Ehara, and Nakatsuji]{kuramoto2005theoretical}
Kuramoto,~K.; Ehara,~M.; Nakatsuji,~H. Theoretical fine spectroscopy with symmetry adapted cluster–configuration interaction general-{R} method: {First}-row {K}-shell ionizations and their satellites. \emph{J. Chem. Phys.} \textbf{2005}, \emph{122}\relax
\mciteBstWouldAddEndPuncttrue
\mciteSetBstMidEndSepPunct{\mcitedefaultmidpunct}
{\mcitedefaultendpunct}{\mcitedefaultseppunct}\relax
\EndOfBibitem
\bibitem[Ehara \latin{et~al.}(2005)Ehara, Hasegawa, and Nakatsuji]{ehara2005sac}
Ehara,~M.; Hasegawa,~J.; Nakatsuji,~H. \emph{Theory and {Applications} of {Computational} {Chemistry}}; Elsevier, 2005; pp 1099--1141\relax
\mciteBstWouldAddEndPuncttrue
\mciteSetBstMidEndSepPunct{\mcitedefaultmidpunct}
{\mcitedefaultendpunct}{\mcitedefaultseppunct}\relax
\EndOfBibitem
\bibitem[Lisini and Decleva(1992)Lisini, and Decleva]{lisini1992calculation}
Lisini,~A.; Decleva,~P. Calculation of dynamical correlation effects by quasidegenerate perturbation theory. {An} application to photoionization spectra. \emph{Chem. Phys.} \textbf{1992}, \emph{168}, 1--13\relax
\mciteBstWouldAddEndPuncttrue
\mciteSetBstMidEndSepPunct{\mcitedefaultmidpunct}
{\mcitedefaultendpunct}{\mcitedefaultseppunct}\relax
\EndOfBibitem
\bibitem[Lisini \latin{et~al.}(1993)Lisini, Decleva, and Fronzoni]{lisini1993quasidegenerate}
Lisini,~A.; Decleva,~P.; Fronzoni,~G. Quasidegenerate perturbation theory applied to the calculation of excitation spectra. \emph{Chem. Phys.} \textbf{1993}, \emph{171}, 159--170\relax
\mciteBstWouldAddEndPuncttrue
\mciteSetBstMidEndSepPunct{\mcitedefaultmidpunct}
{\mcitedefaultendpunct}{\mcitedefaultseppunct}\relax
\EndOfBibitem
\bibitem[Fronzoni \latin{et~al.}(1999)Fronzoni, De~Alti, and Decleva]{fronzoni1999theoretical}
Fronzoni,~G.; De~Alti,~G.; Decleva,~P. Theoretical {C1s} and {O1s} core shake-up spectra of {CO} by highly correlated {QDPTCI} approach. \emph{J. Phys. B} \textbf{1999}, \emph{32}, 5357\relax
\mciteBstWouldAddEndPuncttrue
\mciteSetBstMidEndSepPunct{\mcitedefaultmidpunct}
{\mcitedefaultendpunct}{\mcitedefaultseppunct}\relax
\EndOfBibitem
\bibitem[Rehr \latin{et~al.}(2020)Rehr, Vila, Kas, Hirshberg, Kowalski, and Peng]{rehr2020equation}
Rehr,~J.~J.; Vila,~F.~D.; Kas,~J.~J.; Hirshberg,~N.~Y.; Kowalski,~K.; Peng,~B. Equation of motion coupled-cluster cumulant approach for intrinsic losses in x-ray spectra. \emph{J. Chem. Phys.} \textbf{2020}, \emph{152}\relax
\mciteBstWouldAddEndPuncttrue
\mciteSetBstMidEndSepPunct{\mcitedefaultmidpunct}
{\mcitedefaultendpunct}{\mcitedefaultseppunct}\relax
\EndOfBibitem
\bibitem[Vila \latin{et~al.}(2022)Vila, Kowalski, Peng, Kas, and Rehr]{vila_real-time_2020}
Vila,~F.; Kowalski,~K.; Peng,~B.; Kas,~J.; Rehr,~J. Real-{Time} {Equation}-of-{Motion} {CCSD} {Cumulant} {Green}’s {Function}. \emph{J. Chem. Theory Comput.} \textbf{2022}, \emph{18}, 1799--1807\relax
\mciteBstWouldAddEndPuncttrue
\mciteSetBstMidEndSepPunct{\mcitedefaultmidpunct}
{\mcitedefaultendpunct}{\mcitedefaultseppunct}\relax
\EndOfBibitem
\bibitem[Vila \latin{et~al.}(2021)Vila, Kas, Rehr, Kowalski, and Peng]{vila2021equation}
Vila,~F.~D.; Kas,~J.~J.; Rehr,~J.~J.; Kowalski,~K.; Peng,~B. Equation-of-{Motion} {Coupled}-{Cluster} {Cumulant} {Green}’s {Function} for {Excited} {States} and {X}-{Ray} {Spectra}. \emph{Front. Chem.} \textbf{2021}, \emph{9}, 734945\relax
\mciteBstWouldAddEndPuncttrue
\mciteSetBstMidEndSepPunct{\mcitedefaultmidpunct}
{\mcitedefaultendpunct}{\mcitedefaultseppunct}\relax
\EndOfBibitem
\bibitem[Vila \latin{et~al.}(2022)Vila, Kowalski, Peng, Kas, and Rehr]{vila2022real}
Vila,~F.; Kowalski,~K.; Peng,~B.; Kas,~J.; Rehr,~J. Real-{Time} {Equation}-of-{Motion} {CCSD} {Cumulant} {Green}’s {Function}. \emph{J. Chem. Theory Comput.} \textbf{2022}, \emph{18}, 1799--1807\relax
\mciteBstWouldAddEndPuncttrue
\mciteSetBstMidEndSepPunct{\mcitedefaultmidpunct}
{\mcitedefaultendpunct}{\mcitedefaultseppunct}\relax
\EndOfBibitem
\bibitem[Sankari \latin{et~al.}(2006)Sankari, Ehara, Nakatsuji, De~Fanis, Aksela, Sorensen, Piancastelli, Kukk, and Ueda]{sankari2006high}
Sankari,~R.; Ehara,~M.; Nakatsuji,~H.; De~Fanis,~A.; Aksela,~H.; Sorensen,~S.; Piancastelli,~M.; Kukk,~E.; Ueda,~K. High resolution {O} 1s photoelectron shake-up satellite spectrum of {H2O}. \emph{Chem. Phys. Lett.} \textbf{2006}, \emph{422}, 51--57\relax
\mciteBstWouldAddEndPuncttrue
\mciteSetBstMidEndSepPunct{\mcitedefaultmidpunct}
{\mcitedefaultendpunct}{\mcitedefaultseppunct}\relax
\EndOfBibitem
\bibitem[Schirmer \latin{et~al.}(1987)Schirmer, Angonoa, Svensson, Nordfors, and Gelius]{schirmer1987high}
Schirmer,~J.; Angonoa,~G.; Svensson,~S.; Nordfors,~D.; Gelius,~U. High-energy photoelectron {C} 1s and {O} 1s shake-up spectra of {CO}. \emph{J. Phys. B} \textbf{1987}, \emph{20}, 6031\relax
\mciteBstWouldAddEndPuncttrue
\mciteSetBstMidEndSepPunct{\mcitedefaultmidpunct}
{\mcitedefaultendpunct}{\mcitedefaultseppunct}\relax
\EndOfBibitem
\bibitem[Angonoa \latin{et~al.}(1987)Angonoa, Walter, and Schirmer]{angonoa1987theoretical}
Angonoa,~G.; Walter,~O.; Schirmer,~J. Theoretical {K}-shell ionization spectra of {N2} and {CO} by a fourth-order {Green}’s function method. \emph{J. Chem. Phys.} \textbf{1987}, \emph{87}, 6789--6801\relax
\mciteBstWouldAddEndPuncttrue
\mciteSetBstMidEndSepPunct{\mcitedefaultmidpunct}
{\mcitedefaultendpunct}{\mcitedefaultseppunct}\relax
\EndOfBibitem
\bibitem[Ehara \latin{et~al.}(2006)Ehara, Kuramoto, Nakatsuji, Hoshino, Tanaka, Kitajima, Tanaka, De~Fanis, Tamenori, and Ueda]{ehara2006c1s}
Ehara,~M.; Kuramoto,~K.; Nakatsuji,~H.; Hoshino,~M.; Tanaka,~T.; Kitajima,~M.; Tanaka,~H.; De~Fanis,~A.; Tamenori,~Y.; Ueda,~K. C1s and {O1s} photoelectron satellite spectra of {CO} with symmetry-dependent vibrational excitations. \emph{J. Chem. Phys.} \textbf{2006}, \emph{125}\relax
\mciteBstWouldAddEndPuncttrue
\mciteSetBstMidEndSepPunct{\mcitedefaultmidpunct}
{\mcitedefaultendpunct}{\mcitedefaultseppunct}\relax
\EndOfBibitem
\bibitem[Kuramoto \latin{et~al.}(2005)Kuramoto, Ehara, Nakatsuji, Kitajima, Tanaka, De~Fanis, Tamenori, and Ueda]{kuramoto2005c}
Kuramoto,~K.; Ehara,~M.; Nakatsuji,~H.; Kitajima,~M.; Tanaka,~H.; De~Fanis,~A.; Tamenori,~Y.; Ueda,~K. C 1s and {O} 1s photoelectron spectra of formaldehyde with satellites: theory and experiment. \emph{J. Electron Spectros. Relat. Phenomena} \textbf{2005}, \emph{142}, 253--259\relax
\mciteBstWouldAddEndPuncttrue
\mciteSetBstMidEndSepPunct{\mcitedefaultmidpunct}
{\mcitedefaultendpunct}{\mcitedefaultseppunct}\relax
\EndOfBibitem
\bibitem[Angonoa and Schirmer(1989)Angonoa, and Schirmer]{angonoa1989theoretical}
Angonoa,~G.; Schirmer,~J. Theoretical {K}-shell photoelectron spectra of {H2CO} and {C2H2}. \emph{J. Mol. Struct. THEOCHEM} \textbf{1989}, \emph{202}, 203--211\relax
\mciteBstWouldAddEndPuncttrue
\mciteSetBstMidEndSepPunct{\mcitedefaultmidpunct}
{\mcitedefaultendpunct}{\mcitedefaultseppunct}\relax
\EndOfBibitem
\bibitem[Pathak \latin{et~al.}(2023)Pathak, Panyala, Peng, Bauman, Mutlu, Rehr, Vila, and Kowalski]{pathak2023real}
Pathak,~H.; Panyala,~A.; Peng,~B.; Bauman,~N.~P.; Mutlu,~E.; Rehr,~J.~J.; Vila,~F.~D.; Kowalski,~K. Real-time equation-of-motion coupled-cluster cumulant green’s function method: {Heterogeneous} parallel implementation based on the tensor algebra for many-body methods infrastructure. \emph{J. Chem. Theory Comput.} \textbf{2023}, \emph{19}, 2248--2257\relax
\mciteBstWouldAddEndPuncttrue
\mciteSetBstMidEndSepPunct{\mcitedefaultmidpunct}
{\mcitedefaultendpunct}{\mcitedefaultseppunct}\relax
\EndOfBibitem
\bibitem[Hedin(1965)]{hedin1965new}
Hedin,~L. New method for calculating the one-particle {Green}'s function with application to the electron-gas problem. \emph{Phys. Rev.} \textbf{1965}, \emph{139}, A796\relax
\mciteBstWouldAddEndPuncttrue
\mciteSetBstMidEndSepPunct{\mcitedefaultmidpunct}
{\mcitedefaultendpunct}{\mcitedefaultseppunct}\relax
\EndOfBibitem
\bibitem[Onida \latin{et~al.}(2002)Onida, Reining, and Rubio]{onida_electronic_2002}
Onida,~G.; Reining,~L.; Rubio,~A. Electronic excitations: density-functional versus many-body {Green}'s-function approaches. \emph{Rev. Mod. Phys.} \textbf{2002}, \emph{74}, 601--659\relax
\mciteBstWouldAddEndPuncttrue
\mciteSetBstMidEndSepPunct{\mcitedefaultmidpunct}
{\mcitedefaultendpunct}{\mcitedefaultseppunct}\relax
\EndOfBibitem
\bibitem[Martin \latin{et~al.}(2016)Martin, Reining, and Ceperley]{martin_interacting_2016}
Martin,~R.~M.; Reining,~L.; Ceperley,~D.~M. \emph{Interacting {Electrons}}; 2016\relax
\mciteBstWouldAddEndPuncttrue
\mciteSetBstMidEndSepPunct{\mcitedefaultmidpunct}
{\mcitedefaultendpunct}{\mcitedefaultseppunct}\relax
\EndOfBibitem
\bibitem[Golze \latin{et~al.}(2019)Golze, Dvorak, and Rinke]{golze_gw_2019}
Golze,~D.; Dvorak,~M.; Rinke,~P. The {GW} {Compendium}: {A} {Practical} {Guide} to {Theoretical} {Photoemission} {Spectroscopy}. \emph{Front. Chem.} \textbf{2019}, \emph{7}\relax
\mciteBstWouldAddEndPuncttrue
\mciteSetBstMidEndSepPunct{\mcitedefaultmidpunct}
{\mcitedefaultendpunct}{\mcitedefaultseppunct}\relax
\EndOfBibitem
\bibitem[Van~Setten \latin{et~al.}(2015)Van~Setten, Caruso, Sharifzadeh, Ren, Scheffler, Liu, Lischner, Lin, Deslippe, Louie, and {others}]{van2015gw}
Van~Setten,~M.~J.; Caruso,~F.; Sharifzadeh,~S.; Ren,~X.; Scheffler,~M.; Liu,~F.; Lischner,~J.; Lin,~L.; Deslippe,~J.~R.; Louie,~S.~G.; {others} {GW} 100: {Benchmarking} {G} 0 {W} 0 for molecular systems. \emph{J. Chem. Theory Comput.} \textbf{2015}, \emph{11}, 5665--5687\relax
\mciteBstWouldAddEndPuncttrue
\mciteSetBstMidEndSepPunct{\mcitedefaultmidpunct}
{\mcitedefaultendpunct}{\mcitedefaultseppunct}\relax
\EndOfBibitem
\bibitem[Bruneval \latin{et~al.}(2021)Bruneval, Dattani, and van Setten]{bruneval2021gw}
Bruneval,~F.; Dattani,~N.; van Setten,~M.~J. The {GW} miracle in many-body perturbation theory for the ionization potential of molecules. \emph{Front. Chem.} \textbf{2021}, \emph{9}, 749779\relax
\mciteBstWouldAddEndPuncttrue
\mciteSetBstMidEndSepPunct{\mcitedefaultmidpunct}
{\mcitedefaultendpunct}{\mcitedefaultseppunct}\relax
\EndOfBibitem
\bibitem[Langreth(1970)]{langreth1970singularities}
Langreth,~D.~C. Singularities in the x-ray spectra of metals. \emph{Phys. Rev. B} \textbf{1970}, \emph{1}, 471\relax
\mciteBstWouldAddEndPuncttrue
\mciteSetBstMidEndSepPunct{\mcitedefaultmidpunct}
{\mcitedefaultendpunct}{\mcitedefaultseppunct}\relax
\EndOfBibitem
\bibitem[Aryasetiawan \latin{et~al.}(1996)Aryasetiawan, Hedin, and Karlsson]{aryasetiawan1996multiple}
Aryasetiawan,~F.; Hedin,~L.; Karlsson,~K. Multiple plasmon satellites in {Na} and {Al} spectral functions from ab initio cumulant expansion. \emph{Phys. Rev. Lett.} \textbf{1996}, \emph{77}, 2268\relax
\mciteBstWouldAddEndPuncttrue
\mciteSetBstMidEndSepPunct{\mcitedefaultmidpunct}
{\mcitedefaultendpunct}{\mcitedefaultseppunct}\relax
\EndOfBibitem
\bibitem[Zhou \latin{et~al.}(2015)Zhou, Kas, Sponza, Reshetnyak, Guzzo, Giorgetti, Gatti, Sottile, Rehr, and Reining]{zhou_dynamical_2015}
Zhou,~J.~S.; Kas,~J.~J.; Sponza,~L.; Reshetnyak,~I.; Guzzo,~M.; Giorgetti,~C.; Gatti,~M.; Sottile,~F.; Rehr,~J.~J.; Reining,~L. Dynamical effects in electron spectroscopy. \emph{J. Chem. Phys.} \textbf{2015}, \emph{143}, 184109\relax
\mciteBstWouldAddEndPuncttrue
\mciteSetBstMidEndSepPunct{\mcitedefaultmidpunct}
{\mcitedefaultendpunct}{\mcitedefaultseppunct}\relax
\EndOfBibitem
\bibitem[Vigil-Fowler \latin{et~al.}(2016)Vigil-Fowler, Louie, and Lischner]{vigil2016dispersion}
Vigil-Fowler,~D.; Louie,~S.~G.; Lischner,~J. Dispersion and line shape of plasmon satellites in one, two, and three dimensions. \emph{Phys. Rev. B} \textbf{2016}, \emph{93}, 235446\relax
\mciteBstWouldAddEndPuncttrue
\mciteSetBstMidEndSepPunct{\mcitedefaultmidpunct}
{\mcitedefaultendpunct}{\mcitedefaultseppunct}\relax
\EndOfBibitem
\bibitem[McClain \latin{et~al.}(2016)McClain, Lischner, Watson, Matthews, Ronca, Louie, Berkelbach, and Chan]{mcclain2016spectral}
McClain,~J.; Lischner,~J.; Watson,~T.; Matthews,~D.~A.; Ronca,~E.; Louie,~S.~G.; Berkelbach,~T.~C.; Chan,~G. K.-L. Spectral functions of the uniform electron gas via coupled-cluster theory and comparison to the {GW} and related approximations. \emph{Phys. Rev. B} \textbf{2016}, \emph{93}, 235139\relax
\mciteBstWouldAddEndPuncttrue
\mciteSetBstMidEndSepPunct{\mcitedefaultmidpunct}
{\mcitedefaultendpunct}{\mcitedefaultseppunct}\relax
\EndOfBibitem
\bibitem[Zhou \latin{et~al.}(2018)Zhou, Gatti, Kas, Rehr, and Reining]{zhou2018cumulant}
Zhou,~J.~S.; Gatti,~M.; Kas,~J.; Rehr,~J.; Reining,~L. Cumulant {Green}'s function calculations of plasmon satellites in bulk sodium: {Influence} of screening and the crystal environment. \emph{Phys. Rev. B} \textbf{2018}, \emph{97}, 035137\relax
\mciteBstWouldAddEndPuncttrue
\mciteSetBstMidEndSepPunct{\mcitedefaultmidpunct}
{\mcitedefaultendpunct}{\mcitedefaultseppunct}\relax
\EndOfBibitem
\bibitem[Vos \latin{et~al.}(1999)Vos, Kheifets, Weigold, Canney, Holm, Aryasetiawan, and Karlsson]{vos1999determination}
Vos,~M.; Kheifets,~A.; Weigold,~E.; Canney,~S.; Holm,~B.; Aryasetiawan,~F.; Karlsson,~K. Determination of the energy-momentum densities of aluminium by electron momentum spectroscopy. \emph{J. Phys. Condens. Matter} \textbf{1999}, \emph{11}, 3645\relax
\mciteBstWouldAddEndPuncttrue
\mciteSetBstMidEndSepPunct{\mcitedefaultmidpunct}
{\mcitedefaultendpunct}{\mcitedefaultseppunct}\relax
\EndOfBibitem
\bibitem[Vos \latin{et~al.}(2002)Vos, Kheifets, Sashin, Weigold, Usuda, and Aryasetiawan]{vos2002quantitative}
Vos,~M.; Kheifets,~A.; Sashin,~V.; Weigold,~E.; Usuda,~M.; Aryasetiawan,~F. Quantitative measurement of the spectral function of aluminum and lithium by electron momentum spectroscopy. \emph{Phys. Rev. B} \textbf{2002}, \emph{66}, 155414\relax
\mciteBstWouldAddEndPuncttrue
\mciteSetBstMidEndSepPunct{\mcitedefaultmidpunct}
{\mcitedefaultendpunct}{\mcitedefaultseppunct}\relax
\EndOfBibitem
\bibitem[Kas \latin{et~al.}(2016)Kas, Rehr, and Curtis]{kas2016particle}
Kas,~J.; Rehr,~J.; Curtis,~J. Particle-hole cumulant approach for inelastic losses in x-ray spectra. \emph{Phys. Rev. B} \textbf{2016}, \emph{94}, 035156\relax
\mciteBstWouldAddEndPuncttrue
\mciteSetBstMidEndSepPunct{\mcitedefaultmidpunct}
{\mcitedefaultendpunct}{\mcitedefaultseppunct}\relax
\EndOfBibitem
\bibitem[Guzzo \latin{et~al.}(2011)Guzzo, Lani, Sottile, Romaniello, Gatti, Kas, Rehr, Silly, Sirotti, and Reining]{guzzo2011valence}
Guzzo,~M.; Lani,~G.; Sottile,~F.; Romaniello,~P.; Gatti,~M.; Kas,~J.~J.; Rehr,~J.~J.; Silly,~M.~G.; Sirotti,~F.; Reining,~L. Valence electron photoemission spectrum of semiconductors: {Ab} initio description of multiple satellites. \emph{Phys. Rev. Lett.} \textbf{2011}, \emph{107}, 166401\relax
\mciteBstWouldAddEndPuncttrue
\mciteSetBstMidEndSepPunct{\mcitedefaultmidpunct}
{\mcitedefaultendpunct}{\mcitedefaultseppunct}\relax
\EndOfBibitem
\bibitem[Kheifets \latin{et~al.}(2003)Kheifets, Sashin, Vos, Weigold, and Aryasetiawan]{kheifets2003spectral}
Kheifets,~A.; Sashin,~V.; Vos,~M.; Weigold,~E.; Aryasetiawan,~F. Spectral properties of quasiparticles in silicon: {A} test of many-body theory. \emph{Phys. Rev. B} \textbf{2003}, \emph{68}, 233205\relax
\mciteBstWouldAddEndPuncttrue
\mciteSetBstMidEndSepPunct{\mcitedefaultmidpunct}
{\mcitedefaultendpunct}{\mcitedefaultseppunct}\relax
\EndOfBibitem
\bibitem[Lischner \latin{et~al.}(2015)Lischner, Pálsson, Vigil-Fowler, Nemsak, Avila, Asensio, Fadley, and Louie]{lischner2015satellite}
Lischner,~J.; Pálsson,~G.; Vigil-Fowler,~D.; Nemsak,~S.; Avila,~J.; Asensio,~M.; Fadley,~C.; Louie,~S.~G. Satellite band structure in silicon caused by electron-plasmon coupling. \emph{Phys. Rev. B} \textbf{2015}, \emph{91}, 205113\relax
\mciteBstWouldAddEndPuncttrue
\mciteSetBstMidEndSepPunct{\mcitedefaultmidpunct}
{\mcitedefaultendpunct}{\mcitedefaultseppunct}\relax
\EndOfBibitem
\bibitem[Caruso \latin{et~al.}(2015)Caruso, Lambert, and Giustino]{caruso2015band}
Caruso,~F.; Lambert,~H.; Giustino,~F. Band structures of plasmonic polarons. \emph{Phys. Rev. Lett.} \textbf{2015}, \emph{114}, 146404\relax
\mciteBstWouldAddEndPuncttrue
\mciteSetBstMidEndSepPunct{\mcitedefaultmidpunct}
{\mcitedefaultendpunct}{\mcitedefaultseppunct}\relax
\EndOfBibitem
\bibitem[Gumhalter \latin{et~al.}(2016)Gumhalter, Kovač, Caruso, Lambert, and Giustino]{gumhalter2016combined}
Gumhalter,~B.; Kovač,~V.; Caruso,~F.; Lambert,~H.; Giustino,~F. On the combined use of {GW} approximation and cumulant expansion in the calculations of quasiparticle spectra: {The} paradigm of {Si} valence bands. \emph{Phys. Rev. B} \textbf{2016}, \emph{94}, 035103\relax
\mciteBstWouldAddEndPuncttrue
\mciteSetBstMidEndSepPunct{\mcitedefaultmidpunct}
{\mcitedefaultendpunct}{\mcitedefaultseppunct}\relax
\EndOfBibitem
\bibitem[Lischner \latin{et~al.}(2013)Lischner, Vigil-Fowler, and Louie]{lischner2013physical}
Lischner,~J.; Vigil-Fowler,~D.; Louie,~S.~G. Physical {Origin} of {Satellites} in {Photoemission} of {Doped} {Graphene}: {An} {Ab} {Initio} {GW} {Plus} {Cumulant} {Study}. \emph{Phys. Rev. Lett.} \textbf{2013}, \emph{110}, 146801\relax
\mciteBstWouldAddEndPuncttrue
\mciteSetBstMidEndSepPunct{\mcitedefaultmidpunct}
{\mcitedefaultendpunct}{\mcitedefaultseppunct}\relax
\EndOfBibitem
\bibitem[Guzzo \latin{et~al.}(2014)Guzzo, Kas, Sponza, Giorgetti, Sottile, Pierucci, Silly, Sirotti, Rehr, and Reining]{guzzo2014multiple}
Guzzo,~M.; Kas,~J.~J.; Sponza,~L.; Giorgetti,~C.; Sottile,~F.; Pierucci,~D.; Silly,~M.~G.; Sirotti,~F.; Rehr,~J.~J.; Reining,~L. Multiple satellites in materials with complex plasmon spectra: {From} graphite to graphene. \emph{Phys. Rev. B} \textbf{2014}, \emph{89}, 085425\relax
\mciteBstWouldAddEndPuncttrue
\mciteSetBstMidEndSepPunct{\mcitedefaultmidpunct}
{\mcitedefaultendpunct}{\mcitedefaultseppunct}\relax
\EndOfBibitem
\bibitem[Nakamura \latin{et~al.}(2016)Nakamura, Nohara, Yosimoto, and Nomura]{nakamura2016ab}
Nakamura,~K.; Nohara,~Y.; Yosimoto,~Y.; Nomura,~Y. Ab initio {GW} plus cumulant calculation for isolated band systems: {Application} to organic conductor ({TMTSF}) 2 {PF} 6 and transition-metal oxide {SrVO} 3. \emph{Phys. Rev. B} \textbf{2016}, \emph{93}, 085124\relax
\mciteBstWouldAddEndPuncttrue
\mciteSetBstMidEndSepPunct{\mcitedefaultmidpunct}
{\mcitedefaultendpunct}{\mcitedefaultseppunct}\relax
\EndOfBibitem
\bibitem[Loos \latin{et~al.}(2024)Loos, Marie, and Ammar]{loos2024cumulant}
Loos,~P.-F.; Marie,~A.; Ammar,~A. Cumulant {Green}'s function methods for molecules. \emph{Faraday Discuss.} \textbf{2024}, \relax
\mciteBstWouldAddEndPunctfalse
\mciteSetBstMidEndSepPunct{\mcitedefaultmidpunct}
{}{\mcitedefaultseppunct}\relax
\EndOfBibitem
\bibitem[Golze \latin{et~al.}(2018)Golze, Wilhelm, Van~Setten, and Rinke]{golze2018core}
Golze,~D.; Wilhelm,~J.; Van~Setten,~M.~J.; Rinke,~P. Core-level binding energies from {GW}: {An} efficient full-frequency approach within a localized basis. \emph{J. Chem. Theory Comput.} \textbf{2018}, \emph{14}, 4856--4869\relax
\mciteBstWouldAddEndPuncttrue
\mciteSetBstMidEndSepPunct{\mcitedefaultmidpunct}
{\mcitedefaultendpunct}{\mcitedefaultseppunct}\relax
\EndOfBibitem
\bibitem[Golze \latin{et~al.}(2020)Golze, Keller, and Rinke]{golze2020accurate}
Golze,~D.; Keller,~L.; Rinke,~P. Accurate absolute and relative core-level binding energies from {GW}. \emph{J. Phys. Chem. Lett.} \textbf{2020}, \emph{11}, 1840--1847\relax
\mciteBstWouldAddEndPuncttrue
\mciteSetBstMidEndSepPunct{\mcitedefaultmidpunct}
{\mcitedefaultendpunct}{\mcitedefaultseppunct}\relax
\EndOfBibitem
\bibitem[Li \latin{et~al.}(2022)Li, Jin, Rinke, Yang, and Golze]{li2022benchmark}
Li,~J.; Jin,~Y.; Rinke,~P.; Yang,~W.; Golze,~D. Benchmark of {GW} methods for core-level binding energies. \emph{J. Chem. Theory Comput.} \textbf{2022}, \emph{18}, 7570--7585\relax
\mciteBstWouldAddEndPuncttrue
\mciteSetBstMidEndSepPunct{\mcitedefaultmidpunct}
{\mcitedefaultendpunct}{\mcitedefaultseppunct}\relax
\EndOfBibitem
\bibitem[Golze \latin{et~al.}(2022)Golze, Hirvensalo, Hernández-León, Aarva, Etula, Susi, Rinke, Laurila, and Caro]{golze2022accurate}
Golze,~D.; Hirvensalo,~M.; Hernández-León,~P.; Aarva,~A.; Etula,~J.; Susi,~T.; Rinke,~P.; Laurila,~T.; Caro,~M.~A. Accurate computational prediction of core-electron binding energies in carbon-based materials: {A} machine-learning model combining density-functional theory and {GW}. \emph{Chem. Mater.} \textbf{2022}, \emph{34}, 6240--6254\relax
\mciteBstWouldAddEndPuncttrue
\mciteSetBstMidEndSepPunct{\mcitedefaultmidpunct}
{\mcitedefaultendpunct}{\mcitedefaultseppunct}\relax
\EndOfBibitem
\bibitem[Zhu and Chan(2021)Zhu, and Chan]{zhu2021all}
Zhu,~T.; Chan,~G. K.-L. All-electron {Gaussian}-based {G} 0 {W} 0 for valence and core excitation energies of periodic systems. \emph{J. Chem. Theory Comput.} \textbf{2021}, \emph{17}, 727--741\relax
\mciteBstWouldAddEndPuncttrue
\mciteSetBstMidEndSepPunct{\mcitedefaultmidpunct}
{\mcitedefaultendpunct}{\mcitedefaultseppunct}\relax
\EndOfBibitem
\bibitem[Mejia-Rodriguez \latin{et~al.}(2021)Mejia-Rodriguez, Kunitsa, Apra, and Govind]{mejia2021scalable}
Mejia-Rodriguez,~D.; Kunitsa,~A.; Apra,~E.; Govind,~N. Scalable molecular {GW} calculations: {Valence} and core spectra. \emph{J. Chem. Theory Comput.} \textbf{2021}, \emph{17}, 7504--7517\relax
\mciteBstWouldAddEndPuncttrue
\mciteSetBstMidEndSepPunct{\mcitedefaultmidpunct}
{\mcitedefaultendpunct}{\mcitedefaultseppunct}\relax
\EndOfBibitem
\bibitem[Mejia-Rodriguez \latin{et~al.}(2022)Mejia-Rodriguez, Kunitsa, Aprà, and Govind]{mejia2022basis}
Mejia-Rodriguez,~D.; Kunitsa,~A.; Aprà,~E.; Govind,~N. Basis set selection for molecular core-level gw calculations. \emph{J. Chem. Theory Comput.} \textbf{2022}, \emph{18}, 4919--4926\relax
\mciteBstWouldAddEndPuncttrue
\mciteSetBstMidEndSepPunct{\mcitedefaultmidpunct}
{\mcitedefaultendpunct}{\mcitedefaultseppunct}\relax
\EndOfBibitem
\bibitem[Bruneval and Förster(2024)Bruneval, and Förster]{bruneval2024fully}
Bruneval,~F.; Förster,~A. Fully {Dynamic} {G3W2} {Self}-{Energy} for {Finite} {Systems}: {Formulas} and {Benchmark}. \emph{J. Chem. Theory Comput.} \textbf{2024}, \emph{20}, 3218--3230\relax
\mciteBstWouldAddEndPuncttrue
\mciteSetBstMidEndSepPunct{\mcitedefaultmidpunct}
{\mcitedefaultendpunct}{\mcitedefaultseppunct}\relax
\EndOfBibitem
\bibitem[Aoki and Ohno(2018)Aoki, and Ohno]{aoki2018accurate}
Aoki,~T.; Ohno,~K. Accurate quasiparticle calculation of x-ray photoelectron spectra of solids. \emph{J. Condens. Matter Phys.} \textbf{2018}, \emph{30}, 21LT01\relax
\mciteBstWouldAddEndPuncttrue
\mciteSetBstMidEndSepPunct{\mcitedefaultmidpunct}
{\mcitedefaultendpunct}{\mcitedefaultseppunct}\relax
\EndOfBibitem
\bibitem[van Setten \latin{et~al.}(2018)van Setten, Costa, Vines, and Illas]{van2018assessing}
van Setten,~M.~J.; Costa,~R.; Vines,~F.; Illas,~F. Assessing {GW} approaches for predicting core level binding energies. \emph{J. Chem. Theory Comput.} \textbf{2018}, \emph{14}, 877--883\relax
\mciteBstWouldAddEndPuncttrue
\mciteSetBstMidEndSepPunct{\mcitedefaultmidpunct}
{\mcitedefaultendpunct}{\mcitedefaultseppunct}\relax
\EndOfBibitem
\bibitem[Keller \latin{et~al.}(2020)Keller, Blum, Rinke, and Golze]{keller2020relativistic}
Keller,~L.; Blum,~V.; Rinke,~P.; Golze,~D. Relativistic correction scheme for core-level binding energies from {GW}. \emph{J. Chem. Phys.} \textbf{2020}, \emph{153}\relax
\mciteBstWouldAddEndPuncttrue
\mciteSetBstMidEndSepPunct{\mcitedefaultmidpunct}
{\mcitedefaultendpunct}{\mcitedefaultseppunct}\relax
\EndOfBibitem
\bibitem[Duchemin and Blase(2020)Duchemin, and Blase]{duchemin2020robust}
Duchemin,~I.; Blase,~X. Robust analytic-continuation approach to many-body {GW} calculations. \emph{J. Chem. Theory Comput.} \textbf{2020}, \emph{16}, 1742--1756\relax
\mciteBstWouldAddEndPuncttrue
\mciteSetBstMidEndSepPunct{\mcitedefaultmidpunct}
{\mcitedefaultendpunct}{\mcitedefaultseppunct}\relax
\EndOfBibitem
\bibitem[Galleni \latin{et~al.}(2022)Galleni, Sajjadian, Conard, Escudero, Pourtois, and van Setten]{galleni2022modeling}
Galleni,~L.; Sajjadian,~F.~S.; Conard,~T.; Escudero,~D.; Pourtois,~G.; van Setten,~M.~J. Modeling X-ray photoelectron spectroscopy of macromolecules using GW. \emph{J. Phys. Chem. Lett.} \textbf{2022}, \emph{13}, 8666--8672\relax
\mciteBstWouldAddEndPuncttrue
\mciteSetBstMidEndSepPunct{\mcitedefaultmidpunct}
{\mcitedefaultendpunct}{\mcitedefaultseppunct}\relax
\EndOfBibitem
\bibitem[Galleni \latin{et~al.}(2024)Galleni, Escudero, Pourtois, and van Setten]{galleni2024c1s}
Galleni,~L.; Escudero,~D.; Pourtois,~G.; van Setten,~M.~J. The C1s core levels of polycyclic aromatic hydrocarbons and styrenic polymers: A first-principles study. \emph{J. Chem. Phys.} \textbf{2024}, \emph{160}\relax
\mciteBstWouldAddEndPuncttrue
\mciteSetBstMidEndSepPunct{\mcitedefaultmidpunct}
{\mcitedefaultendpunct}{\mcitedefaultseppunct}\relax
\EndOfBibitem
\bibitem[Tölle and Kin-Lic~Chan(2024)Tölle, and Kin-Lic~Chan]{tolle2024ab}
Tölle,~J.; Kin-Lic~Chan,~G. {AB}-{G0W0}: {A} practical {G0W0} method without frequency integration based on an auxiliary boson expansion. \emph{J. Chem. Phys.} \textbf{2024}, \emph{160}\relax
\mciteBstWouldAddEndPuncttrue
\mciteSetBstMidEndSepPunct{\mcitedefaultmidpunct}
{\mcitedefaultendpunct}{\mcitedefaultseppunct}\relax
\EndOfBibitem
\bibitem[Panadés-Barrueta and Golze(2023)Panadés-Barrueta, and Golze]{panades2023accelerating}
Panadés-Barrueta,~R.~L.; Golze,~D. Accelerating core-level {GW} calculations by combining the contour deformation approach with the analytic continuation of {W}. \emph{J. Chem. Theory Comput.} \textbf{2023}, \emph{19}, 5450--5464\relax
\mciteBstWouldAddEndPuncttrue
\mciteSetBstMidEndSepPunct{\mcitedefaultmidpunct}
{\mcitedefaultendpunct}{\mcitedefaultseppunct}\relax
\EndOfBibitem
\bibitem[Guzzo(2012)]{guzzo2012dynamical}
Guzzo,~M. Dynamical correlation in solids: a perspective in photoelectron spectroscopy. {PhD} {Thesis}, Ecole Polytechnique X, 2012\relax
\mciteBstWouldAddEndPuncttrue
\mciteSetBstMidEndSepPunct{\mcitedefaultmidpunct}
{\mcitedefaultendpunct}{\mcitedefaultseppunct}\relax
\EndOfBibitem
\bibitem[Perdew \latin{et~al.}(1996)Perdew, Burke, and Ernzerhof]{perdew1996generalized}
Perdew,~J.~P.; Burke,~K.; Ernzerhof,~M. Generalized gradient approximation made simple. \emph{Phys. Rev. Lett.} \textbf{1996}, \emph{77}, 3865\relax
\mciteBstWouldAddEndPuncttrue
\mciteSetBstMidEndSepPunct{\mcitedefaultmidpunct}
{\mcitedefaultendpunct}{\mcitedefaultseppunct}\relax
\EndOfBibitem
\bibitem[Gelius \latin{et~al.}(1970)Gelius, Heden, Hedman, Lindberg, Manne, Nordberg, Nordling, and Siegbahn]{gelius1970molecular}
Gelius,~U.; Heden,~P.; Hedman,~J.; Lindberg,~B.; Manne,~R.; Nordberg,~R.; Nordling,~C.; Siegbahn,~K. Molecular spectroscopy by means of {ESCA} {III}. {Carbon} compounds. \emph{Physica Scripta} \textbf{1970}, \emph{2}, 70\relax
\mciteBstWouldAddEndPuncttrue
\mciteSetBstMidEndSepPunct{\mcitedefaultmidpunct}
{\mcitedefaultendpunct}{\mcitedefaultseppunct}\relax
\EndOfBibitem
\bibitem[Lunell \latin{et~al.}(1978)Lunell, Svensson, Malmqvist, Gelius, Basilier, and Siegbahn]{lunell1978theoretical}
Lunell,~S.; Svensson,~S.; Malmqvist,~P.~A.; Gelius,~U.; Basilier,~E.; Siegbahn,~K. A theoretical and experimental study of the carbon 1s shake-up structure of benzene. \emph{Chem. Phys. Lett.} \textbf{1978}, \emph{54}, 420--424\relax
\mciteBstWouldAddEndPuncttrue
\mciteSetBstMidEndSepPunct{\mcitedefaultmidpunct}
{\mcitedefaultendpunct}{\mcitedefaultseppunct}\relax
\EndOfBibitem
\bibitem[Hedin(1999)]{hedin1999correlation}
Hedin,~L. On correlation effects in electron spectroscopies and the {GW} approximation. \emph{J. Phys. Condens. Matter} \textbf{1999}, \emph{11}, R489\relax
\mciteBstWouldAddEndPuncttrue
\mciteSetBstMidEndSepPunct{\mcitedefaultmidpunct}
{\mcitedefaultendpunct}{\mcitedefaultseppunct}\relax
\EndOfBibitem
\bibitem[Kas \latin{et~al.}(2014)Kas, Rehr, and Reining]{kas_cumulant_2014}
Kas,~J.~J.; Rehr,~J.~J.; Reining,~L. Cumulant expansion of the retarded one-electron {Green} function. \emph{Phys. Rev. B} \textbf{2014}, \emph{90}, 085112\relax
\mciteBstWouldAddEndPuncttrue
\mciteSetBstMidEndSepPunct{\mcitedefaultmidpunct}
{\mcitedefaultendpunct}{\mcitedefaultseppunct}\relax
\EndOfBibitem
\bibitem[Landau(1944)]{landau1944energy}
Landau,~L.~D. On the energy loss of fast particles by ionization. \emph{J. Phys.} \textbf{1944}, \emph{8}, 201--205\relax
\mciteBstWouldAddEndPuncttrue
\mciteSetBstMidEndSepPunct{\mcitedefaultmidpunct}
{\mcitedefaultendpunct}{\mcitedefaultseppunct}\relax
\EndOfBibitem
\bibitem[Mahan(1982)]{mahan1982core}
Mahan,~G. Core-hole {Green}'s function: {Dispersion} theory. \emph{Phys. Rev. B} \textbf{1982}, \emph{25}, 5021\relax
\mciteBstWouldAddEndPuncttrue
\mciteSetBstMidEndSepPunct{\mcitedefaultmidpunct}
{\mcitedefaultendpunct}{\mcitedefaultseppunct}\relax
\EndOfBibitem
\bibitem[Tzavala \latin{et~al.}(2020)Tzavala, Kas, Reining, and Rehr]{tzavala2020nonlinear}
Tzavala,~M.; Kas,~J.~J.; Reining,~L.; Rehr,~J.~J. Nonlinear response in the cumulant expansion for core-level photoemission. \emph{Phys. Rev. Res.} \textbf{2020}, \emph{2}, 033147\relax
\mciteBstWouldAddEndPuncttrue
\mciteSetBstMidEndSepPunct{\mcitedefaultmidpunct}
{\mcitedefaultendpunct}{\mcitedefaultseppunct}\relax
\EndOfBibitem
\bibitem[Sun \latin{et~al.}(2020)Sun, Zhang, Banerjee, Bao, Barbry, Blunt, Bogdanov, Booth, Chen, Cui, and {others}]{sun2020recent}
Sun,~Q.; Zhang,~X.; Banerjee,~S.; Bao,~P.; Barbry,~M.; Blunt,~N.~S.; Bogdanov,~N.~A.; Booth,~G.~H.; Chen,~J.; Cui,~Z.-H.; {others} Recent developments in the {PySCF} program package. \emph{J. Chem. Phys.} \textbf{2020}, \emph{153}\relax
\mciteBstWouldAddEndPuncttrue
\mciteSetBstMidEndSepPunct{\mcitedefaultmidpunct}
{\mcitedefaultendpunct}{\mcitedefaultseppunct}\relax
\EndOfBibitem
\bibitem[Van~Meer \latin{et~al.}(2014)Van~Meer, Gritsenko, and Baerends]{van2014physical}
Van~Meer,~R.; Gritsenko,~O.; Baerends,~E. Physical meaning of virtual {Kohn}–{Sham} orbitals and orbital energies: an ideal basis for the description of molecular excitations. \emph{J. Chem. Theory Comput.} \textbf{2014}, \emph{10}, 4432--4441\relax
\mciteBstWouldAddEndPuncttrue
\mciteSetBstMidEndSepPunct{\mcitedefaultmidpunct}
{\mcitedefaultendpunct}{\mcitedefaultseppunct}\relax
\EndOfBibitem
\bibitem[Barth and Cederbaum(1981)Barth, and Cederbaum]{barth1981many}
Barth,~A.; Cederbaum,~L. Many-body theory of core-valence excitations. \emph{Phys. Rev. A} \textbf{1981}, \emph{23}, 1038\relax
\mciteBstWouldAddEndPuncttrue
\mciteSetBstMidEndSepPunct{\mcitedefaultmidpunct}
{\mcitedefaultendpunct}{\mcitedefaultseppunct}\relax
\EndOfBibitem
\bibitem[Wenzel \latin{et~al.}(2014)Wenzel, Wormit, and Dreuw]{wenzel2014calculating}
Wenzel,~J.; Wormit,~M.; Dreuw,~A. Calculating core-level excitations and {X}-ray absorption spectra of medium-sized closed-shell molecules with the algebraic-diagrammatic construction scheme for the polarization propagator. \emph{J. Chem. Theory Comput.} \textbf{2014}, \emph{35}, 1900--1915\relax
\mciteBstWouldAddEndPuncttrue
\mciteSetBstMidEndSepPunct{\mcitedefaultmidpunct}
{\mcitedefaultendpunct}{\mcitedefaultseppunct}\relax
\EndOfBibitem
\bibitem[Wenzel \latin{et~al.}(2014)Wenzel, Wormit, and Dreuw]{wenzel2014calculating2}
Wenzel,~J.; Wormit,~M.; Dreuw,~A. Calculating {X}-ray absorption spectra of open-shell molecules with the unrestricted algebraic-diagrammatic construction scheme for the polarization propagator. \emph{J. Chem. Theory Comput.} \textbf{2014}, \emph{10}, 4583--4598\relax
\mciteBstWouldAddEndPuncttrue
\mciteSetBstMidEndSepPunct{\mcitedefaultmidpunct}
{\mcitedefaultendpunct}{\mcitedefaultseppunct}\relax
\EndOfBibitem
\bibitem[Coriani and Koch(2015)Coriani, and Koch]{coriani2015communication}
Coriani,~S.; Koch,~H. Communication: {X}-ray absorption spectra and core-ionization potentials within a core-valence separated coupled cluster framework. \emph{J. Chem. Phys.} \textbf{2015}, \emph{143}\relax
\mciteBstWouldAddEndPuncttrue
\mciteSetBstMidEndSepPunct{\mcitedefaultmidpunct}
{\mcitedefaultendpunct}{\mcitedefaultseppunct}\relax
\EndOfBibitem
\bibitem[Coriani and Koch(2016)Coriani, and Koch]{coriani2016erratum}
Coriani,~S.; Koch,~H. Erratum:“{Communication}: {X}-ray absorption spectra and core-ionization potentials within a core-valence separated coupled cluster framework”[{J}. {Chem}. {Phys}. 143, 181103 (2015)]. \emph{J. Chem. Phys.} \textbf{2016}, \emph{145}\relax
\mciteBstWouldAddEndPuncttrue
\mciteSetBstMidEndSepPunct{\mcitedefaultmidpunct}
{\mcitedefaultendpunct}{\mcitedefaultseppunct}\relax
\EndOfBibitem
\bibitem[Aryasetiawan and Gunnarsson(1998)Aryasetiawan, and Gunnarsson]{aryasetiawan1998gw}
Aryasetiawan,~F.; Gunnarsson,~O. The {GW} method. \emph{Rep. Prog. Phys.} \textbf{1998}, \emph{61}, 237\relax
\mciteBstWouldAddEndPuncttrue
\mciteSetBstMidEndSepPunct{\mcitedefaultmidpunct}
{\mcitedefaultendpunct}{\mcitedefaultseppunct}\relax
\EndOfBibitem
\bibitem[Bruneval(2005)]{bruneval2005exchange}
Bruneval,~F. Exchange and correlation in the electronic structure of solids, from silicon to cuprous oxide: {GW} approximation and beyond. \emph{PhD Thesis} \textbf{2005}, \relax
\mciteBstWouldAddEndPunctfalse
\mciteSetBstMidEndSepPunct{\mcitedefaultmidpunct}
{}{\mcitedefaultseppunct}\relax
\EndOfBibitem
\bibitem[Blum \latin{et~al.}(2009)Blum, Gehrke, Hanke, Havu, Havu, Ren, Reuter, and Scheffler]{blum2009ab}
Blum,~V.; Gehrke,~R.; Hanke,~F.; Havu,~P.; Havu,~V.; Ren,~X.; Reuter,~K.; Scheffler,~M. Ab initio molecular simulations with numeric atom-centered orbitals. \emph{Comput. Phys. Commun.} \textbf{2009}, \emph{180}, 2175--2196\relax
\mciteBstWouldAddEndPuncttrue
\mciteSetBstMidEndSepPunct{\mcitedefaultmidpunct}
{\mcitedefaultendpunct}{\mcitedefaultseppunct}\relax
\EndOfBibitem
\bibitem[Azizi \latin{et~al.}(2023)Azizi, Wilhelm, Golze, Giantomassi, Panad{\'e}s-Barrueta, Delesma, Buccheri, Gulans, Rinke, Draxl, \latin{et~al.} others]{azizi2023time}
Azizi,~M.; Wilhelm,~J.; Golze,~D.; Giantomassi,~M.; Panad{\'e}s-Barrueta,~R.~L.; Delesma,~F.~A.; Buccheri,~A.; Gulans,~A.; Rinke,~P.; Draxl,~C.; others Time-frequency component of the GreenX library: minimax grids for efficient RPA and GW calculations. \emph{J. Open Source Softw.} \textbf{2023}, \emph{8}, 5570\relax
\mciteBstWouldAddEndPuncttrue
\mciteSetBstMidEndSepPunct{\mcitedefaultmidpunct}
{\mcitedefaultendpunct}{\mcitedefaultseppunct}\relax
\EndOfBibitem
\bibitem[Azizi \latin{et~al.}(2024)Azizi, Wilhelm, Golze, Delesma, Panad{\'e}s-Barrueta, Rinke, Giantomassi, and Gonze]{azizi2024validation}
Azizi,~M.; Wilhelm,~J.; Golze,~D.; Delesma,~F.~A.; Panad{\'e}s-Barrueta,~R.~L.; Rinke,~P.; Giantomassi,~M.; Gonze,~X. Validation of the GreenX library time-frequency component for efficient GW and RPA calculations. \emph{Phys. Rev. B} \textbf{2024}, \emph{109}, 245101\relax
\mciteBstWouldAddEndPuncttrue
\mciteSetBstMidEndSepPunct{\mcitedefaultmidpunct}
{\mcitedefaultendpunct}{\mcitedefaultseppunct}\relax
\EndOfBibitem
\bibitem[Frigo and Johnson(2005)Frigo, and Johnson]{frigo2005design}
Frigo,~M.; Johnson,~S.~G. The {Design} and {Implementation} of {FFTW3}. \emph{Proc. IEEE} \textbf{2005}, \emph{93}, 216--231\relax
\mciteBstWouldAddEndPuncttrue
\mciteSetBstMidEndSepPunct{\mcitedefaultmidpunct}
{\mcitedefaultendpunct}{\mcitedefaultseppunct}\relax
\EndOfBibitem
\bibitem[Ren \latin{et~al.}(2012)Ren, Rinke, Blum, Wieferink, Tkatchenko, Sanfilippo, Reuter, and Scheffler]{ren2012resolution}
Ren,~X.; Rinke,~P.; Blum,~V.; Wieferink,~J.; Tkatchenko,~A.; Sanfilippo,~A.; Reuter,~K.; Scheffler,~M. Resolution-of-identity approach to {Hartree}–{Fock}, hybrid density functionals, {RPA}, {MP2} and {GW} with numeric atom-centered orbital basis functions. \emph{New J. Phys.} \textbf{2012}, \emph{14}, 053020\relax
\mciteBstWouldAddEndPuncttrue
\mciteSetBstMidEndSepPunct{\mcitedefaultmidpunct}
{\mcitedefaultendpunct}{\mcitedefaultseppunct}\relax
\EndOfBibitem
\bibitem[Vahtras \latin{et~al.}(1993)Vahtras, Almlöf, and Feyereisen]{vahtras1993integral}
Vahtras,~O.; Almlöf,~J.; Feyereisen,~M. Integral approximations for {LCAO}-{SCF} calculations. \emph{Chem. Phys. Lett.} \textbf{1993}, \emph{213}, 514--518\relax
\mciteBstWouldAddEndPuncttrue
\mciteSetBstMidEndSepPunct{\mcitedefaultmidpunct}
{\mcitedefaultendpunct}{\mcitedefaultseppunct}\relax
\EndOfBibitem
\bibitem[Tkatchenko and Scheffler(2009)Tkatchenko, and Scheffler]{tkatchenko2009accurate}
Tkatchenko,~A.; Scheffler,~M. Accurate {Molecular} {Van} {Der} {Waals} {Interactions} from {Ground}-{State} {Electron} {Density} and {Free}-{Atom} {Reference} {Data}. \emph{Phys. Rev. Lett.} \textbf{2009}, \emph{102}, 073005\relax
\mciteBstWouldAddEndPuncttrue
\mciteSetBstMidEndSepPunct{\mcitedefaultmidpunct}
{\mcitedefaultendpunct}{\mcitedefaultseppunct}\relax
\EndOfBibitem
\bibitem[Dunning~Jr(1989)]{dunning1989gaussian}
Dunning~Jr,~T.~H. Gaussian basis sets for use in correlated molecular calculations. {I}. {The} atoms boron through neon and hydrogen. \emph{J. Chem. Phys.} \textbf{1989}, \emph{90}, 1007--1023\relax
\mciteBstWouldAddEndPuncttrue
\mciteSetBstMidEndSepPunct{\mcitedefaultmidpunct}
{\mcitedefaultendpunct}{\mcitedefaultseppunct}\relax
\EndOfBibitem
\bibitem[Kendall \latin{et~al.}(1992)Kendall, Dunning, and Harrison]{kendall1992electron}
Kendall,~R.~A.; Dunning,~T.~H.; Harrison,~R.~J. Electron affinities of the first-row atoms revisited. {Systematic} basis sets and wave functions. \emph{J. Chem. Phys.} \textbf{1992}, \emph{96}, 6796--6806\relax
\mciteBstWouldAddEndPuncttrue
\mciteSetBstMidEndSepPunct{\mcitedefaultmidpunct}
{\mcitedefaultendpunct}{\mcitedefaultseppunct}\relax
\EndOfBibitem
\bibitem[Ambroise \latin{et~al.}(2021)Ambroise, Dreuw, and Jensen]{ambroise2021probing}
Ambroise,~M.~A.; Dreuw,~A.; Jensen,~F. Probing basis set requirements for calculating core ionization and core excitation spectra using correlated wave function methods. \emph{J. Chem. Theory Comput.} \textbf{2021}, \emph{17}, 2832--2842\relax
\mciteBstWouldAddEndPuncttrue
\mciteSetBstMidEndSepPunct{\mcitedefaultmidpunct}
{\mcitedefaultendpunct}{\mcitedefaultseppunct}\relax
\EndOfBibitem
\bibitem[Liu \latin{et~al.}(2020)Liu, Kloppenburg, Yao, Ren, Appel, Kanai, and Blum]{liu2020all}
Liu,~C.; Kloppenburg,~J.; Yao,~Y.; Ren,~X.; Appel,~H.; Kanai,~Y.; Blum,~V. All-electron ab initio {Bethe}-{Salpeter} equation approach to neutral excitations in molecules with numeric atom-centered orbitals. \emph{J. Chem. Phys.} \textbf{2020}, \emph{152}\relax
\mciteBstWouldAddEndPuncttrue
\mciteSetBstMidEndSepPunct{\mcitedefaultmidpunct}
{\mcitedefaultendpunct}{\mcitedefaultseppunct}\relax
\EndOfBibitem
\bibitem[Yao \latin{et~al.}(2022)Yao, Golze, Rinke, Blum, and Kanai]{yao2022all}
Yao,~Y.; Golze,~D.; Rinke,~P.; Blum,~V.; Kanai,~Y. All-electron {BSE}@ {GW} method for {K}-edge core electron excitation energies. \emph{J. Chem. Theory Comput.} \textbf{2022}, \emph{18}, 1569--1583\relax
\mciteBstWouldAddEndPuncttrue
\mciteSetBstMidEndSepPunct{\mcitedefaultmidpunct}
{\mcitedefaultendpunct}{\mcitedefaultseppunct}\relax
\EndOfBibitem
\bibitem[nom()]{nomad_reference}
All data will be published on NOMAD upon revision.\relax
\mciteBstWouldAddEndPunctfalse
\mciteSetBstMidEndSepPunct{\mcitedefaultmidpunct}
{}{\mcitedefaultseppunct}\relax
\EndOfBibitem
\bibitem[Sun(2015)]{sun2015libcint}
Sun,~Q. Libcint: An efficient general integral library for g aussian basis functions. \emph{J. Comput. Chem.} \textbf{2015}, \emph{36}, 1664--1671\relax
\mciteBstWouldAddEndPuncttrue
\mciteSetBstMidEndSepPunct{\mcitedefaultmidpunct}
{\mcitedefaultendpunct}{\mcitedefaultseppunct}\relax
\EndOfBibitem
\bibitem[Sun \latin{et~al.}(2018)Sun, Berkelbach, Blunt, Booth, Guo, Li, Liu, McClain, Sayfutyarova, Sharma, \latin{et~al.} others]{sun2018pyscf}
Sun,~Q.; Berkelbach,~T.~C.; Blunt,~N.~S.; Booth,~G.~H.; Guo,~S.; Li,~Z.; Liu,~J.; McClain,~J.~D.; Sayfutyarova,~E.~R.; Sharma,~S.; others PySCF: the Python-based simulations of chemistry framework. \emph{Wiley Interdiscip. Rev. Comput. Mol. Sci.} \textbf{2018}, \emph{8}, e1340\relax
\mciteBstWouldAddEndPuncttrue
\mciteSetBstMidEndSepPunct{\mcitedefaultmidpunct}
{\mcitedefaultendpunct}{\mcitedefaultseppunct}\relax
\EndOfBibitem
\bibitem[Bruneval \latin{et~al.}(2015)Bruneval, Hamed, and Neaton]{bruneval2015systematic}
Bruneval,~F.; Hamed,~S.~M.; Neaton,~J.~B. A systematic benchmark of the ab initio {Bethe}-{Salpeter} equation approach for low-lying optical excitations of small organic molecules. \emph{J. Chem. Phys.} \textbf{2015}, \emph{142}\relax
\mciteBstWouldAddEndPuncttrue
\mciteSetBstMidEndSepPunct{\mcitedefaultmidpunct}
{\mcitedefaultendpunct}{\mcitedefaultseppunct}\relax
\EndOfBibitem
\bibitem[Adamo and Barone(1999)Adamo, and Barone]{adamo1999toward}
Adamo,~C.; Barone,~V. Toward reliable density functional methods without adjustable parameters: {The} {PBE0} model. \emph{J. Chem. Phys.} \textbf{1999}, \emph{110}, 6158--6170\relax
\mciteBstWouldAddEndPuncttrue
\mciteSetBstMidEndSepPunct{\mcitedefaultmidpunct}
{\mcitedefaultendpunct}{\mcitedefaultseppunct}\relax
\EndOfBibitem
\bibitem[Ernzerhof and Scuseria(1999)Ernzerhof, and Scuseria]{ernzerhof1999assessment}
Ernzerhof,~M.; Scuseria,~G.~E. Assessment of the {Perdew}–{Burke}–{Ernzerhof} exchange-correlation functional. \emph{J. Chem. Phys.} \textbf{1999}, \emph{110}, 5029--5036\relax
\mciteBstWouldAddEndPuncttrue
\mciteSetBstMidEndSepPunct{\mcitedefaultmidpunct}
{\mcitedefaultendpunct}{\mcitedefaultseppunct}\relax
\EndOfBibitem
\bibitem[Atalla \latin{et~al.}(2013)Atalla, Yoon, Caruso, Rinke, and Scheffler]{atalla2013hybrid}
Atalla,~V.; Yoon,~M.; Caruso,~F.; Rinke,~P.; Scheffler,~M. Hybrid density functional theory meets quasiparticle calculations: {A} consistent electronic structure approach. \emph{Phys. Rev. B.} \textbf{2013}, \emph{88}, 165122\relax
\mciteBstWouldAddEndPuncttrue
\mciteSetBstMidEndSepPunct{\mcitedefaultmidpunct}
{\mcitedefaultendpunct}{\mcitedefaultseppunct}\relax
\EndOfBibitem
\bibitem[Leang \latin{et~al.}(2012)Leang, Zahariev, and Gordon]{leang2012benchmarking}
Leang,~S.~S.; Zahariev,~F.; Gordon,~M.~S. Benchmarking the performance of time-dependent density functional methods. \emph{J. Chem. Phys.} \textbf{2012}, \emph{136}\relax
\mciteBstWouldAddEndPuncttrue
\mciteSetBstMidEndSepPunct{\mcitedefaultmidpunct}
{\mcitedefaultendpunct}{\mcitedefaultseppunct}\relax
\EndOfBibitem
\bibitem[Laurent and Jacquemin(2013)Laurent, and Jacquemin]{laurent2013td}
Laurent,~A.~D.; Jacquemin,~D. {TD}-{DFT} benchmarks: a review. \emph{Int. J. Quantum Chem.} \textbf{2013}, \emph{113}, 2019--2039\relax
\mciteBstWouldAddEndPuncttrue
\mciteSetBstMidEndSepPunct{\mcitedefaultmidpunct}
{\mcitedefaultendpunct}{\mcitedefaultseppunct}\relax
\EndOfBibitem
\bibitem[Baerends \latin{et~al.}(2013)Baerends, Gritsenko, and Van~Meer]{baerends2013kohn}
Baerends,~E.; Gritsenko,~O.; Van~Meer,~R. The {Kohn}–{Sham} gap, the fundamental gap and the optical gap: the physical meaning of occupied and virtual {Kohn}–{Sham} orbital energies. \emph{Phys. Chem. Chem. Phys.} \textbf{2013}, \emph{15}, 16408--16425\relax
\mciteBstWouldAddEndPuncttrue
\mciteSetBstMidEndSepPunct{\mcitedefaultmidpunct}
{\mcitedefaultendpunct}{\mcitedefaultseppunct}\relax
\EndOfBibitem
\bibitem[Gritsenko and Baerends(2004)Gritsenko, and Baerends]{gritsenko2004asymptotic}
Gritsenko,~O.; Baerends,~E.~J. Asymptotic correction of the exchange–correlation kernel of time-dependent density functional theory for long-range charge-transfer excitations. \emph{J. Chem. Phys.} \textbf{2004}, \emph{121}, 655--660\relax
\mciteBstWouldAddEndPuncttrue
\mciteSetBstMidEndSepPunct{\mcitedefaultmidpunct}
{\mcitedefaultendpunct}{\mcitedefaultseppunct}\relax
\EndOfBibitem
\bibitem[Schambeck \latin{et~al.}(2024)Schambeck, Golze, and Wilhelm]{schambeck2024solving}
Schambeck,~M.; Golze,~D.; Wilhelm,~J. Solving multi-pole challenges in the {GW100} benchmark enables precise low-scaling {GW} calculations. \emph{Phys. Rev. B} \textbf{2024}, \relax
\mciteBstWouldAddEndPunctfalse
\mciteSetBstMidEndSepPunct{\mcitedefaultmidpunct}
{}{\mcitedefaultseppunct}\relax
\EndOfBibitem
\bibitem[Maggio and Kresse(2017)Maggio, and Kresse]{maggio2017gw}
Maggio,~E.; Kresse,~G. {GW} vertex corrected calculations for molecular systems. \emph{J. Chem. Theory Comput.} \textbf{2017}, \emph{13}, 4765--4778\relax
\mciteBstWouldAddEndPuncttrue
\mciteSetBstMidEndSepPunct{\mcitedefaultmidpunct}
{\mcitedefaultendpunct}{\mcitedefaultseppunct}\relax
\EndOfBibitem
\bibitem[Lewis and Berkelbach(2019)Lewis, and Berkelbach]{lewis2019vertex}
Lewis,~A.~M.; Berkelbach,~T.~C. Vertex corrections to the polarizability do not improve the {GW} approximation for the ionization potential of molecules. \emph{J. Chem. Theory Comput.} \textbf{2019}, \emph{15}, 2925--2932\relax
\mciteBstWouldAddEndPuncttrue
\mciteSetBstMidEndSepPunct{\mcitedefaultmidpunct}
{\mcitedefaultendpunct}{\mcitedefaultseppunct}\relax
\EndOfBibitem
\bibitem[Nordfors \latin{et~al.}(1988)Nordfors, Nilsson, Mårtensson, Svensson, Gelius, and Lunell]{nordfors1988experimental}
Nordfors,~D.; Nilsson,~A.; Mårtensson,~N.; Svensson,~S.; Gelius,~U.; Lunell,~S. Experimental and {INDO}/{CI} calculated gas phase {C1} s shake-up spectra of {C6H6}, {C6H5OH}, and {C6H5CH2OH}. \emph{J. Chem. Phys.} \textbf{1988}, \emph{88}, 2630--2636\relax
\mciteBstWouldAddEndPuncttrue
\mciteSetBstMidEndSepPunct{\mcitedefaultmidpunct}
{\mcitedefaultendpunct}{\mcitedefaultseppunct}\relax
\EndOfBibitem
\bibitem[Brena \latin{et~al.}(2005)Brena, Carniato, and Luo]{brena2005functional}
Brena,~B.; Carniato,~S.; Luo,~Y. Functional and basis set dependence of {K}-edge shake-up spectra of molecules. \emph{J. Chem. Phys.} \textbf{2005}, \emph{122}\relax
\mciteBstWouldAddEndPuncttrue
\mciteSetBstMidEndSepPunct{\mcitedefaultmidpunct}
{\mcitedefaultendpunct}{\mcitedefaultseppunct}\relax
\EndOfBibitem
\bibitem[Rocco \latin{et~al.}(2008)Rocco, Haeming, Batchelor, Fink, Schöll, and Umbach]{rocco2008electronic}
Rocco,~M.; Haeming,~M.; Batchelor,~D.; Fink,~R.; Schöll,~A.; Umbach,~E. Electronic relaxation effects in condensed polyacenes: {A} high-resolution photoemission study. \emph{J. Chem. Phys.} \textbf{2008}, \emph{129}\relax
\mciteBstWouldAddEndPuncttrue
\mciteSetBstMidEndSepPunct{\mcitedefaultmidpunct}
{\mcitedefaultendpunct}{\mcitedefaultseppunct}\relax
\EndOfBibitem
\bibitem[Colson \latin{et~al.}(1968)Colson, Hanson, Kopelman, and Robinson]{colson1968direct}
Colson,~S.; Hanson,~D.; Kopelman,~R.; Robinson,~G. Direct observation of the entire exciton band of the first excited singlet states of crystalline benzene and naphthalene. \emph{J. Chem. Phys.} \textbf{1968}, \emph{48}, 2215--2231\relax
\mciteBstWouldAddEndPuncttrue
\mciteSetBstMidEndSepPunct{\mcitedefaultmidpunct}
{\mcitedefaultendpunct}{\mcitedefaultseppunct}\relax
\EndOfBibitem
\bibitem[Glöckner and Wolf(1969)Glöckner, and Wolf]{glockner1969fluoreszenzspektrum}
Glöckner,~E.; Wolf,~H. Das {Fluoreszenzspektrum} von {Anthracen}-{Kristallen}. \emph{Z. Naturforsch. A} \textbf{1969}, \emph{24}, 943--951\relax
\mciteBstWouldAddEndPuncttrue
\mciteSetBstMidEndSepPunct{\mcitedefaultmidpunct}
{\mcitedefaultendpunct}{\mcitedefaultseppunct}\relax
\EndOfBibitem
\bibitem[Prikhotko \latin{et~al.}(1969)Prikhotko, Skorobogatko, and Tsikora]{prikhotko1969spectral}
Prikhotko,~A.; Skorobogatko,~A.; Tsikora,~L. Spectral investigations of pentacene crystals. \emph{Opt. Spectrosc.} \textbf{1969}, \emph{26}, 524\relax
\mciteBstWouldAddEndPuncttrue
\mciteSetBstMidEndSepPunct{\mcitedefaultmidpunct}
{\mcitedefaultendpunct}{\mcitedefaultseppunct}\relax
\EndOfBibitem
\bibitem[Mizuno \latin{et~al.}(1989)Mizuno, Matsui, and Sloan]{mizuno1989exciton}
Mizuno,~K.-i.; Matsui,~A.; Sloan,~G.~J. Exciton-phonon interaction in tetracene single crystals under pressure. \emph{Chem. Phys.} \textbf{1989}, \emph{131}, 423--433\relax
\mciteBstWouldAddEndPuncttrue
\mciteSetBstMidEndSepPunct{\mcitedefaultmidpunct}
{\mcitedefaultendpunct}{\mcitedefaultseppunct}\relax
\EndOfBibitem
\bibitem[Jentzsch \latin{et~al.}(1998)Jentzsch, Juepner, Brzezinka, and Lau]{jentzsch1998efficiency}
Jentzsch,~T.; Juepner,~H.; Brzezinka,~K.-W.; Lau,~A. Efficiency of optical second harmonic generation from pentacene films of different morphology and structure. \emph{Thin solid films} \textbf{1998}, \emph{315}, 273--280\relax
\mciteBstWouldAddEndPuncttrue
\mciteSetBstMidEndSepPunct{\mcitedefaultmidpunct}
{\mcitedefaultendpunct}{\mcitedefaultseppunct}\relax
\EndOfBibitem
\bibitem[Tölle and Kin-Lic~Chan(2023)Tölle, and Kin-Lic~Chan]{tolle2023exact}
Tölle,~J.; Kin-Lic~Chan,~G. Exact relationships between the {GW} approximation and equation-of-motion coupled-cluster theories through the quasi-boson formalism. \emph{J. Chem. Phys.} \textbf{2023}, \emph{158}\relax
\mciteBstWouldAddEndPuncttrue
\mciteSetBstMidEndSepPunct{\mcitedefaultmidpunct}
{\mcitedefaultendpunct}{\mcitedefaultseppunct}\relax
\EndOfBibitem
\bibitem[Lange and Berkelbach(2018)Lange, and Berkelbach]{lange2018relation}
Lange,~M.~F.; Berkelbach,~T.~C. On the relation between equation-of-motion coupled-cluster theory and the {GW} approximation. \emph{J. Chem. Theory Comput.} \textbf{2018}, \emph{14}, 4224--4236\relax
\mciteBstWouldAddEndPuncttrue
\mciteSetBstMidEndSepPunct{\mcitedefaultmidpunct}
{\mcitedefaultendpunct}{\mcitedefaultseppunct}\relax
\EndOfBibitem
\bibitem[Quintero-Monsebaiz \latin{et~al.}(2022)Quintero-Monsebaiz, Monino, Marie, and Loos]{quintero2022connections}
Quintero-Monsebaiz,~R.; Monino,~E.; Marie,~A.; Loos,~P.-F. Connections between many-body perturbation and coupled-cluster theories. \emph{J. Chem. Phys.} \textbf{2022}, \emph{157}\relax
\mciteBstWouldAddEndPuncttrue
\mciteSetBstMidEndSepPunct{\mcitedefaultmidpunct}
{\mcitedefaultendpunct}{\mcitedefaultseppunct}\relax
\EndOfBibitem
\end{mcitethebibliography}
\end{document}